\documentclass[letterpaper, 11pt]{article}
\pdfoutput=1
\usepackage{jheppub}

\usepackage{graphicx}
\usepackage{amsmath, amssymb}
\usepackage{amsfonts}
\usepackage{amsbsy}
\usepackage{bbding}
\usepackage{wasysym}
\usepackage{stmaryrd}
\usepackage{bm}
\usepackage{verbatim}

\usepackage{tikz}
\usepackage{caption}
\usepackage{subcaption}
\usepackage{rotating}

\usepackage{scalefnt}
\usepackage{youngtab}  
\usepackage{enumerate}
\usetikzlibrary{shapes.geometric}
\usetikzlibrary{calc,shadows}

\makeatletter

\newcommand{\Rmnum}[1]{\expandafter\@slowromancap\romannumeral #1@}
\makeatother

\newcommand{\ba}{\begin{align}}
\newcommand{\ea}{\end{align}}
\newcommand{\bea}{\begin{eqnarray}}
\newcommand{\eea}{\end{eqnarray}}
\newcommand{\be}{\begin{eqnarray}}
\newcommand{\ee}{\end{eqnarray}}
\newcommand{\nn}{\nonumber}
\newcommand{\bn}{\begin{enumerate}}
\newcommand{\en}{\end{enumerate}}

\def\half{\frac{1}{2}}

\def\IZ{\mathbb{Z}}

\def\CC{{\cal C}}

\def\CF{{\cal F}}

\def\CI{{\cal I}}

\def\CL{{\cal L}}
\def\CM{{\cal M}}
\def\CN{{\cal N}}

\def\CS{{\cal S}}
\def\CT{{\cal T}}
\def\CU{{\cal U}}

\def\a{\alpha}
\def\b{\beta}

\def\s{\sigma}

\def\Tr{{\rm Tr}}
\def\tr{{\rm tr}}

\def\vev#1{\langle #1 \rangle}
\def\vec#1{\bm{#1}}

\title{New N=1 Dualities from M5-branes and Outer-automorphism Twists}

\author{Prarit Agarwal}
\author{and Jaewon Song}

\affiliation{Department of Physics\\ University of California, San Diego \\La Jolla, CA 92093, USA}

\emailAdd{pagarwal@physics.ucsd.edu}
\emailAdd{jsong@physics.ucsd.edu}

\abstract
{
We generalize recent construction of four-dimensional $\CN=1$ SCFT from wrapping six-dimensional $\CN=(2, 0)$ theory on a Riemann surface to the case of $D$-type with outer-automorphism twists. This construction allows us to build various dual theories for a class of $\CN=1$ quiver theories of $SO$-$USp$ type. In particular, we find there are five dual frames to $SO(2N)/USp(2N-2)/G_2$ gauge theories with $(4N-4)/(4N)/8$ fundamental flavors, where three of them being non-Lagrangian. We check the dualities by computing the anomaly coefficients and the superconformal indices. In the process we verify that the index of $D_4$ theory on a certain three punctured sphere with $\IZ_2$ and $\IZ_3$ twist lines exhibits expected symmetry enhancement from $G_2 \times USp(6)$ to $E_7$. 
}

\preprint{
UCSD-PTH-13-13
}

\begin{document}
\maketitle
\flushbottom

\section{Introduction}
Supersymmetric gauge theories have been extremely fruitful in our endeavor to uncover the rich structure of quantum field theory. One of the most remarkable phenomenon discovered in supersymmetric gauge theory is Seiberg duality \cite{Seiberg:1994pq} where two different UV gauge theories flow to the same fixed-point in the IR . The original example studied by Seiberg was $\CN=1$ SQCD with $SU(N)$ gauge group, which was subsequently generalized to $SO(N)$ gauge groups by Intriligator-Seiberg \cite{Intriligator:1995id} and to $USp(2N)$ gauge groups\footnote{In this paper we use the notation $USp(2N) = C_N$ for the symplectic groups so that $USp(2)=SU(2)$.} by Intriligator-Pouliot \cite{Intriligator:1995ne}. 

Recently, a new dual description to $SU(N)$ SQCD has been found by Gadde-Maruyoshi-Tachikawa-Yan (GMTY) \cite{Gadde:2013fma}. Their new dual theory involves coupling two copies of the so-called $T_N$ theory. The new theory can be thought of as a generalization of the (multiple) self-duality of Csaki-Schmaltz-Skiba-Terning \cite{Csaki:1997cu} from $SU(2)$ to $SU(N)$. The main component they used was the $T_N$ theory which arises from wrapping $N$ coincident M5-branes or $A_{N-1}$ six-dimensional $\CN=(2, 0)$ theory on a 3-punctured sphere \cite{Gaiotto:2009we}. 

One of the objectives of this paper is to generalize the GMTY duality to the $SO/USp$ theories thereby adding more dual theories in addition to the ones found in \cite{Intriligator:1995id,Intriligator:1995ne}. Moreover, we will show that there is not just one new dual theory but three more dual descriptions to each theory. From this, we argue there are five different theories in the UV that flow to the same superconformal theory in the IR. 

We also find new dual theories for the $G_2$ gauge theory with 8 fundamentals. $G_2$ is the simplest group with a trivial center and hence QCD with a $G_2$ gauge group provide us with an opportunity to study the role of the center of a gauge group in confinement \cite{Holland:2003jy}. A dual for $G_2$ QCD with $5$ flavors was discussed in  \cite{Giddings:1995ns, Pesando:1995bq} while for $5 < N_f < 12$, a magnetic theory with an $SU(N_f -3)$ gauge group was found by Pouliot \cite{Pouliot:1995zc}.  The duality frames discovered in this paper are either non-Lagrangian or based on $Spin(8)$ gauge group and hence constitute a new class of magnetic theories.

Two dual frames among five have Lagrangian descriptions. The `electric theory' $\CU$ is the original SQCD with certain number of flavors and the `magnetic theory' $\CU_{c1}$ is also an SQCD with the same number of flavors\footnote{Except for the $G_2$ case where both the gauge group and matter contents changed.} but also has mesons coupled through a superpotential. 
Three non-Lagrangian dual theories can be categorized into `swap' theories $\CU_s$ following the nomenclature of \cite{Gadde:2013fma}, and Argyres-Seiberg type $\CU_{as}$ since it can be thought of as $\CN=1$ version of the dualities found in \cite{Argyres:2007cn}, and the crossing type $\CU_{c2}$. 

Our discussion is motivated from the six-dimensional construction of $\CN=1$ superconformal field theories. It is an extension of the so-called the $\CN=2$ theories of class $\CS$ \cite{Gaiotto:2009hg, Gaiotto:2009we}. A class $\CS$ theory is constructed by compactifying the six-dimensional $\CN=(2, 0)$ theory of type $\Gamma = A, D, E$ on a Riemann surface $\CC$ with a partial topological twist. This gives rise to $\CN=2$ theory in 4-dimensions labeled by $\CC$ called the UV curve. Since any (negatively curved) Riemann surface can be decomposed in terms of pair of pants or 3-punctured sphere, it is natural to associate a 4d theory to a 3-punctured sphere and regard it as a building block for the 4d theory. The 4-dimensional theory associated to $\CC$ has to be the same regardless of how we decompose the Riemann surface. The statement of duality is equivalent to saying that the different pair-of-pants decompositions give rise to the same 4-dimensional theory. 

In order to write down various dual theories, one needs to identify the theory corresponding to the various different types of three punctured spheres. This has been extensively studied, for example in \cite{Chacaltana:2010ks, Chacaltana:2011ze, Chacaltana:2012ch, Chacaltana:2013oka}, from which they find new $\CN=2$ SCFTs and dualities. The class $\CS$ construction for the $D_N$ type was first studied in \cite{Tachikawa:2009rb} and the effect of outer-automorphism twists has been studied in \cite{Tachikawa:2010vg}. 

One can generalize this construction to $\CN=1$ theory. The simplest way is to give mass to the chiral adjoints in the $\CN=2$ vector multiplets. In the IR, the massive chiral adjoints will be integrated out and we land on a SCFT \cite{Maruyoshi:2009uk, Benini:2009mz}. One can construct more general theories by requiring non-baryonic $U(1)_\CF$ to be conserved. This gives rise to a new class of $\CN=1$ SCFTs generically not the same as the mass-deformed $\CN=2$ theories in class $\CS$ \cite{Bah:2011je, Bah:2012dg}. This class of theories are subsequently generalized to include the Riemann surface with punctures in \cite{Gadde:2013fma, Xie:2013gma, Bah:2013aha} so that the theory can have larger global symmetries. Further studies of $\CN=1$ class $\CS$ theories have been done in \cite{Maruyoshi:2013hja, Bonelli:2013pva, Xie:2013rsa, Yonekura:2013mya, Bah:2013qya}. In this paper, we generalize this construction to the case of $\Gamma = D_N$ series with outer-automorphism twists. 

This construction requires extra data beyond the choice of the Riemann surface, namely the degree of the normal bundles $\CL(p) \oplus \CL(q) \to \CC_{g, n}$ with $p+q=2g-2+n$. This stems from the fact that we have one parameter ways to twist the 6d $\CN=(2, 0)$ theory while preserving $\CN=1$ SUSY in 4-dimensions.\footnote{For the purpose of preserving supersymmetry, the rank 2 bundle $E \to \CC_{g, n}$ is not necessarily given by a sum of two line bundles. The only necessary condition is to have $\textrm{det} E$ equal to the canonical bundle $K_{\CC_{g, n}}$. But here we restrict ourselves to the case where the rank 2 bundle is given by a direct sum.} 
The punctures also have to be more general than the $\CN=2$ counterpart. In our case, we put $\IZ_2$ valued `color' to the punctures in addition to the usual $\CN=2$ data. In order to realize SQCD of gauge group $SO(2N)/USp(2N-2)$ with $(4N-4)/(4N)$ fundamental quarks,\footnote{The number of flavors here is counted by the number of chiral multiplets. This is in contrast with the $SU(N)$ theory, which has both quarks and anti-quarks.} we put the $\Gamma=D_N$ theory on a 4-punctured sphere with two twisted full punctures with each color and two twisted null punctures with each color and choose the normal bundle to be $(p, q)=(1, 1)$. For the case of $G_2$ theory, start with $\Gamma=D_4$ with 4 punctures, but also with $\IZ_3$ twist line running between two $USp(4)$ punctures of each color. We also need two twisted null punctures of each color as well. 
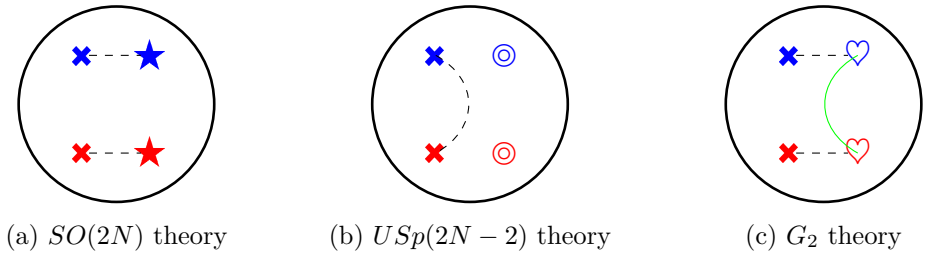
\begin{figure}[h]
\centering
\begin{subfigure}[b]{1.8in}
\centering
\begin{tikzpicture}[scale=1.3, every node/.style={transform shape}]
\draw[line width=1] (0, 0) circle (1.0);
\draw[text = blue,font=\small ] (0.35 ,0.5) node {\FiveStar};
\draw[text = red,font=\small] (0.35 ,-0.5) node {\FiveStar};
\draw[text = blue,font=\tiny ] (-0.35 ,0.5) node {\XSolidBold};
\draw[text = red,font=\tiny ] (-0.35 ,-0.5) node {\XSolidBold};
\draw[dashed] (-0.28,0.5) -- (0.25,0.5);
\draw[dashed] (-0.28,-0.5) -- (0.25,-0.5);
\end{tikzpicture}
\caption{$SO(2N)$ theory}
\end{subfigure}
\begin{subfigure}[b]{1.8in}
\centering
\begin{tikzpicture}[scale=1.3, every node/.style={transform shape}]
\draw[line width=1] (0, 0) circle (1.0);
\draw[text = blue,font=\small ] (0.35 ,0.5) node {$\varocircle$};
\draw[text = red,font=\small] (0.35 ,-0.5) node {$\varocircle$};
\draw[text = blue,font=\tiny ] (-0.35 ,0.5) node {\XSolidBold};
\draw[text = red,font=\tiny ] (-0.35 ,-0.5) node {\XSolidBold};
\draw[dashed] (-0.35,0.5)  .. controls (0.1,0.25) and (0.1,-0.25) .. (-0.35 ,-0.5);
\end{tikzpicture}
\caption{$USp(2N-2)$ theory}
\end{subfigure}
\begin{subfigure}[b]{1.8in}
\centering
\begin{tikzpicture}[scale=1.3, every node/.style={transform shape}]
\draw[line width=1] (0, 0) circle (1.0);
\draw[text = blue,font=\small ] (0.35 ,0.5) node {$\heartsuit$};
\draw[text = red,font=\small] (0.35 ,-0.5) node {$\heartsuit$};
\draw[text = blue,font=\tiny ] (-0.35 ,0.5) node {\XSolidBold};
\draw[text = red,font=\tiny ] (-0.35 ,-0.5) node {\XSolidBold};
\draw[dashed] (-0.28,0.5) -- (0.25,0.5);
\draw[dashed] (-0.28,-0.5) -- (0.25,-0.5);
\draw[draw=green] (0.35,0.5)  .. controls (-0.1,0.25) and (-0.1,-0.25) .. (0.35 ,-0.5);
\end{tikzpicture}
\caption{$G_2$ theory}
\end{subfigure}
\caption{The UV curves realizing SQCDs in this paper. The symbol {\tiny{\XSolidBold}} denotes twisted null puncture, $\varocircle$ the full puncture having $SO(2N)$ flavor symmetry, {\small{\FiveStar}} the twisted full puncture having $USp(2N-2)$ flavor symmetry and $\heartsuit$ denotes $USp(4)$ puncture. The dashed line and the green solid line denote $\IZ_2$ and $\IZ_3$ twist line respectively.}
\end{figure}

The notion of pair-of-pants decomposition needs an extra ingredient because of the normal bundles. It can be realized by putting colors to the pair-of-pants itself. It turns out there are five different colored pair-of-pants decompositions for our setup, thereby giving five dual frames to the SQCD.\footnote{Actually there is one more in terms of colored pair-of-pants decomposition, but it is identical to one of five upon inverting the color.} The list of dual theories we find are summarized as follows. 
For the $SO(2N)$ theory, the five dual frames are:
\begin{itemize}
\item $\CU^{SO}$: $SO(2N)$ with $4N-4$ fundamentals (vectors)
\item $\CU^{SO}_{c1}$: $SO(2N)$ with $4N-4$ fundamentals and mesons \cite{Intriligator:1995id}
\item Three non-Lagrangian duals: $\CU^{SO}_{s}$, $\CU^{SO}_{as}$, $\CU^{SO}_{c2}$
\end{itemize}
and for the $USp(2N-2)$ theory:
\begin{itemize}
\item $\CU^{Sp}$: $USp(2N-2)$ with $4N$ fundamentals
\item $\CU^{Sp}_{c1}$: $USp(2N-2)$ with $4N$ fundamentals and mesons \cite{Intriligator:1995ne}
\item Three non-Lagrangian duals: $\CU^{Sp}_{s}$, $\CU^{Sp}_{as}$, $\CU^{Sp}_{c2}$ 
\end{itemize}
and for the $G_2$ theory: 
\begin{itemize}
\item $\CU^{G_2}$: $G_2$ with $8$ fundamentals
\item $\CU^{G_2}_{c1}$: $Spin(8)$ with 6 quarks in $8_V$ and $8_S$ and mesons
\item Three non-Lagrangian duals: $\CU^{G_2}_{s}$, $\CU^{G_2}_{as}$, $\CU^{G_2}_{c2}$ 
\end{itemize}
Three out of five dual theories are non-Lagrangian. We will explain these non-Lagrangian duals in detail in later sections.

We provide evidence to these dualities through computing the anomaly coefficients and the superconformal indices. 
In order to compute the superconformal index of $G_2$ theory, we also discuss $\CN=2$ index with $\IZ_3$ twist line and $G_2$ puncture. Especially, we find that the theory with $UV$ curve given by three punctured sphere with $USp(6)$, $G_2$ and twisted null punctures has enhanced $E_7$ flavor symmetry as expected in \cite{Tachikawa:2010vg} where it was identified as the theory of Minahan-Nemeschansky \cite{Minahan:1996cj}. 

The paper is organized as follows. In section \ref{sec:N1M5}, we review construction of the $\CN=1$ theories of class $\CS$ from which we construct our dual theories. We will also discuss the effect of outer-automorphism twist in the setup. In section \ref{sec:SO}, we propose dualities of $SO(2N)$ gauge theories and check the 't Hooft anomaly coefficients. In section \ref{sec:Sp}, we discuss the dualities of $USp(2N-2)$ gauge theories. In the section \ref{sec:G2}, we discuss the dualities of $G_2$ gauge theory. Finally, in section \ref{sec:index}, we check our duality proposals by computing the superconformal index. In the appendix, we derive certain chiral ring relations for the $T_{SO(2N)}$ and the twisted $\tilde{T}_{SO(2N)}$ blocks, which are necessary in other sections. 

\section{Constructing $\CN=1$ theory from M5-branes} \label{sec:N1M5}
In this section, we review the construction of 4d $\CN=1$ theories from 6d perspective due to \cite{Maruyoshi:2009uk, Benini:2009mz, Bah:2011je, Bah:2012dg, Gadde:2013fma, Xie:2013gma, Bah:2013aha}. From this, we propose several dual theories based on different ways of gluing the 3-punctured spheres.  

\subsubsection*{Setup and Data}
In order to obtain an $\CN=1$ SCFT from M5-branes dubbed the theories of class $\CS$, we need the following data: 
\begin{itemize}
\item Choice of the `gauge' group $\Gamma = A_n, D_n, E_{6, 7, 8}$. 
\item Riemann surface $\CC_{g, n}$ of genus $g$ and $n$ punctures. We call it a UV-curve. 
\item Choice of two normal bundles $\CL_1(p), \CL_2(q)$ of degree $p, q$ over $\CC_{g, n}$ such that $p+q = 2g-2+n$. 
\item The choice of appropriate boundary condition on each punctures. 
\end{itemize}
The choice of $\Gamma$ labels the 6-dimensional $\CN=(2, 0)$ theory and we compactify the 6d theory on $\CC_{g, n}$ to obtain the $\CN=1$ theory in 4-dimension. When compactifying the theory, we have to perform partial topological twist in order to preserve any supersymmetry. It turns out that there is an integer parameter family of different ways to twist the theory while preserving 4 supercharges. This can be understood as the choice of the normal bundles $\CL_1 (p) \oplus \CL_2(q) \to \CC_{g, n}$. The total space of this rank-2 bundle becomes Calabi-Yau 3-fold if it satisfies $p+q = 2g-2+n$. 

The data on a puncture is specified by the following conditions which are all equivalent: 
\begin{itemize}
\item $\frac{1}{4}$-BPS boundary condition of $\CN=4, d=4$ SYM theory.
\item Choice of the singular boundary condition of a generalized Hitchin equation on $\CC_{g, n}$
	\be \label{eq:ghitchin}
	&{ }& D_{\bar{z}} \Phi_{1} = D_{\bar{z}} \Phi_{2} = 0 \ , \nn \\
	&{ }& [ \Phi_1 , \Phi_2 ]  = 0 \ , \\
	&{ }& F_{z\bar{z}} + [ \Phi_1, \Phi_1^* ] + [ \Phi_2, \Phi_2^* ] = 0 \nn \ . 
	\ee
\item Choice of the singular boundary condition of a generalized Nahm's equation. 
\end{itemize}
When one of $p$ or $q$ is zero, then we go back to the $\CN=2$ theories of class $\CS$ \cite{Gaiotto:2009hg, Gaiotto:2009we}. In this case, the data on the puncture is specified by a $\half$-BPS boundary condition of $\CN=4, d=4$ SYM theory, or the embedding of $SU(2)$ group to $\Gamma$. Equivalently, one of the Higgs field $\Phi_{1, 2}$ vanishes and we get the ordinary Hitchin equation. When $\Gamma = A_{n-1}$ it is labeled by a Young tableau with $n$ boxes. 
\begin{figure}[h]
	\centering
	\includegraphics[width=2.3in]{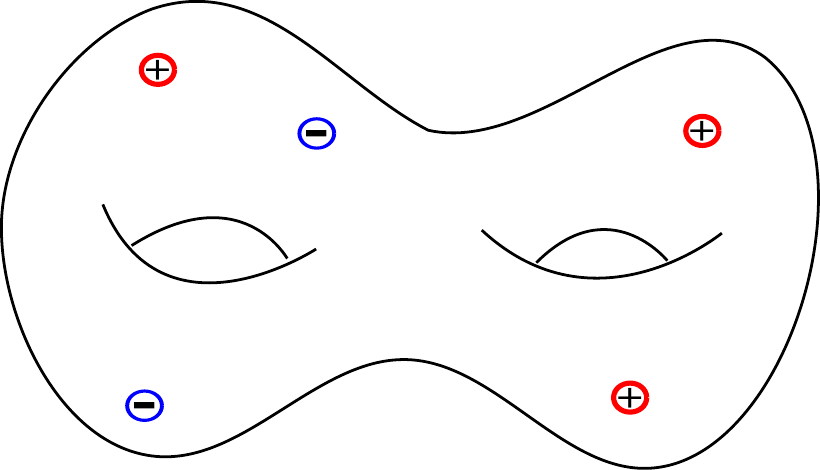}
	\caption{A choice of UV curve with colored punctures. Here we suppressed the labeling $\rho$ for each punctures.}
\end{figure}

\subsubsection*{Colored $\CN=2$ punctures}
Generally, $\CN=1$ puncture will involve both $\Phi_1$ and $\Phi_2$ in \eqref{eq:ghitchin} developing singularities at the same point where the punctures sits. Throughout the paper we restrict ourselves to the case where only one of them develops a singularity at a given point. In a sense this makes our system $\CN=2$-like near the puncture. We will label each puncture by a color $\s = \pm$ along with the choice of embedding $\rho: SU(2) \to \Gamma$. We will call them as colored $\CN=2$ punctures. 

When the group $\Gamma$ admits an outer-automorphism (when the corresponding Dynkin diagram is symmetric under a discrete action $o$), we can twist the punctures accordingly \cite{Tachikawa:2009rb, Tachikawa:2010vg, Chacaltana:2012zy}. 
\begin{table}[h] 
\begin{center}
\begin{tabular}{c|ccccc}
	$\Gamma$ & 	$A_{2n-1}$ & 	$A_{2n}$ & 	$D_{n+1}$ & 	$D_4$ & 	$E_6$ \\
	\hline
	$o$ &		$\IZ_2$ & 		$\IZ_2$ &		$\IZ_2$ &		$\IZ_3$ &	$\IZ_2$ \\
	$G$	&		$B_n$ &		$C_n$ &	 	$C_n$ & 		$G_2$ &	$F_4$ \\
	$G^\vee$  &	$C_n$ &		$B_n$ &	 	$C_n$ & 		$G_2$ &	$F_4$
\end{tabular}
\end{center}
\caption{The group $\Gamma$ changes to $G$ under the outer-automorphism twist $o$. It is given by the Langlands-dual of the $G^\vee$ which is the subgroup of $\Gamma$ invariant under $o$. }
\label{table:GammaG}
\end{table}
Once we twist the puncture, the punctures are no longer labeled by the $SU(2)$ embedding into $\Gamma$ but into $G$, see Table \ref{table:GammaG}. 
Another thing to notice here is that the number of twisted punctures cannot be arbitrary, but is required to be such that the product of monodromies around the punctures should be equal to one.\footnote{We thank Yuji Tachikawa for bringing this to our attention.} For example, we need to have even numbers of $\IZ_2$-twisted punctures. In the case with $\IZ_3$ punctures, we could also have odd number of $\IZ_3$-twisted punctures as in the figure \ref{fig:E7}.
\begin{figure}[h]
	\begin{center}
	\begin{subfigure}[b]{2.8in}
		\centering
		\includegraphics[width=1.8in]{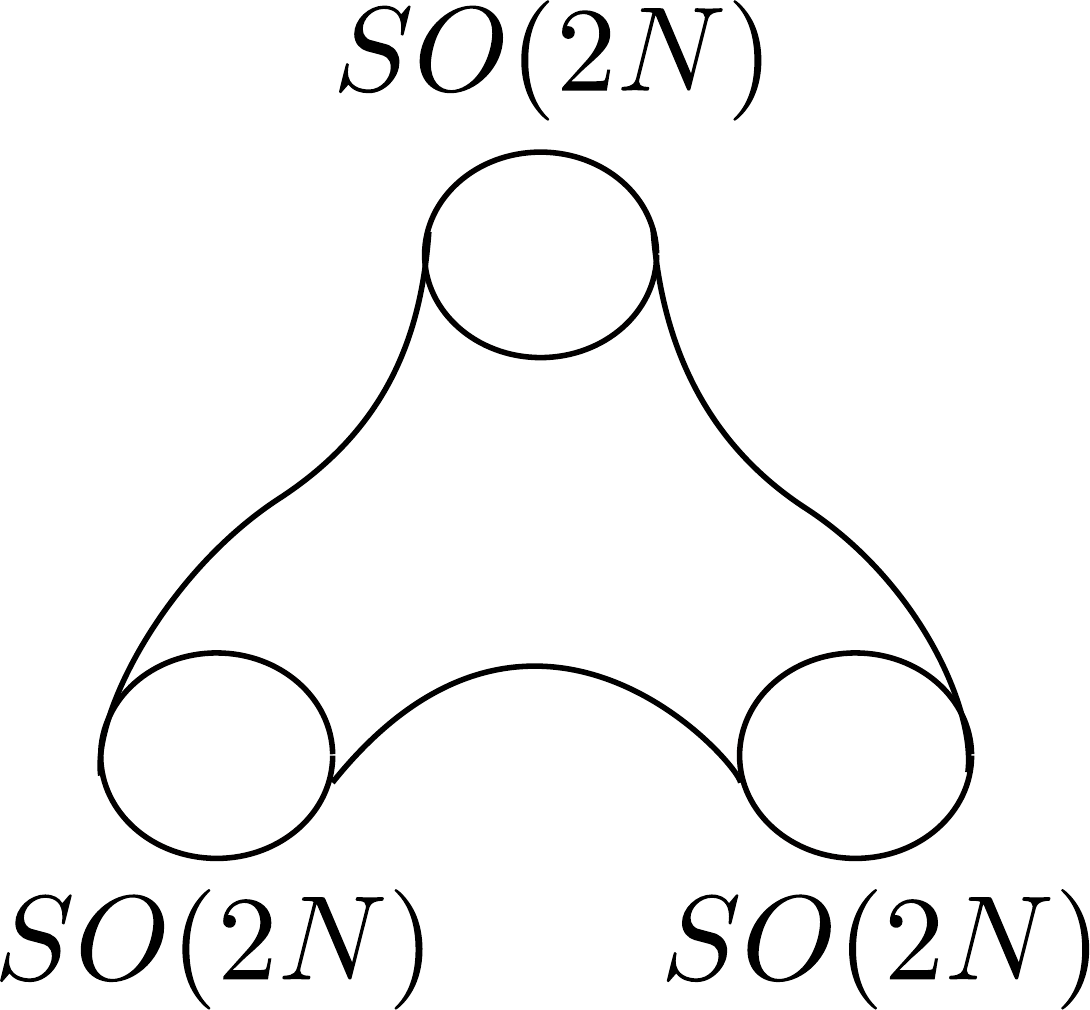}
		\caption{$D_N$ theory on sphere with 3 untwisted punctures}
	\end{subfigure}
	\quad \quad
	\begin{subfigure}[b]{2.8in}
		\centering
		\includegraphics[width=2.15in]{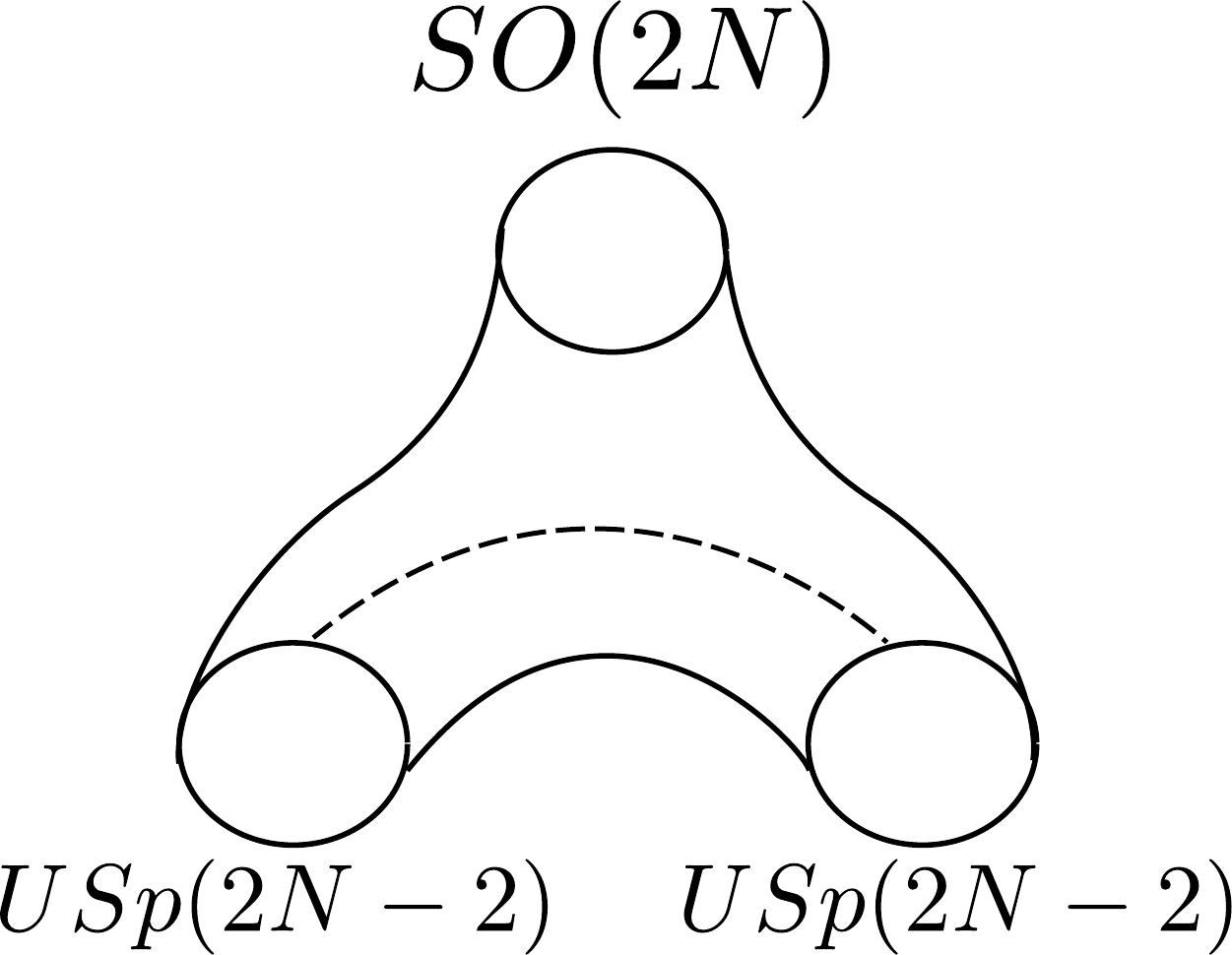}
		\caption{$D_N$ theory on sphere with 1 untwisted and 2 twisted punctures}
	\end{subfigure}
	\caption{By twisting the punctures of $D_N$ theory, we get twisted punctures having the $C_{N-1}$ flavor symmetry.}
	\end{center}
\end{figure}

\subsubsection*{Colored pair-of-pants decomposition}
\begin{figure}[h]
\centering
\includegraphics[width=4.5in]{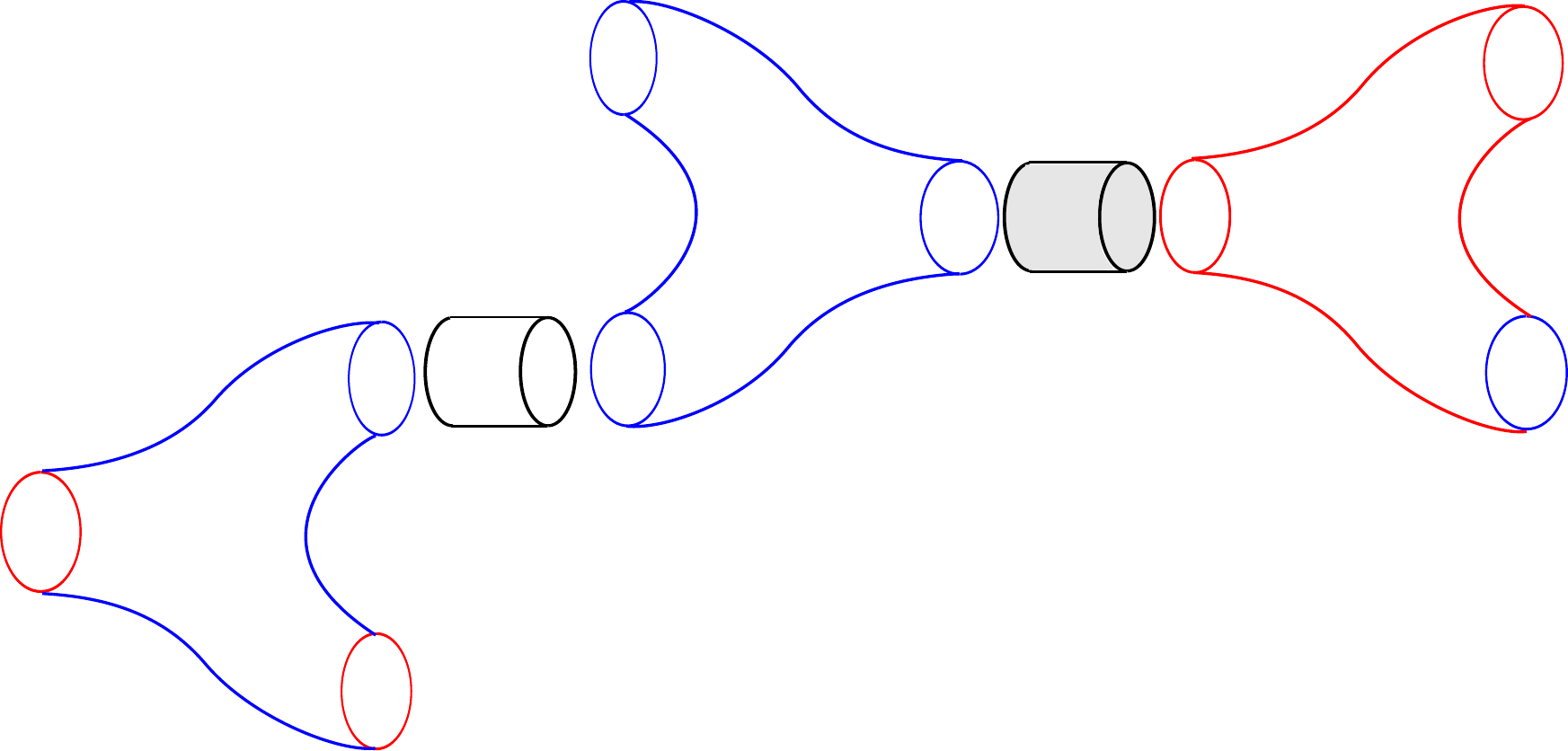}
\caption{An example of colored pair-of-pants decomposition. Here red/blue means $\s=\pm$ respectively. Three red punctures and two blue punctures with $p=1, q=2$. Grey tube denotes $\CN=1$ vector, white tube denotes $\CN=2$ vector multiplet. We have 3 mesons associated to the blue puncture on the right and two red punctures on the left.}
\label{fig:pDecompEx}
\end{figure}

For a given such configuration, we can have various different dual frames by considering different pair-of-pants decompositions. On each pair of pants, we also label it by a color $\s=\pm$. The number of the pair-of-pants labelled by $+$ and by $-$ are given by the degree of line bundles $p$ and $q$ respectively. Now, for a given pair of pants, we have the following data:
\bn
\item	The choice of color $\s^p$ of the pair of pants itself. 
\item $(\rho^p_i, \s^p_i)$ where $\rho^p_i : SU(2) \to G$ labels the $SU(2)$ embedding in $G$ and $\s^p_i$ denotes a coloring for each punctures $i=1, 2, 3$. 
\en
When we glue two pair of pants, we gauge the flavor symmetry associated to punctures we glue. When the $\s^p$ of two pair of pants are the same, we gauge it using the $\CN=2$ vector multiplets, and when the $\s^p$ are different, we glue it through $\CN=1$ vector multiplet. Note that when we glue two punctures, we can always choose the coloring of the punctures as the same as the pair of pants that we are gluing. See figure \ref{fig:pDecompEx} for an illustration of the construction. 

Now, for a given colored pair of pants with color $\s^p$, we identify the building block as follows. When all the punctures have the same color as the pair of pants itself, we identify the theory as the same one as $\CN=2$ theory. For example, when all the punctures are (untwisted) full punctures, then we get $T_{\Gamma}$ theory.\footnote{For $\Gamma=A_{N-1}$, it is usually called as $T_N$ theory.} When a full puncture has a different sign from the pair of pants, we add a `meson' field that transforms as an adjoint of $\Gamma$ associated to the puncture. Moreover we add a superpotential term for the meson field: $W= \Tr (M \mu)$, where $\mu$ is an operator  associated to the puncture. The operator $\mu$ transforms under the adjoint representation of $\Gamma$ and has the conformal weight $\Delta = 2$. 

For a theory in class $\CS$, we have $U(1)_\CF$ global symmetry in addition to the $\CN=1$ superconformal symmetry and the global symmetry labeled by the punctures. Suppose we have only maximal punctures meaning $\rho$ is given by the trivial embedding and has the full global symmetry $G$. 
We define the $U(1)_\CF$ global charge to be
\be \label{eq:U1Fdef}
 \CF = \sum_i \s_i J_i \ , 
\ee
where $J_i$ are the global $U(1)$ charge at each pair of pants. Note that each pair-of-pants or three punctured sphere describes $\CN=2$ superconformal theory. It has $SU(2)_R \times U(1)_R$ R-symmetry which is broken down to $U(1)_R \times U(1)_{J_i}$ upon coupling to $\CN=1$ vectors. The coloring $\s_i$ labels the choice of the sign of $U(1)_{J_i}$ charge we can make.  

The color $\s_i^p$ of a puncture tells us the charges of the operator $\mu_i^p$. We assign $U(1)_\CF$ of $\mu$ to be $2\s_i^p$. When we have a meson field $M_i^p$, the $U(1)_\CF$ charge for the $\mu_i^p$ is reversed to $-2\s_i^p$ and the meson has charge $2\s_i^p$ instead. In addition to the operators corresponding to the punctures, we also have `internal' operators $\mu_i$ associated to the punctures glued via cylinders in the pair-of-pants decompositions. The $U(1)_\CF$ charge for $\mu_i$ is given by $2 \s_i$.  When the gluing is done through $\CN=2$ vectors, we also have an adjoint chiral multiplet $\phi$. The $U(1)_\CF$ charge for $\phi$ is $-2 \s_i$, so that the $\CN=2$ superpotential term $W = \Tr (\phi \mu + \phi \tilde{\mu} )$ preserves the $U(1)_\CF$ where $\tilde{\mu}$ is the operator corresponds to the other glued puncture. For the $\CN=1$ gluing, we can have a superpotential term $W = \Tr (\mu \tilde{\mu})$ which is exactly marginal.  

This global symmetry is not anomalous and in general not baryonic. The true R-charge in the IR will mix with $U(1)_\CF$ charge in the UV. Therefore one needs to perform a-maximization \cite{Intriligator:2003jj} to obtain the correct $R$-charge. 

\subsubsection*{Non-maximal punctures via Higgsing}
If the labeling of the punctures $\rho$ is non-maximal, we `Higgs' a maximal puncture down to a non-maximal one in the following ways: For the puncture with the same color as the color of the pair-of-pants $\s_p$, we give vev to the moment map $\vev{\mu} = \rho(\sigma^+)$, and for the puncture with different color $\s_p$, we give vev to the meson $\vev{M} = \rho(\sigma^+)$ where $\rho$ is the embedding of $SU(2)$ into $G$ which labels the puncture itself. For the latter case, this yields the superpotential $W = \tr \rho(\s^+) \mu + \tr M' \mu'$ where $\mu'$ are the components of $\mu$ which commute with $\rho^T$ and $M'$ are the mesonic fluctuations around its vev. Higgsing breaks the global symmetry from $G$ down to the commutant $G_F$ of $\rho(SU(2))$ in $G$. 

When some of the punctures are non-maximal, it shifts the $U(1)_\CF$ of \eqref{eq:U1Fdef} by a certain amount, if the color of the puncture is different from the pair-of-pants. The shifted $U(1)_\CF$ is given by 
\be
 \CF = \sum_i \left( \s_i J_i  + 2 \sum_{p, \s^p_i = -\s_i} \s^p_i \rho_i^p (\s_3) \right) \ ,
\ee  
where $(\rho_i^p, \s_i^p)$ labels the punctures and their colors.

\subsubsection*{$\CN=1$ Dualities from colored pair-of-pants decompositions} 
As we discussed above, the different pair of pants decomposition describes different dual frames. Additional ingredient here is the assignment of color $\s_i^p$ for each pair of pants. This adds another choices on the top of the pair of pants decomposition and it makes the duality structure richer than the $\CN=2$ counterpart. We call it colored pair-of-pants decomposition. 

\begin{figure}[h]
\centering
\begin{subfigure}[b]{1.0in}
\centering
\begin{tikzpicture}[scale=1.3, every node/.style={transform shape}]
\draw[color=blue, line width=1] (0,0) arc (-60:240:0.6cm);
\draw (0.1,0.3) .. controls (-0.07,0) and (-0.07,-0.5) .. (0.1,-0.8);
\draw (-0.7,0.3) .. controls (-0.53,0) and (-0.53,-0.5) .. (-0.7,-0.8);
\draw[color=red, line width=1] (0,-0.5) arc (60:-240:0.6cm);
\draw[text = blue,font=\small ] (0.05 ,0.7) node {\FiveStar};
\draw[text = red,font=\small] (0.05 ,-1.2) node {\FiveStar};
\draw[text = blue,font=\tiny ] (-0.65 ,0.5) node {\XSolidBold};
\draw[text = red,font=\tiny ] (-0.65 ,-1) node {\XSolidBold};
\draw[dashed] (-0.65,0.5) -- (0.05,0.7);
\draw[dashed] (-0.65,-1) -- (0.05,-1.2);
\draw[text = black,font=\tiny ] (0.05 ,0.45) node {A};
\draw[text = black,font=\tiny ] (0.0 ,-0.90) node {B};
\end{tikzpicture}
\caption{$\CU^{SO}$}
\end{subfigure}
\begin{subfigure}[b]{1.0in}
\centering
\begin{tikzpicture}[scale=1.3, every node/.style={transform shape}]
\draw[color=blue, line width=1] (0,0) arc (-60:240:0.6cm);
\draw (0.1,0.3) .. controls (-0.07,0) and (-0.07,-0.5) .. (0.1,-0.8);
\draw (-0.7,0.3) .. controls (-0.53,0) and (-0.53,-0.5) .. (-0.7,-0.8);
\draw[color=red, line width=1] (0,-0.5) arc (60:-240:0.6cm);
\draw[text = black,font=\small, color=red ] (0.05 ,0.7) node {\FiveStar};
\draw[text = black,font=\tiny ] (0.05 ,0.45) node {B};
\draw[text = black,font=\small, color=blue ] (0.05 ,-1.2) node {\FiveStar};
\draw[text = black,font=\tiny ] (0.0 ,-0.90) node {A};
\draw[text = black,font=\tiny, color=blue ] (-0.65 ,0.5) node {\XSolidBold};
\draw[text = black,font=\tiny, color=red ] (-0.65 ,-1) node {\XSolidBold};
\draw[dashed] (-0.65,0.5) -- (0.05,0.7);
\draw[dashed] (-0.65,-1) -- (0.05,-1.2);
\end{tikzpicture}
\caption{$\CU^{SO}_{c1}$}
\end{subfigure}
\begin{subfigure}[b]{1.0in}
\centering
\begin{tikzpicture}[scale=1.3, every node/.style={transform shape}]
\draw[color=red, line width=1] (0,0) arc (-60:240:0.6cm);
\draw (0.1,0.3) .. controls (-0.07,0) and (-0.07,-0.5) .. (0.1,-0.8);
\draw (-0.7,0.3) .. controls (-0.53,0) and (-0.53,-0.5) .. (-0.7,-0.8);
\draw[color=blue, line width=1] (0,-0.5) arc (60:-240:0.6cm);
\draw[text = blue,font=\small ] (0.05 ,0.7) node {\FiveStar};
\draw[text = red,font=\small] (0.05 ,-1.2) node {\FiveStar};
\draw[text = blue,font=\tiny ] (-0.65 ,0.5) node {\XSolidBold};
\draw[text = red,font=\tiny ] (-0.65 ,-1) node {\XSolidBold};
\draw[dashed] (-0.65,0.5) -- (0.05,0.7);
\draw[dashed] (-0.65,-1) -- (0.05,-1.2);
\draw[text = black,font=\tiny ] (0.05 ,0.45) node {A};
\draw[text = black,font=\tiny ] (0.0 ,-0.90) node {B};
\end{tikzpicture}
\caption{$\CU^{SO}_{s}$}
\end{subfigure}
\begin{subfigure}[b]{1.0in}
\centering
\begin{tikzpicture}[scale=1.3, every node/.style={transform shape}]
\draw[color=blue, line width=1] (0,0) arc (-60:240:0.6cm);
\draw (0.1,0.3) .. controls (-0.07,0) and (-0.07,-0.5) .. (0.1,-0.8);
\draw (-0.7,0.3) .. controls (-0.53,0) and (-0.53,-0.5) .. (-0.7,-0.8);
\draw[color=red, line width=1] (0,-0.5) arc (60:-240:0.6cm);
\draw[text = black,font=\tiny, color=red ] (0.05 ,0.7) node {\XSolidBold};
\draw[text = black,font=\small, color=blue ] (0.05 ,-1.2) node {\FiveStar};
\draw[text = black,font=\tiny ] (0.0 ,-0.90) node {A};
\draw[text = black,font=\tiny, color=blue ] (-0.65 ,0.5) node {\XSolidBold};
\draw[text = black,font=\small, color=red ] (-0.65 ,-1) node {\FiveStar};
\draw[text = black,font=\tiny ] (-0.50 ,-0.75) node {B};
\draw[dashed] (-0.65,0.5) -- (0.05,0.7);
\draw[dashed] (-0.65,-1) -- (0.05,-1.2);
\end{tikzpicture}
\caption{$\CU^{SO}_{as}$}
\end{subfigure}
\begin{subfigure}[b]{1.0in}
\centering
\begin{tikzpicture}[scale=1.3, every node/.style={transform shape}]
\draw[color=red, line width=1] (0,0) arc (-60:240:0.6cm);
\draw (0.1,0.3) .. controls (-0.07,0) and (-0.07,-0.5) .. (0.1,-0.8);
\draw (-0.7,0.3) .. controls (-0.53,0) and (-0.53,-0.5) .. (-0.7,-0.8);
\draw[color=blue, line width=1] (0,-0.5) arc (60:-240:0.6cm);
\draw[text = black,font=\small, color=red ] (0.05 ,0.7) node {\FiveStar};
\draw[text = black,font=\tiny ] (0.05 ,0.45) node {B};
\draw[text = black,font=\small, color=blue ] (0.05 ,-1.2) node {\FiveStar};
\draw[text = black,font=\tiny ] (0.0 ,-0.90) node {A};
\draw[text = black,font=\tiny, color=blue ] (-0.65 ,0.5) node {\XSolidBold};
\draw[text = black,font=\tiny, color=red ] (-0.65 ,-1) node {\XSolidBold};
\draw[dashed] (-0.65,0.5) -- (0.05,0.7);
\draw[dashed] (-0.65,-1) -- (0.05,-1.2);
\end{tikzpicture}
\caption{$\CU^{SO}_{c2}$}
\end{subfigure}
\caption{Colored pair-of-pants decompositions for a 4-punctured sphere with two twisted full punctures and two twisted null punctures of each color. The degrees of normal bundles are $(p, q)=(1, 1)$. 
Each subscript stands for: crossing-type 1, swap, Argyres-Seiberg type, crossing-type 2. The first two dual frames have Lagrangian descriptions. The theory $\CU^{SO}_{c1}$ turns out to be identical to the dual theory of \cite{Intriligator:1995id}. The latter three theories are all non-Lagrangian theories. The theory $\CU^{SO}_{s}$ is an $SO$ version of \cite{Gadde:2013fma}. 
}
\label{fig:coloredPoP}
\end{figure}
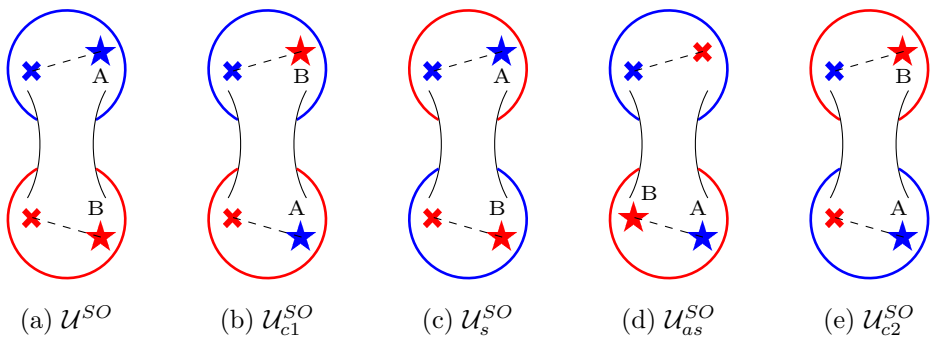

The SQCD with $SO(2N)$ gauge group and $4N-4$ vectors can be realized by choosing the normal bundles and the UV curve to be $\CL(1)\oplus \CL(1) \to \CC_{g=0, n=4}$. Two of the punctures are twisted maximal ones having $USp(2N-2)$ flavor symmetries with each color, and we also put two twisted punctures with no flavor symmetry with each color. Since we have 4 distinct punctures and two distinct pair-of-pants, there are many more dual frames compared to the case of $\CN=2$ theory. See figure \ref{fig:coloredPoP}.  

One can also consider having other type of punctures to realize $USp(2N-2)$ gauge theories or $G_2$ gauge theory. The colored pair-of-pants decompositions will be almost the same as this example. There are five dual frames, one of them being the electric gauge theory. There is one Lagrangian dual which we denote as crossing 1 and three non-Lagrangian theories which we name as swap, Argyres-Seiberg type and the crossing 2 type. This fact will be universal regardless of the choice of the gauge group, as it can be easily read off from the geometry. In the later sections, we study each theories in more detail. 

\section{Dualities for $SO(2N)$ gauge theory} \label{sec:SO}
In this section, we study dualities for the $SO(2N)$ gauge theory with $4N-4$ vectors. 

\subsection{$T_{SO(2N)}$ and $\tilde{T}_{SO(2N)}$ theory and Higgsing}
For a class $\CS$ theory of type $\Gamma$, the most basic building block is $T_\Gamma$ which is given by wrapping the 6d theory on a three punctured sphere with 3 maximal punctures. The theory has $\Gamma_A \times \Gamma_B \times \Gamma_C$ global symmetry, and has special dimension 2 operators $\mu_{A, B, C}$ that transform under the adjoint of $\Gamma_{A, B, C}$ respectively. These operators satisfy a chiral ring relation 
\be \label{eq:chiral}
 \tr \mu_A^2 = \tr \mu_B^2 =  \tr \mu_C^2 . 
\ee
This relation is proved in \cite{Benini:2009mz} for the $\Gamma = SU(N)$ where the theory is usually called as $T_N$. 
We will mainly use the twisted $\tilde{T}_{SO(2N)}$ theory to construct various theories of interest. It has $SO(2N) \times USp(2N-2) \times USp(2N-2)$ global symmetry. We prove the chiral ring relation \eqref{eq:chiral} for the $T_{D_n}$ and the twisted $\tilde{T}_{D_n}$ in appendix \ref{app:chiralring}. 
\begin{figure}[h]
	\begin{center}
	\begin{subfigure}[b]{2.5in}
		\centering
		\includegraphics[width=1.8in]{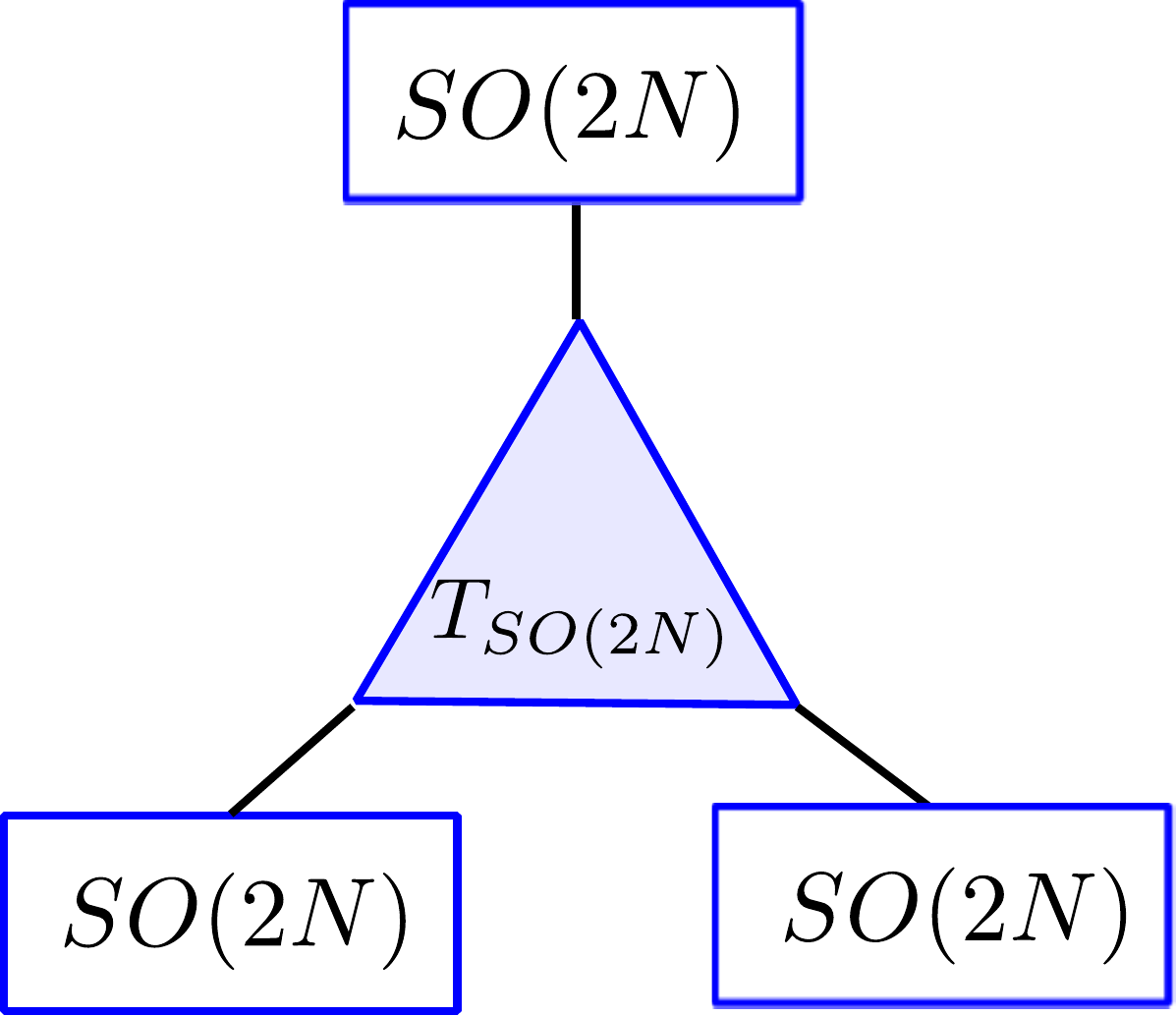}
		\caption{$T_{SO(2N)}$}
	\end{subfigure}
	\quad \quad
	\begin{subfigure}[b]{2.5in}
		\centering
		\includegraphics[width=1.8in]{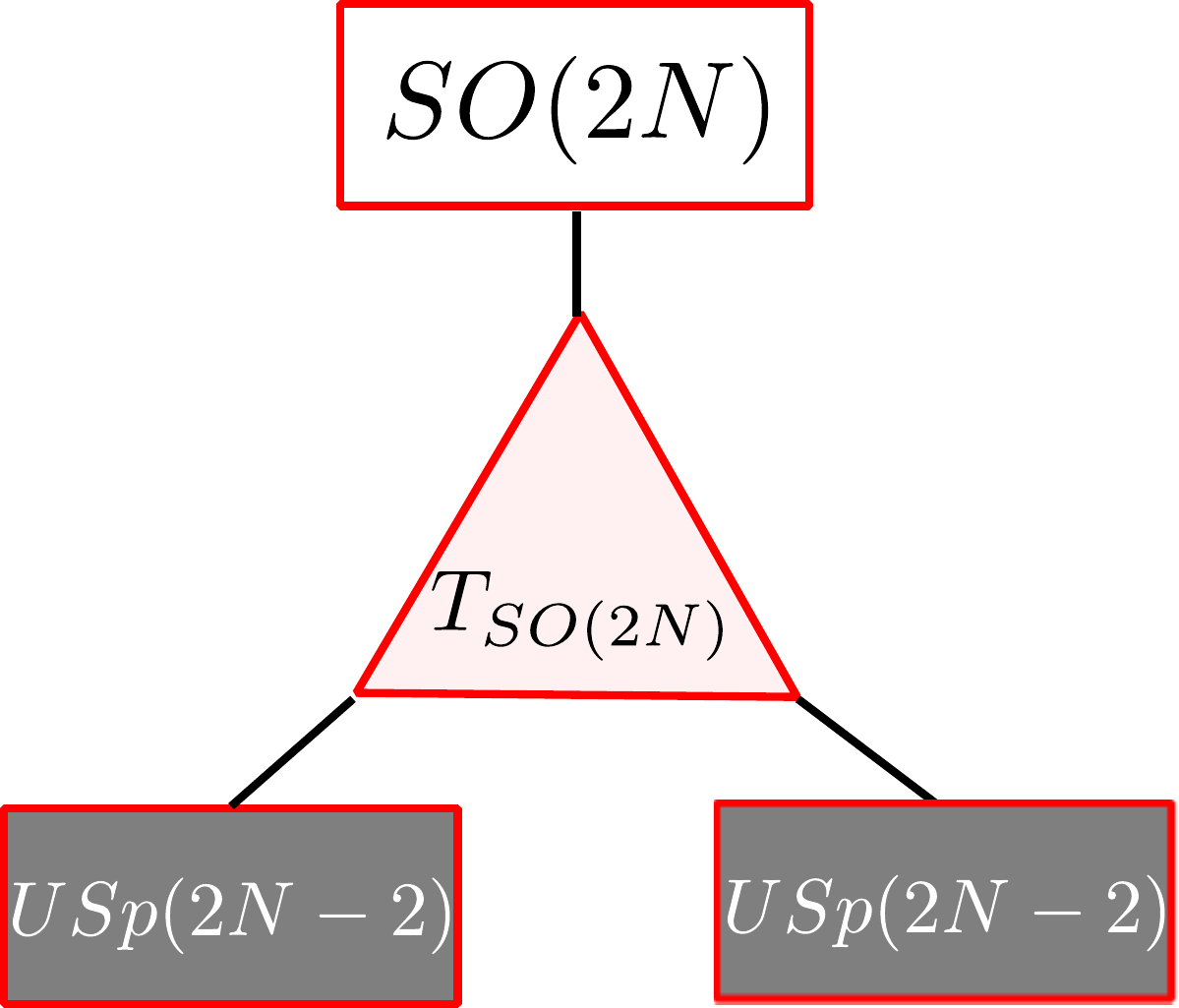}
		\caption{$\tilde{T}_{SO(2N)}$}
	\end{subfigure}
	\caption{Left: $T_{SO(2N)}$ theory, Right: $\tilde{T}_{SO(2N)}$ theory}
	\end{center}
\end{figure}

The number of effective vector multiplets $n_v$ and hypermultiplets $n_h$ for $T_{D_n}$ and $\tilde{T}_{D_n}$ can be computed using the equations (3.16) and (3.19) of \cite{Chacaltana:2012zy}. Each puncture contributes by
\be
 &{ }& n_v (SO(2N)) = \frac{1}{3} N(7-15N+8N^2) \ , \\
 &{ } &n_v (USp(2N-2)) = \frac{1}{6} (-3 + 20N - 30N^2 + 16N^3) \ , \\
 &{ }& n_h (SO(2N)) = n_h (USp(2N-2)) = \frac{2}{3} N(2N-1)(2N-2) \ .
\ee
There is also a contribution from the bulk 
\be
 n_v (g=0) &=& - \frac{4}{3} (2N-2)N(2N-1) - N \ , \\
 n_h (g=0) &=& - \frac{4}{3} (2N-2)N(2N-1) \ , 
\ee
from which we can compute the $n_v, n_h$ for $T_{D_n}$ and $\tilde{T}_{D_n}$ to get
\be
 n_v (T_{D_n}) &=& \frac{1}{3} N(10 - 21N + 8N^2 ) \ , \\
 n_v (\tilde{T}_{D_n}) &=& -1 + \frac{16}{3}N - 7N^2 + \frac{8}{3} N^3 \ , \\ 
 n_h (T_{D_n}) &=& n_h (\tilde{T}_{D_n}) = \frac{4}{3} n(n-1)(2n-1) \ . 
\ee
We will use these formula in later sections to compute the anomaly coefficients. 

\subsubsection*{Higgsing the $\tilde{T}_{SO(2N)}$ theory}
From the $\tilde{T}_{SO(2N)}$ theory, we can obtain other building blocks by partially closing the full puncture to a one with smaller global symmetries. The $SU(2)$ embedding $\rho: SU(2) \to G$ where $G=SO(2N)$ or $G=USp(2N-2)$ induces a decomposition of adjoint representations into the representations of $SU(2)$ and its commutant $G_F$ 
\be \label{eq:AdjDecom}
 \textrm{adj} = \bigoplus_j R_j \otimes V_j \ , 
\ee
where $V_j$ is the spin-$j$ representation of SU(2) and $R_j$ are the representations of the flavor symmetry $G_F$ associated to the puncture. 

For example, when we close one of the twisted puncture having $USp(2N-2)$ completely to have no global symmetry, we obtain a free theory with bifundamental of $SO(2N)$-$USp(2N-2)$. More concretely, we give vev to the operator $\mu$ associated to the puncture as
\begin{equation}\label{eq:vevNull}
\langle \mu \rangle = \rho_{\varnothing} (\sigma^+) = \sum_{\alpha} E_\alpha^+ \ ,  
\end{equation}
where $\alpha$ are the simple roots of $USp(2N-2)$ and $E_\alpha^+$ are the corresponding raising operators. \footnote{We will be cavalier about our notations denoting the Lie groups and Lie algebras.} The $\rho_{\varnothing}$ denotes the principal embedding of $SU(2)$ into $USp(2N-2)$, and $\s^+ = \s_1 + i \s_2$ where $\s_i$ are the Pauli matrices. This embedding leaves no flavor symmetry at all. 
Under this embedding the adjoint representation of $USp(2N-2)$ decomposes as
\begin{equation}
{\tiny\yng(2)} = \bigoplus_{k=1}^{N-1} V_{2k-1} \ , 
\end{equation}
where $V_j$ is the spin-$j$ representation of $SU(2)$. The dimension of the nilpotent orbit of $\rho(\sigma^+)$ then gives us the number of free half-hyper multiplets produced in the process. 
Thus we find that after Higgsing, the theory flows to an $SO(2N)$-$USp(2N-2)$ bifundamental along with $2(N-1)^2$ free half-hypermultiplets. See for example section 2 of \cite{Chacaltana:2012zy}.

\subsection{Dualities for $SO(2N)$-coupled $\tilde{T}_{SO(2N)}$ theories}
Before going into the SQCD, let us consider the theory that does not have a known Lagrangian description. 
Consider a theory realized by the UV curve given by 4 punctured sphere with two red and blue colors each. Choose all the punctures to be the twisted maximal ones having $USp(2N-2)$ flavor symmetries. We decompose it as two pair-of-pants with red and blue colors and arrange all the punctures to lie in the same color as the pair-of-pants. Each pair-of-pants gives $\tilde{T}_{SO(2N)}$ block. Let us call the red punctures to be $A, B$ and blue punctures to be $C, D$. 
\begin{figure}[h]
\centering
\includegraphics[width=2.6in]{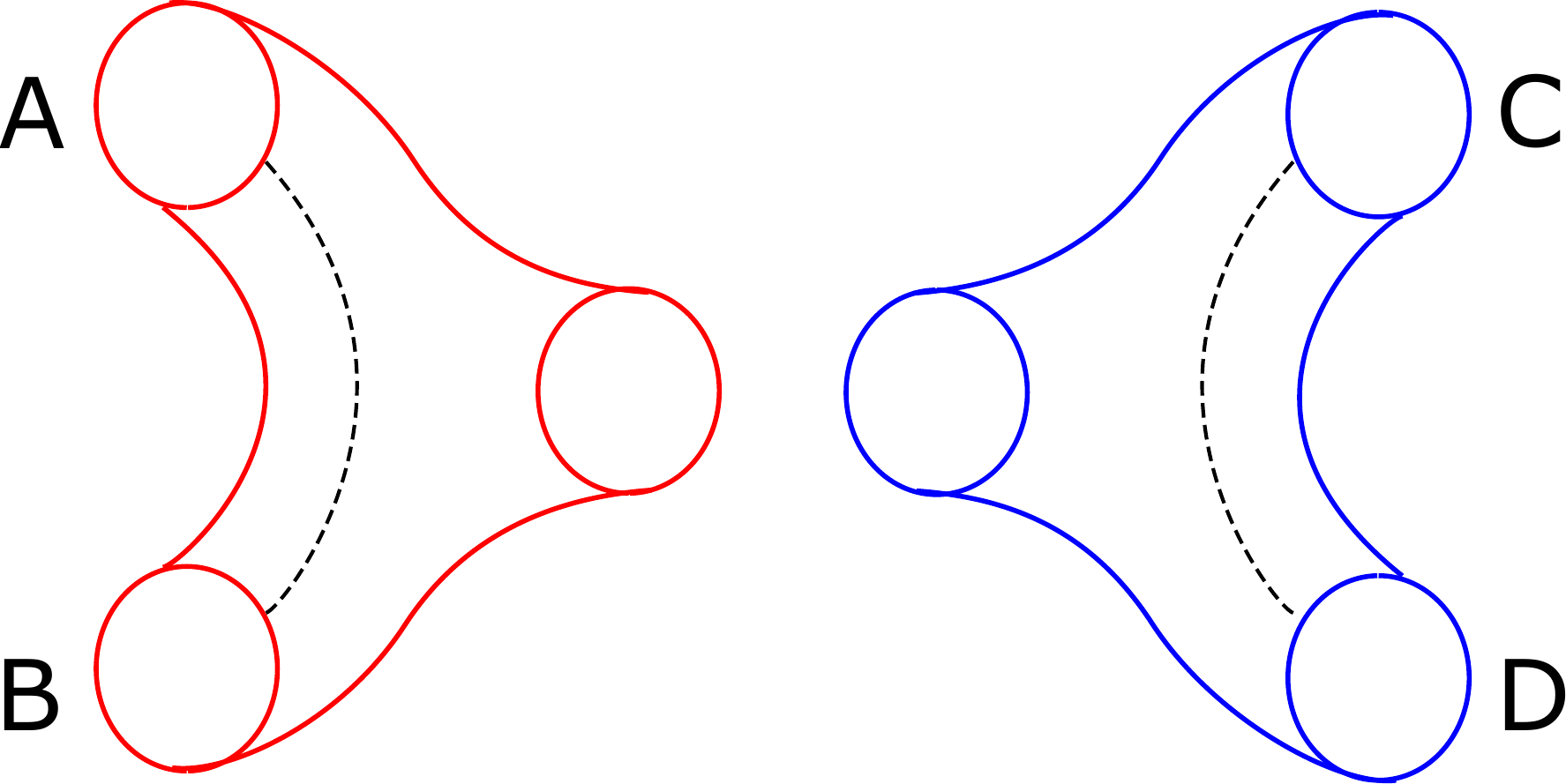}
\caption{Coupling two copies of $\tilde{T}_{SO(2N)}$ theories}
\end{figure}

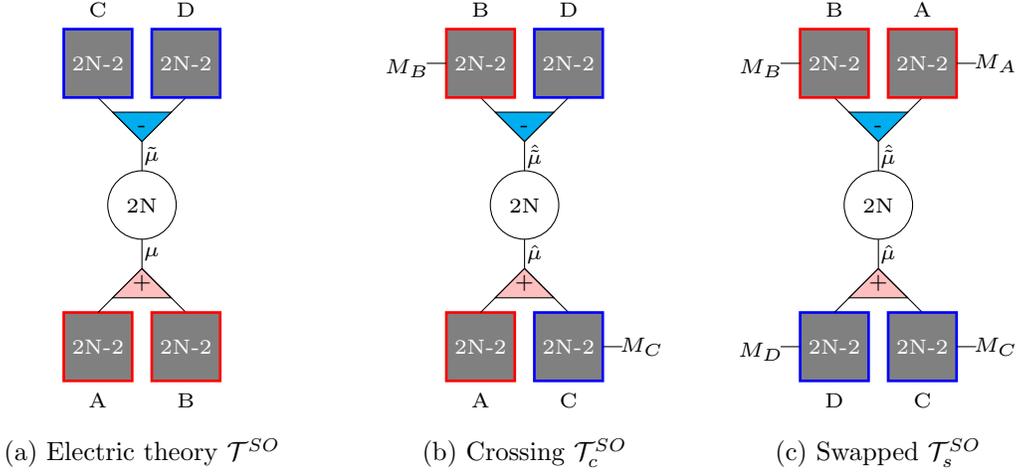
\begin{figure}[ht]
\centering
\begin{subfigure}[b]{1.8in}
\centering
\begin{tikzpicture}[scale=1.3, every node/.style={transform shape}]
\filldraw[fill=gray, draw=red, line width=1]  (0,0) rectangle (0.70,0.70);
\draw[text = white, font=\tiny] (0.35,0.35) node {2N-2};
\draw[text= black, font=\tiny] (0.35,-0.2) node{A};
\draw (0.35,0.7)-- ++(0.45,0.45);

\filldraw[fill=pink, draw=black] (0.50,0.85)--(0.80,1.15)--(1.10,0.85) -- cycle;
\draw[text= black, font=\tiny] (0.80,1.0) node{+};

\draw (0.80,1.15)-- ++(0.45,-0.45);
\filldraw[fill=gray, draw=red, line width=1]  (0.90,0) rectangle ++(0.70,0.70);
\draw[text = white, font=\tiny] (1.25,0.35) node {2N-2};
\draw[text= black, font=\tiny] (1.25,-0.2) node{B};
\draw (0.80,1.15) -- (0.80,1.45);
\draw[text=black,font=\tiny] (0.90,1.30) node{$\mu$};
\filldraw[fill=white, draw=black]  (0.80,1.80) circle (0.35 cm);
\draw[text = black, font=\tiny] (0.80,1.80) node {2N};
\draw (0.80,2.15) --(0.80, 2.45);
\draw[text=black,font=\tiny] (0.90,2.30) node{$\tilde\mu$};
\draw (0.80,2.45) -- ++(-0.45,0.45);
\draw(0.80,2.45) -- ++(0.45,0.45);
\filldraw[fill=cyan, draw=black] (0.80,2.45)--++ (0.30,0.30)--++(-0.60,0) -- cycle;
\draw[text= black, font=\tiny] (0.80,2.60) node{-};
\filldraw[fill=gray, draw=blue, line width=1]  (0,2.9) rectangle ++(0.70,0.70);
\draw[text = white, font=\tiny] (0.35,3.25) node {2N-2};
\draw[text= black, font=\tiny] (0.35,3.8) node{C};
\filldraw[fill=gray, draw=blue, line width=1]  (0.9,2.9) rectangle ++(0.70,0.70);
\draw[text = white, font=\tiny] (1.25,3.25) node {2N-2};
\draw[text= black, font=\tiny] (1.25,3.8) node{D};
\end{tikzpicture}
\caption{Electric theory $\CT^{SO}$}
\label{fig:soe}
\end{subfigure}
\quad
\begin{subfigure}[b]{1.8in}
\centering
\begin{tikzpicture}[scale=1.3, every node/.style={transform shape}]
\filldraw[fill=gray, draw=red, line width=1]  (0,0) rectangle (0.70,0.70);
\draw[text = white, font=\tiny] (0.35,0.35) node {2N-2};
\draw[text= black, font=\tiny] (0.35,-0.2) node{A};
\draw (0.35,0.7)-- ++(0.45,0.45);
\filldraw[fill=pink, draw=black] (0.50,0.85)--(0.80,1.15)--(1.10,0.85) -- cycle;
\draw[text= black, font=\tiny] (0.80,1.0) node{+};
\draw (0.80,1.15)-- ++(0.45,-0.45);
\filldraw[fill=gray, draw=blue, line width=1]  (0.90,0) rectangle ++(0.70,0.70);
\draw[text = white, font=\tiny] (1.25,0.35) node {2N-2};
\draw[text= black, font=\tiny] (1.25,-0.2) node{C};
\draw[text = black, font=\tiny]  (1.6,0.35)--++(0.2,0) ++(0.2,0)node{$M_C$};
\draw (0.80,1.15) -- (0.80,1.45);
\draw[text=black,font=\tiny] (0.90,1.30) node{$\hat\mu$};
\filldraw[fill=white, draw=black]  (0.80,1.80) circle (0.35 cm);
\draw[text = black, font=\tiny] (0.80,1.80) node {2N};
\draw (0.80,2.15) --(0.80, 2.45);
\draw[text=black,font=\tiny] (0.90,2.30) node{$\hat{\tilde\mu}$};
\draw (0.80,2.45) -- ++(-0.45,0.45);
\draw(0.80,2.45) -- ++(0.45,0.45);
\filldraw[fill=cyan, draw=black] (0.80,2.45)--++ (0.30,0.30)--++(-0.60,0) -- cycle;
\draw[text= black, font=\tiny] (0.80,2.60) node{-};
\filldraw[fill=gray, draw=red, line width=1]  (0,2.9) rectangle ++(0.70,0.70);
\draw[text = white, font=\tiny] (0.35,3.25) node {2N-2};
\draw[text= black, font=\tiny] (0.35,3.8) node{B};
\draw[text = black, font=\tiny]  (0,3.25)--++(-0.2,0) ++(-0.2,-0.05) node{$M_B$};
\filldraw[fill=gray, draw=blue, line width=1]  (0.9,2.9) rectangle ++(0.70,0.70);
\draw[text = white, font=\tiny] (1.25,3.25) node {2N-2};
\draw[text= black, font=\tiny] (1.25,3.8) node{D};
\end{tikzpicture}
\caption{Crossing $\CT^{SO}_c$}
\label{fig:somm}
\end{subfigure}
\begin{subfigure}[b]{1.8in}
\centering
\begin{tikzpicture}[scale=1.3, every node/.style={transform shape}]
\filldraw[fill=gray, draw=blue, line width=1]  (0,0) rectangle (0.70,0.70);
\draw[text = white, font=\tiny] (0.35,0.35) node {2N-2};
\draw[text= black, font=\tiny] (0.35,-0.2) node{D};
\draw[text = black, font=\tiny]  (0,0.35)--++(-0.2,0) ++(-0.2,-0.05) node{$M_D$};
\draw (0.35,0.7)-- ++(0.45,0.45);

\filldraw[fill=pink, draw=black] (0.50,0.85)--(0.80,1.15)--(1.10,0.85) -- cycle;
\draw[text= black, font=\tiny] (0.80,1.0) node{+};
\draw (0.80,1.15)-- ++(0.45,-0.45);
\filldraw[fill=gray, draw=blue, line width=1]  (0.90,0) rectangle ++(0.70,0.70);
\draw[text = white, font=\tiny] (1.25,0.35) node {2N-2};
\draw[text= black, font=\tiny] (1.25,-0.2) node{C};
\draw[text = black, font=\tiny]  (1.6,0.35)--++(0.2,0) ++(0.2,0)node{$M_C$};
\draw (0.80,1.15) -- (0.80,1.45);
\draw[text=black,font=\tiny] (0.90,1.30) node{$\hat\mu$};
\filldraw[fill=white, draw=black]  (0.80,1.80) circle (0.35 cm);
\draw[text = black, font=\tiny] (0.80,1.80) node {2N};
\draw (0.80,2.15) --(0.80, 2.45);
\draw[text=black,font=\tiny] (0.90,2.30) node{$\hat{\tilde\mu}$};
\draw (0.80,2.45) -- ++(-0.45,0.45);
\draw(0.80,2.45) -- ++(0.45,0.45);
\filldraw[fill=cyan, draw=black] (0.80,2.45)--++ (0.30,0.30)--++(-0.60,0) -- cycle;
\draw[text= black, font=\tiny] (0.80,2.60) node{-};
\filldraw[fill=gray, draw=red, line width=1]  (0,2.9) rectangle ++(0.70,0.70);
\draw[text = white, font=\tiny] (0.35,3.25) node {2N-2};
\draw[text= black, font=\tiny] (0.35,3.8) node{B};
\draw[text = black, font=\tiny]  (0,3.25)--++(-0.2,0) ++(-0.2,-0.05) node{$M_B$};
\filldraw[fill=gray, draw=red, line width=1]  (0.9,2.9) rectangle ++(0.70,0.70);
\draw[text = white, font=\tiny] (1.25,3.25) node {2N-2};
\draw[text= black, font=\tiny] (1.25,3.8) node{A};
\draw[text = black, font=\tiny]  (1.6,3.25)--++(0.2,0) ++(0.2,0)node{$M_A$};\
\end{tikzpicture}
\caption{Swapped $\CT^{SO}_s$}
\label{fig:somm2}
\end{subfigure}
\caption{Dual frames of the two $\tilde{T}_{SO}$ blocks coupled by $SO$ gauge group. The red/blue color means $\s = +/-$ respectively. }
\label{fig:soso}
\end{figure}

This construction realizes two $\tilde{T}_{SO(2N)}$ blocks coupled along their $SO(2N)$ puncture by an $\mathcal{N}=1$ vector multiplet and a superpotential given by 
\begin{equation}
W= c \tr \mu \tilde \mu \ . 
\end{equation} 
Here $\mu$ is the dimension 2 operator transforming in the adjoint representation of the $SO(2N)$ flavor symmetry of $\tilde{T}_{SO(2N)}$ while $\tilde{\mu}$ is its counterpart coming from the other $\tilde{T}_{SO(2N)}$ block. 
The $U(1)_{\CF}$ charge for $\mu$ is $+2$ while $\tilde \mu$ has $-2$. The $U(1)_\CF$ charges of the operators are determined by the color choice $\s$ for each punctures as described in section \ref{sec:N1M5}. Diagrammatically we can represent this theory as in figure \ref{fig:soe}. We will call this theory as $\CT^{SO}$. 

This theory can also be obtained by starting from two $\tilde{T}_{SO(2N)}$ blocks coupled along with their $SO(2N)$ flavor symmetry by an $\mathcal{N}=2$ vector multiplet and then integrating out the adjoint chiral in the vector multiplet by giving it mass and then flowing to the IR. Since the operators $\mu$ and $\tilde\mu$ both have $R$-charge $1$, the operator $\mu \tilde{\mu}$ is marginal. 

A dual of this theory can be obtained by exchanging the punctures labeled $B$ and $C$. We will also have to integrate in mesons $M_B$ and $M_C$ that transform in the adjoint representation of $USp(2N-2)_B$ and $USp(2N-2)_C$ respectively \cite{Gadde:2013fma}. The superpotential in the dual theory is given by 
\begin{equation} \label{eq:WTsoCrossing}
W=\hat c \tr \hat\mu\hat{\tilde\mu} + \tr  ~\Omega M_B \Omega \hat\mu_B  + \tr ~ \Omega M_C \Omega \hat\mu_C \ , 
\end{equation}
where $\Omega$ is the $USp(2N-2)$ invariant antisymmetric form. 
We now have the dual operators $\hat\mu_B, \hat\mu_C$ for the punctures $B, C$ which have their $U(1)_\CF$ charges reversed, and also meson operators $M_B, M_C$ which has the same $U(1)_\CF$ charges as $\mu_B, \mu_C$. We depict this theory by figure \ref{fig:somm}.

We can further exchange punctures $A$ and $D$ to obtain a third theory which is dual to the previous two. The superpotential now becomes
\begin{equation} \label{eq:WTsoSwap}
W=\hat c \tr \hat\mu\hat{\tilde\mu} + \tr M_A\Omega\hat\mu_A\Omega + \tr\Omega M_B \Omega\hat\mu_B + \tr\Omega M_C\Omega\hat\mu_C + \tr M_D\Omega\hat\mu_D\Omega \ , 
\end{equation}
with extra meson fields $M_A$ and $M_D$. See the figure \ref{fig:somm2}. 

One can also derive these dualities starting from $\CN=2$ S-duality and giving mass to the adjoint chiral multiplet in the $\CN=2$ vector multiplet and integrating it out and then flowing to the IR. Then by using the chiral ring relation derived in the appendix \ref{app:chiralring} and integrating in the mesons, one can reproduce the superpotentials \eqref{eq:WTsoCrossing}, \eqref{eq:WTsoSwap}. We refer to the section 2.2.4 of \cite{Gadde:2013fma} for details. 

Following the nomenclature used in \cite{Gadde:2013fma}, we refer to the dual theories obtained above as being in the ``crossing frame" $\CT_c^{SO}$ and the ``swapped frame" $\CT_s^{SO}$ respectively. These three duality frames will be the basis of the dualities discussed in this section. 

\subsection{Dualities for $SO(2N)$ SQCD}
Now, let us move on to discuss dualities for the theory with UV Lagrangian descriptions. 

\subsubsection*{Intriligator-Seiberg duality}
By partially closing the punctures $A$ and $D$ in the electric theory $\CT^{SO}$, we reduce it to SQCD with gauge group $SO(2N)$ and $N_f = 4N-4$ fundamental (vector) flavors. Partial closing of the puncture is implemented by giving appropriate vevs as in \eqref{eq:vevNull} to $\mu_A$ and $\mu_D$. 
Closing the punctures changes the dual theories as well. 
\begin{figure}[h]
\centering
\begin{subfigure}[b]{2.5in}
\centering
\begin{turn}{90}
\begin{tikzpicture}[scale=1.3, every node/.style={transform shape}]
\filldraw[fill=gray, draw=red, line width =1]  (0.30, 0.65) circle (0.05 cm);
\draw (0.35,0.7)-- ++(0.45,0.45);

\filldraw[fill=pink, draw=black] (0.50,0.85)--(0.80,1.15)--(1.10,0.85) -- cycle;
\draw[text= black, font=\tiny] (0.80,1.0) node[rotate =-90]{+};

\draw (0.80,1.15)-- ++(0.45,-0.45);
\filldraw[fill=gray, draw=red, line width =1]  (0.90,0) rectangle ++(0.70,0.70);
\draw[text = white, font=\tiny] (1.25,0.35) node[rotate =-90] {2N-2};
\draw[text= black, font=\tiny] (1.25,-0.2) node[rotate =-90]{B};
\draw (0.80,1.15) -- (0.80,1.45);
\draw[text=black,font=\tiny] (0.90,1.30) node[rotate =-90]{$\mu$};
\filldraw[fill=white, draw=black]  (0.80,1.80) circle (0.35 cm);
\draw[text = black, font=\tiny] (0.80,1.80) node[rotate =-90]{2N};
\draw (0.80,2.15) --(0.80, 2.45);
\draw[text=black,font=\tiny] (0.90,2.30) node[rotate =-90]{$\tilde\mu$};
\draw (0.80,2.45) -- ++(-0.45,0.45);
\draw(0.80,2.45) -- ++(0.45,0.45);
\filldraw[fill=cyan, draw=black] (0.80,2.45)--++ (0.30,0.30)--++(-0.60,0) -- cycle;
\draw[text= black, font=\tiny] (0.80,2.60) node[rotate =-90]{-};
\filldraw[fill=gray, draw=blue, line width =1]  (0,2.9) rectangle ++(0.70,0.70);
\draw[text = white, font=\tiny] (0.35,3.25) node[rotate =-90] {2N-2};
\draw[text= black, font=\tiny] (0.35,3.8) node[rotate =-90]{C};
\filldraw[fill=gray, draw=blue,  line width =1]  (1.30, 2.95) circle (0.05 cm);
\end{tikzpicture}
\end{turn}
\caption{Electric theory $\CU^{SO}$}
\label{fig:isde}
\end{subfigure}
\quad
\begin{subfigure}[b]{2.5in}
\centering
\begin{turn}{90}
\begin{tikzpicture}[scale=1.3, every node/.style={transform shape}]
\filldraw[fill=gray, draw=red, line width =1]  (0.30, 0.65) circle (0.05 cm);
\draw (0.35,0.7)-- ++(0.45,0.45);

\filldraw[fill=pink, draw=black] (0.50,0.85)--(0.80,1.15)--(1.10,0.85) -- cycle;
\draw[text= black, font=\tiny] (0.80,1.0)  node[rotate =-90]{+};

\draw (0.80,1.15)-- ++(0.45,-0.45);
\filldraw[fill=gray, draw=blue, line width =1]  (0.90,0) rectangle ++(0.70,0.70);
\draw[text = white, font=\tiny] (1.25,0.35)  node[rotate =-90] {2N-2};
\draw[text= black, font=\tiny] (1.25,-0.2)  node[rotate =-90]{C};
\draw[text = black, font=\tiny]  (1.6,0.35)--++(0.2,0) ++(0.11,-0.1) node[rotate =-90]{$M_C$};
\draw (0.80,1.15) -- (0.80,1.45);
\draw[text=black,font=\tiny] (0.90,1.30)  node[rotate =-90]{$\hat\mu$};
\filldraw[fill=white, draw=black]  (0.80,1.80) circle (0.35 cm);
\draw[text = black, font=\tiny] (0.80,1.80)  node[rotate =-90] {2N};
\draw (0.80,2.15) --(0.80, 2.45);
\draw[text=black,font=\tiny] (0.90,2.30)  node[rotate =-90]{$\hat{\tilde\mu}$};
\draw (0.80,2.45) -- ++(-0.45,0.45);
\draw(0.80,2.45) -- ++(0.45,0.45);
\filldraw[fill=cyan, draw=black] (0.80,2.45)--++ (0.30,0.30)--++(-0.60,0) -- cycle;
\draw[text= black, font=\tiny] (0.80,2.60)  node[rotate =-90]{-};
\filldraw[fill=gray, draw=red, line width =1]  (0,2.9) rectangle ++(0.70,0.70);
\draw[text = white, font=\tiny] (0.35,3.25)  node[rotate =-90]{2N-2};
\draw[text= black, font=\tiny] (0.35,3.8)  node[rotate =-90]{B};
\draw[text = black, font=\tiny]  (0,3.25)--++(-0.2,0) ++(-0.11,-0.1)  node[rotate =-90]{$M_B$};
\filldraw[fill=gray, draw=blue, line width =1]  (1.30, 2.95) circle (0.05 cm);
\end{tikzpicture}
\end{turn}
\caption{Magnetic theory $\CU^{SO}_{c1}$}
\label{fig:isdm}
\end{subfigure}
\caption{Intriligator-Seiberg duality}
\label{fig:isd}
\end{figure}
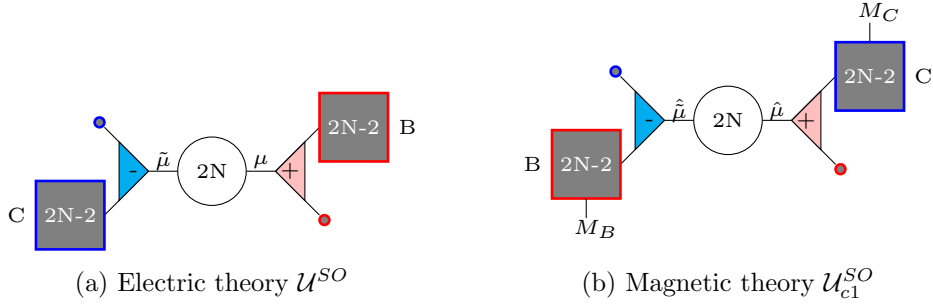
Upon Higgsing, the two copies of $\tilde{T}_{SO(2N)}$ become free bifundamentals of $SO(2N) \times USp(2N-2)_B$ and $SO(2N)\times USp(2N-2)_C$. Therefore, the original theory $\CT$ becomes $SO(2N)$ gauge group with $4N-4$ fundamental(vector) flavors where only the $USp(2N-2)_B \times USp(2N-2)_C \subset SU(4N-4)$ global symmetry is manifest. This is nothing but the usual SQCD. We also have the marginal superpotential
\begin{equation}
 W= c \tr \mu \tilde\mu \ , 
\end{equation} 
where now $\mu_{\alpha\beta} = (Q_{\alpha i} \Omega^{ij} Q_{\beta j})_B$ and  $\tilde{\mu}_{\alpha\beta} = (Q_{\alpha i} \Omega^{ij} Q_{\beta j})_C $ with $\a, \b = 1, \ldots, 2N$ denoting the $SO(2N)$ vector indices and $i, j = 1, \ldots, 2N-2$ denoting the $USp$ indices. 
Here $(Q_B)_{\alpha i}$ is the quark transforming as the bifundamental of $SO(2N) \times USp(2N-2)_B$ while $(Q_C)_{\alpha i}$ is the bifundamental of $SO(2N) \times USp(2N-2)_C$. This superpotential term breaks the global symmetry to $USp(2N-2)_B \times USp(2N-2)_C$. 
We will denote this theory as $\CU^{SO}$. 

Now, let us look at the theory obtained by closing the punctures $A$ and $B$ of crossing frame, $\CT^{SO}_c$. The theory so obtained has two meson fields $M_B, M_C$ each transforming under the adjoint of $USp(2N-2)_B$ and $USp(2N-2)_C$. Also we get superpotential terms as
\be \label{eq:UsoCW}
 W = \hat c \tr \hat \mu \hat{\tilde{\mu}} + \tr M_B \Omega \hat{\mu}_B \Omega  + \tr M_C \Omega \hat{\mu}_C \Omega  \ . 
\ee
 We can write $\hat{\mu}_B$ and $\hat{\mu}_C$ in terms of the fundamental dual quarks $\hat Q$ as $\mu_{B} = \hat{Q}_{B} \hat{{Q}}_{B}$ and $\mu_{C} = \hat{Q}_{C} \hat{{Q}}_{C}$ which are in the adjoint (=symmetric) representations of $USp(2N-2)_{B, C}$. The $\hat\mu$ and $\hat{\tilde{\mu}}$ are given by the dual quark bilinears as $\hat\mu = \hat{Q}_B \Omega \hat{Q}_B$ and $\hat{\tilde{\mu}} = \hat{Q}_C \Omega \hat{Q}_C$ which are in the adjoint of $SO(2N)$. 

The duality frames obtained through this procedure are depicted in figure \ref{fig:isd}. These two duality frames are related to each other by the Intriligator-Seiberg duality \cite{Intriligator:1995id}. 
Applying Intriligator-Seiberg duality to the $SO(2N)$ gauge theory with $4N-4$ fundamentals we find that the magnetic dual is given by the theory with $SO(2N)$ gauge group and $4N-4$ dual quarks $\hat{Q}$ along with mesons and the superpotential term $W= \tr M\hat{Q} \hat{Q}$. In the absence of any other superpotential the global symmetry of this theory would be $SU(4N-4)$ with the mesons transforming in the symmetric representation of $SU(4N-4)$. In terms of  $USp(2N-2)_B \times USp(2N-2)_C \subset SU(4N-4)$, the quarks split into  bifundamentals of $SO(2N) \times USp(2N-2)_B$ and $SO(2N) \times USp(2N-2)_B$ while the meson splits into the following irreducible representations.
\begin{itemize}
\item  symmetric tensor of $USp(2N-2)_B$ : $(M_B)_{i j}$
\item  symmetric tensor of $USp(2N-2)_C$ : $(M_C)_{ i j}$
\item  bifundamental of $USp(2N-2)_B \times USp(2N-2)_C$ : $M_{i j}$
\end{itemize}
Note that $M_{i j}$ is dual to the meson of the electric theory formed by $Q_{B\alpha i} Q_{C\alpha j}$. 
The electric superpotential $\tr \mu \tilde{\mu}$ induces a mass term for $M_{ij}$. 
The dual superpotential of the magnetic theory can be written as  
\be
 W &=&  \tr M \Omega M \Omega + \tr M_B \Omega \hat{Q}_B \hat{Q}_B \Omega + M_C  \Omega \hat{Q}_C \hat{Q}_C \Omega + \tr M \Omega \hat{Q}_B \hat{Q}_C \Omega  \ . 
\ee
Integrating out the massive mesons $M_{ij}$ then gives us the superpotential of \eqref{eq:UsoCW}. We will denote this theory as $\CU^{SO}_{c1}$ since it arises from exchanging the two punctures in the electric theory.

\subsubsection*{Non-Lagrangian dual 1: Swap}
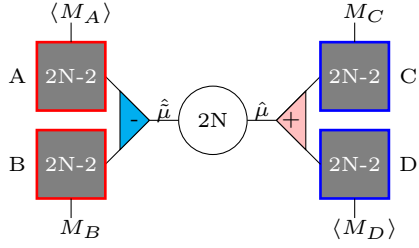
\begin{figure}[h]
\centering
\begin{turn}{90}
\begin{tikzpicture}[scale=1.3, every node/.style={transform shape}]
\filldraw[fill=gray, draw=blue, line width =1]  (0,0) rectangle (0.70,0.70);
\draw[text = white, font=\tiny] (0.35,0.35) node[rotate =-90] {2N-2};
\draw[text= black, font=\tiny] (0.35,-0.2) node[rotate =-90]{D};
\draw[text = black, font=\tiny]  (0,0.35)--++(-0.2,0) ++(-0.11,-0.1) node[rotate =-90]{$\langle M_D \rangle$};
\draw (0.35,0.7)-- ++(0.45,0.45);

\filldraw[fill=pink, draw=black] (0.50,0.85)--(0.80,1.15)--(1.10,0.85) -- cycle;
\draw[text= black, font=\tiny] (0.80,1.0) node[rotate =-90]{+};

\draw (0.80,1.15)-- ++(0.45,-0.45);
\filldraw[fill=gray, draw=blue, line width =1]  (0.90,0) rectangle ++(0.70,0.70);
\draw[text = white, font=\tiny] (1.25,0.35) node[rotate =-90]{2N-2};
\draw[text= black, font=\tiny] (1.25,-0.2) node[rotate =-90]{C};
\draw[text = black, font=\tiny]  (1.6,0.35)--++(0.2,0) ++(0.1,-0.1)node[rotate =-90]{$M_C$};
\draw (0.80,1.15) -- (0.80,1.45);
\draw[text=black,font=\tiny] (0.90,1.30) node[rotate =-90]{$\hat\mu$};
\filldraw[fill=white, draw=black]  (0.80,1.80) circle (0.35 cm);
\draw[text = black, font=\tiny] (0.80,1.80) node[rotate =-90] {2N};
\draw (0.80,2.15) --(0.80, 2.45);
\draw[text=black,font=\tiny] (0.90,2.30) node[rotate =-90]{$\hat{\tilde\mu}$};
\draw (0.80,2.45) -- ++(-0.45,0.45);
\draw(0.80,2.45) -- ++(0.45,0.45);
\filldraw[fill=cyan, draw=black] (0.80,2.45)--++ (0.30,0.30)--++(-0.60,0) -- cycle;
\draw[text= black, font=\tiny] (0.80,2.60) node[rotate =-90]{-};
\filldraw[fill=gray, draw=red, line width =1]  (0,2.9) rectangle ++(0.70,0.70);
\draw[text = white, font=\tiny] (0.35,3.25) node[rotate =-90] {2N-2};
\draw[text= black, font=\tiny] (0.35,3.8) node[rotate =-90]{B};
\draw[text = black, font=\tiny]  (0,3.25)--++(-0.2,0) ++(-0.11,-0.1) node[rotate =-90]{$M_B$};
\filldraw[fill=gray, draw=red, line width =1]  (0.9,2.9) rectangle ++(0.70,0.70);
\draw[text = white, font=\tiny] (1.25,3.25)node[rotate =-90] {2N-2};
\draw[text= black, font=\tiny] (1.25,3.8) node[rotate =-90]{A};
\draw[text = black, font=\tiny]  (1.6,3.25)--++(0.2,0) ++(0.1,-0.1)node[rotate =-90]{$\langle M_A \rangle$};
\end{tikzpicture}
\end{turn}
\caption{Non-Lagrangian dual $\CU^{SO}_s$ of $SO(2N)$ SQCD }
\label{fig:nldis}
\end{figure}

An interesting non-Lagrangian dual (figure \ref{fig:nldis}) to the $SO(2N)$ SQCD is obtained by the Higgsing the swapped theory $\CT_s^{SO}$ of figure \ref{fig:somm2}. In this frame the Higgsing of $USp(2N-2)_A$ and $USp(2N-2)_D$ is implemented through a vev $\rho_{\varnothing}(\sigma^+)$ to the meson fields $M_A$ and $M_D$. The low energy dynamics of this theory can be obtained as follows. With a little abuse of notation, let $M_A$ now represent the fluctuations around the vev $\rho^A(\sigma^+)$. The deformed superpotential now becomes 
\begin{equation} \label{eq:UsoSWa}
\begin{split}
W &= \tr \Omega \rho^A(\sigma^+) \Omega \hat{\mu}_A +  \tr \Omega M_A \Omega\hat{\mu}_A \\
    &= (\hat{\mu}_{A})_{1,-1} + \sum_{j,m} (M_{A})_{j,-m} (\hat{\mu}_{A})_{ j,m} \ , 
    \end{split}
\end{equation}
where we rewrite the components of $(\mu_A)_{ij}$ and $(M_A)_{ij}$ by decomposing into $SU(2)$ representations as in \eqref{eq:AdjDecom}. The indices $j, m$ with $m=-j, -j+1, \ldots, j-1, j$ labels the spin-$j$ representations of $SU(2)$ and $k=1, \ldots, \textrm{dim}R_k$. Since there is no flavor symmetry left here, we do not have any $k$ dependence. 

Since the first term of \eqref{eq:UsoSWa} break the $U(1)_\CF$, we should shift its charge appropriately. Also we want our superpotential term to have $U(1)_R$ charge $2$. In order to achieve this, we shift the $U(1)_{\mathcal{F}}$ flavor symmetry and R-symmetry to
\begin{equation}
\begin{aligned}
\mathcal{F} &= \mathcal{F}_0 + 2\rho^A(\sigma^3) \ , \\
R&=R_0 - \rho^A(\sigma^3) \ , 
\end{aligned}
\end{equation}
where $\mathcal{F}_0$ and $R_0$ are the $U(1)_{\mathcal{F}}$ and R-charges of the fields before Higgsing. 

The $USp(2N-2)_A$ flavor symmetry gets broken and the resulting non-conservation of the associated global currents can be expressed as 
\begin{equation}
\bar{D}^2(J_A)_{j,m} = \delta W = (\hat{\mu}_{A})_{j,m-1} \ . 
\end{equation}
The right-hand side vanishes only if $m=-j$. This implies that the operators $(\hat{\mu}_{A})_{j,m-1}$ are no longer BPS and hence the superpotential terms that couples them to mesonic fields become IR-irrelevant. As a result of this, the fields $(M_{A})_{j,m}$ for $m \neq -j$ decouple. The number of such free fields is $2(N-1)^2$ which is same as the number of free half-hypers obtained from Higgsing $USp(2N-2)_A$ in figure \ref{fig:soe}.  

Repeating the same analysis for $USp(2N-2)_D$ then leads to the following superpotential for our proposed dual 
\begin{equation}
W=\hat c \tr \hat\mu\hat{\tilde\mu} + \tr \Omega M_C \Omega \hat\mu_C + \tr \Omega M_B \Omega \hat \mu_B +\sum_{j} (M_{A})_{j,-j}(\hat{\mu}_{A})_{j,j}+\sum_{j} (M_{D})_{j,-j} (\hat{\mu}_{D})_{j,j} \ , 
\end{equation}
where $j = 1, 3, \ldots, 2N-3$ from which we see $2(N-1)$ gauge singlets. The charges for the $U(1)_{\mathcal{F}}$ and $U(1)_R$ are shifted to
\begin{equation}
\begin{aligned}
\mathcal{F} &= \mathcal{F}_0 + 2\rho^A(\sigma^3)- 2\rho^D(\sigma^3) \ , \\
R&=R_0 - \rho^A(\sigma^3)- \rho^D(\sigma^3) \ . 
\end{aligned}
\end{equation}
We will denote this theory as $U^{SO}_s$. 

\subsubsection*{Non-Lagrangian dual 2: Argyres-Seiberg type dual}
Another type of dual theory to the SQCD can be obtained from Higgsing punctures $B$ and $D$ of the duality frames in figure \ref{fig:soso}. This is possible since the punctures with the same colors are indistinguishable in the non-Lagrangian theory of figure \ref{fig:soso} and therefore their labels can be interchanged.  In the present case we relabel $A \leftrightarrow B$.

\begin{figure}[h]
\centering
\begin{turn}{90}
\begin{tikzpicture}[scale=1.3, every node/.style={transform shape}]
\filldraw[fill=gray, draw=red, line width =1]  (0,0) rectangle (0.70,0.70);
\draw[text = white, font=\tiny] (0.35,0.35) node[rotate=-90] {2N-2};
\draw[text= black, font=\tiny] (0.35,-0.2) node[rotate=-90]{B};
\draw (0.35,0.7)-- ++(0.45,0.45);

\filldraw[fill=pink, draw=black] (0.50,0.85)--(0.80,1.15)--(1.10,0.85) -- cycle;
\draw[text= black, font=\tiny] (0.80,1.0)  node[rotate =-90]{+};

\draw (0.80,1.15)-- ++(0.45,-0.45);
\filldraw[fill=gray, draw=blue, line width =1]  (0.90,0) rectangle ++(0.70,0.70);
\draw[text = white, font=\tiny] (1.25,0.35)  node[rotate =-90] {2N-2};
\draw[text= black, font=\tiny] (1.25,-0.2)  node[rotate =-90]{C};
\draw[text = black, font=\tiny]  (1.6,0.35)--++(0.2,0) ++(0.11,-0.1) node[rotate =-90]{$M_C$};
\draw (0.80,1.15) -- (0.80,1.45);
\draw[text=black,font=\tiny] (0.90,1.30)  node[rotate =-90]{$\hat\mu$};
\filldraw[fill=white, draw=black]  (0.80,1.80) circle (0.35 cm);
\draw[text = black, font=\tiny] (0.80,1.80)  node[rotate =-90] {2N};
\draw (0.80,2.15) --(0.80, 2.45);
\draw[text=black,font=\tiny] (0.90,2.30)  node[rotate =-90]{$\hat{\tilde\mu}$};
\draw (0.80,2.45) -- ++(-0.45,0.45);
\draw(0.80,2.45) -- ++(0.45,0.45);
\filldraw[fill=cyan, draw=black] (0.80,2.45)--++ (0.30,0.30)--++(-0.60,0) -- cycle;
\draw[text= black, font=\tiny] (0.80,2.60)  node[rotate =-90]{-};
\filldraw[fill=gray, draw=red, line width =1]  (0,2.9) rectangle ++(0.70,0.70);
\draw[text = white, font=\tiny] (0.35,3.25)  node[rotate =-90]{2N-2};
\draw[text= black, font=\tiny] (0.35,3.8)  node[rotate =-90]{A};
\draw[text = black, font=\tiny]  (0,3.25)--++(-0.2,0) ++(-0.11,-0.1)  node[rotate =-90]{$\langle M_A \rangle$};
\filldraw[fill=gray, draw=blue, line width =1]  (1.30, 2.95) circle (0.05 cm);
\end{tikzpicture}
\end{turn}
\caption{Argyres-Seiberg type dual  $\CU^{SO}_{as}$ of $SO(2N)$ SQCD }
\label{fig:ASdso}
\end{figure}
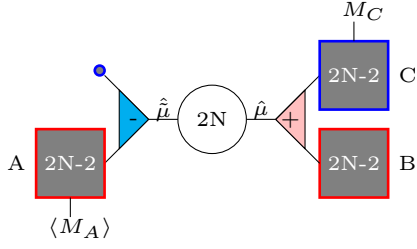

Higgsing the frames $\CT^{SO}$ and $\CT^{SO}_{s}$ give us the theories $\CU^{SO}$ and $\CU^{SO}_s$ respectively. However an Argyres-Seiberg type dual, $\CU^{SO}_{as}$, is obtained upon closing the afore mentioned punctures in $\CT^{SO}_c$ (see figure \ref{fig:ASdso}). Firstly, Higgsing the puncture $D$ will make the theory $\tilde{T}_{SO(2N)}$ on the upper sphere to be the theory of bifundamentals. Therefore we have $\tilde{T}_{SO(2N)}$ theory with $SO(2N)$ flavor symmetry gauged and coupled to $2N-2$ fundamentals (vectors). 

The punctures $A$ and $C$ have different colors from their pair of pants. Therefore we will have meson field $M_A$ and $M_C$ coupled through
\be
 W =  \hat c \tr \hat \mu \hat{\tilde{\mu}}+\tr M_A \Omega (QQ)_A \Omega + \tr M_C \Omega \hat{\mu}_C \Omega \ , 
\ee
where we replaced the operator $\hat{\mu}_B$ by the product of the quarks $(Q_{\a i} Q_{\b i})_A$. In order to Higgs the puncture $A$, we give vev to the meson field $\vev{M_B} = \rho_\varnothing$.  We can now consider low energy fluctuations around this vacuum and repeat the analysis of the previous subsection. The vev for the meson gives a mass to one of the quark bifundamentals which should be integrated out. The resulting low energy theory consists of $2N-3$ fundamentals coupled to a $\tilde{T}_{SO(2N)}$ block along with $N-1$ gauge singlets $(M_{A})_{j,-j}$ and mesons $M_C$ coupled through the superpotential
\be
 W= \hat c \tr \hat \mu \hat{\tilde{\mu}}+\sum_{j} (M_{A})_{j,-j} (\hat{\mu}_{A})_{ j,j} + \tr M_C \Omega \hat{\mu}_C \Omega \ , 
\ee
The $R$- and $\CF$-charges are shifted to   
\be
\begin{aligned}
\mathcal{F} &= \mathcal{F}_0 + 2\rho^A(\sigma^3) \ , \\
R&=R_0 - \rho^A(\sigma^3) \ . 
\end{aligned} 
\ee 

One interesting aspect of this dual description compared to the $\CN=2$ counterpart is that this dual theory has the same gauge group as the electric one. In the $\CN=2$ case, this type of duality changes the gauge group to be $SU(2)$ subgroup of $T_{\Gamma}$ \cite{Argyres:2007cn, Chacaltana:2010ks}, whereas in the present case the gauge group is still $SO(2N)$ unbroken.

\subsubsection*{Non-Lagrangian dual 3: Crossing type}
\begin{figure}[h]
\centering
\begin{turn}{90}
\begin{tikzpicture}[scale=1.3, every node/.style={transform shape}]
\filldraw[fill=gray, draw=red, line width =1]  (0,0) rectangle (0.70,0.70);
\draw[text = white, font=\tiny] (0.35,0.35) node[rotate =-90] {2N-2};
\draw[text= black, font=\tiny] (0.35,-0.2) node[rotate =-90]{A};
\draw[text = black, font=\tiny]  (0,0.35)--++(-0.2,0) ++(-0.11,-0.1) node[rotate =-90]{$\langle M_A \rangle$};
\draw (0.35,0.7)-- ++(0.45,0.45);

\filldraw[fill=cyan, draw=black] (0.50,0.85)--(0.80,1.15)--(1.10,0.85) -- cycle;
\draw[text= black, font=\tiny] (0.80,1.0) node[rotate =-90]{-};

\draw (0.80,1.15)-- ++(0.45,-0.45);
\filldraw[fill=gray, draw=blue, line width =1]  (0.90,0) rectangle ++(0.70,0.70);
\draw[text = white, font=\tiny] (1.25,0.35) node[rotate =-90]{2N-2};
\draw[text= black, font=\tiny] (1.25,-0.2) node[rotate =-90]{C};
\draw (0.80,1.15) -- (0.80,1.45);
\draw[text=black,font=\tiny] (0.90,1.30) node[rotate =-90]{$\hat{\tilde\mu}$};
\filldraw[fill=white, draw=black]  (0.80,1.80) circle (0.35 cm);
\draw[text = black, font=\tiny] (0.80,1.80) node[rotate =-90] {2N};
\draw (0.80,2.15) --(0.80, 2.45);
\draw[text=black,font=\tiny] (0.90,2.30) node[rotate =-90]{$\hat{\mu}$};
\draw (0.80,2.45) -- ++(-0.45,0.45);
\draw(0.80,2.45) -- ++(0.45,0.45);
\filldraw[fill=pink, draw=black] (0.80,2.45)--++ (0.30,0.30)--++(-0.60,0) -- cycle;
\draw[text= black, font=\tiny] (0.80,2.60) node[rotate =-90]{+};
\filldraw[fill=gray, draw=red, line width =1]  (0,2.9) rectangle ++(0.70,0.70);
\draw[text = white, font=\tiny] (0.35,3.25) node[rotate =-90] {2N-2};
\draw[text= black, font=\tiny] (0.35,3.8) node[rotate =-90]{B};
\filldraw[fill=gray, draw=blue, line width =1]  (0.9,2.9) rectangle ++(0.70,0.70);
\draw[text = white, font=\tiny] (1.25,3.25)node[rotate =-90] {2N-2};
\draw[text= black, font=\tiny] (1.25,3.8) node[rotate =-90]{D};
\draw[text = black, font=\tiny]  (1.6,3.25)--++(0.2,0) ++(0.1,-0.1)node[rotate =-90]{$\langle M_D \rangle$};
\end{tikzpicture}
\end{turn}
\caption{Crossing type dual $\CU^{SO}_{c2}$ of $SO(2N)$ SQCD }
\label{fig:cdso}
\end{figure}
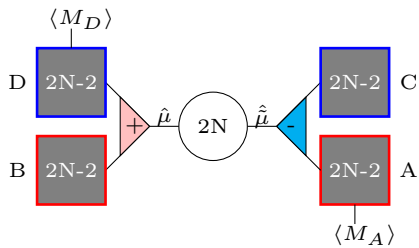

One more dual frame can be obtained from Higgsing $B$ and $C$ punctures of $\CT^{SO}_c$ of figure \ref{fig:somm} and relabeling $A \leftrightarrow B$ and $C \leftrightarrow D$. 
We call this the crossing type dual and denote it by $\CU^{SO}_{c2}$ (see figure \ref{fig:cdso}). It consists of two $\tilde{T}_{SO(2N)}$ blocks coupled to each other along their $SO(2N)$ flavor symmetry. Also there will be mesons $M_A$ and $M_D$ with a vev $\vev{M_A} =\vev{M_D} = \rho_\varnothing$. The low energy superpotential for the theory becomes
\be
 W= \hat c \tr \hat \mu \hat{\tilde{\mu}}+\sum_{j,m} (M_{A})_{j,-m} (\hat{\mu}_{A})_{ j,m} + \sum_{j,m} (M_{D})_{j,-m} (\hat{\mu}_{D})_{ j,m} \ , 
\ee
and shifted $R$- and $\CF$-charges  
\be
\begin{aligned}
\mathcal{F} &= \mathcal{F}_0 + 2\rho^A(\sigma^3) - 2\rho^D(\sigma^3)\ , \\
R&=R_0 - \rho^A(\sigma^3) - \rho^D(\sigma^3) \ .
\end{aligned} 
\ee 

\subsection{'t Hooft anomaly matching}
Now we test our dualities by computing the anomaly coefficients in different dual frames. 

\subsubsection*{Non-Lagrangian duals}
Upon giving a mass to the adjoint chiral superfield in the vector multiplet of an $\mathcal{N}=2$ theory and hence reducing SUSY down to $\mathcal{N}=1$, the residual $U(1)_R$ symmetry that is preserved by this deformation is given by 
\be
R_{\mathcal{N}=1} = \frac{1}{2}R_{\mathcal{N}=2} +I_3 \ , 
\ee
where $I_3$ is the Cartan of  $SU(2)_R$ in the parent theory. 
Thus we can write the $\tr R_{\mathcal{N}=1}$ and $\tr R^3_{\mathcal{N}=1}$ anomalies in terms of the anomalies of the parent $\CN=2$ theory as
\begin{align}
\tr R_{\mathcal{N}=1} &= \frac{1}{2} \tr R_{\mathcal{N}=2} = n_v - n_h \ , 
\end{align}
and
\begin{align}
\tr R^3_{\mathcal{N}=1} &=  \frac{1}{8} \tr R_{\mathcal{N}=2}^3 +  \frac{3}{2} \tr R_{\mathcal{N}=2} I_3^2 
=n_v-\frac{1}{4}n_h \ , 
\end{align}
where $n_v$ is the effective number of vector multiplets and $n_h$ is the effective number of hyper-multiplets in the parent theory.  

It is now straight-forward to check that the $\tr R$ and $\tr R^3$ anomalies of the duality frames shown in figure \ref{fig:soso} match. This is because the mesons have $R$-charge $1$ and hence do not contribute to the $R$-anomalies. Thus all the R-anomalies of these theories are destined to match as a direct consequence of their matching in the $\CN=2$ parent theories. This also implies that the flavor central charge given by $K\delta^{ab} = -3 \tr R T^aT^b$ will only get contributions from the coupled $\tilde{T}_{SO(2N)}$ blocks and hence match in all the three duality frames. 

Let us now consider the matching of $\tr \mathcal{F}T^aT^b$ across the various duality frames. 
The global current $\CF = \sum_i \s_i J_i$ is given by the sum of $J_i$ where the global symmetry $J$ is given by
\be
 J = R_{\CN=2} - 2 I_3 , 
\ee
for the each building block $\tilde{T}_{SO(2N)}$. 
Note that if the corresponding  $\tilde{T}_{SO(2N)}$ block has a $U(1)_\mathcal{F}$-charge $\sigma$ then
\begin{equation}
\tr \mathcal{F}T^aT^b= \sigma \tr R_{\mathcal{N}=2}T^aT^b =- \frac{\sigma}{2} k_{\mathfrak{g}} \ . 
\end{equation}
To begin with, consider the anomaly coefficient for $T^a \in \mathfrak{sp}(2N-2)_A$. Note that $k_{\mathfrak{sp}(2N-2)}$ for $\tilde{T}_{SO(2N)}$ is $4N$ as can be checked by comparing the dual theories of figure \ref{fig:dTSp}. Thus for the electric theory, $\CT^{SO}$ (figure \ref{fig:soe}) we have 
\begin{equation}
\tr \mathcal{F}T^a_AT^b_A= -2N\delta^{ab} \ . 
\end{equation}
This matches trivially to the anomaly coefficient of the theory, $\CT^{SO}_c$ (figure \ref{fig:somm}).  It is much more interesting to compare this with the anomaly coefficient of $\CT^{SO}_s$ (figure \ref{fig:somm2}) which, after taking the contributions of the meson $M_A$ into account, becomes 
\begin{equation}
\begin{split}
\tr\mathcal{F}T^a_AT^b_A &= 2N\delta^{ab} -2\tr_{\textrm{adj}}T_A^aT_A^b \\
&=2N\delta^{ab}-2(2N)\delta^{ab}\\
&=-2N\delta^{ab} \ , 
\end{split}
\end{equation}
which agrees with the original theory. The anomalies of $USp(2N-2)_B$, $USp(2N-2)_C$ and $USp(2N-2)_D$ match in all the duality frames in an analogous manner.

\subsubsection*{Dual theories of $SO(2N)$ SQCD}

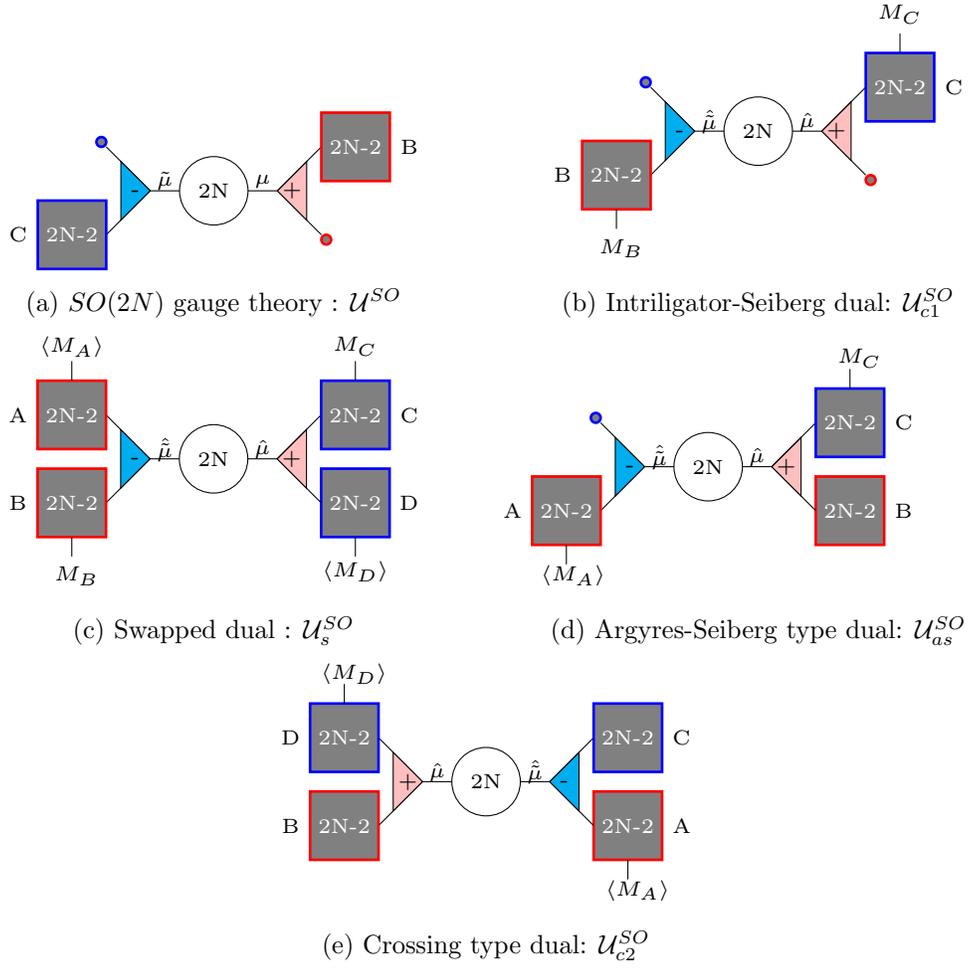
\begin{figure}[h]
\centering
\begin{subfigure}[b]{2.8 in}
\centering
\begin{turn}{90}
\begin{tikzpicture}[scale=1.3, every node/.style={transform shape}]
\filldraw[fill=gray, draw=red, line width =1]  (0.30, 0.65) circle (0.05 cm);
\draw (0.35,0.7)-- ++(0.45,0.45);
\filldraw[fill=pink, draw=black] (0.50,0.85)--(0.80,1.15)--(1.10,0.85) -- cycle;
\draw[text= black, font=\tiny] (0.80,1.0) node[rotate=-90]{+};

\draw (0.80,1.15)-- ++(0.45,-0.45);
\filldraw[fill=gray, draw=red, line width =1]  (0.90,0) rectangle ++(0.70,0.70);
\draw[text = white, font=\tiny] (1.25,0.35) node[rotate=-90] {2N-2};
\draw[text= black, font=\tiny] (1.25,-0.2) node[rotate=-90]{B};
\draw (0.80,1.15) -- (0.80,1.45);
\draw[text=black,font=\tiny] (0.90,1.30) node[rotate=-90]{$\mu$};
\filldraw[fill=white, draw=black]  (0.80,1.80) circle (0.35 cm);
\draw[text = black, font=\tiny] (0.80,1.80) node[rotate=-90] {2N};
\draw (0.80,2.15) --(0.80, 2.45);
\draw[text=black,font=\tiny] (0.90,2.30) node[rotate=-90]{$\tilde\mu$};
\draw (0.80,2.45) -- ++(-0.45,0.45);
\draw(0.80,2.45) -- ++(0.45,0.45);
\filldraw[fill=cyan, draw=black] (0.80,2.45)--++ (0.30,0.30)--++(-0.60,0) -- cycle;
\draw[text= black, font=\tiny] (0.80,2.60) node[rotate=-90]{-};
\filldraw[fill=gray, draw=blue, line width =1]  (0,2.9) rectangle ++(0.70,0.70);
\draw[text = white, font=\tiny] (0.35,3.25) node[rotate=-90] {2N-2};
\draw[text= black, font=\tiny] (0.35,3.8) node[rotate=-90]{C};
\filldraw[fill=gray, draw=blue, line width =1]  (1.30, 2.95) circle (0.05 cm);
\end{tikzpicture}
\end{turn}
\caption{$SO(2N)$ gauge theory : $\CU^{SO}$}
\label{fig:soduality1}
\end{subfigure}
\begin{subfigure}[b]{2.8 in}
\centering
\begin{turn}{90}
\begin{tikzpicture}[scale=1.3, every node/.style={transform shape}]
\filldraw[fill=gray, draw=red, line width =1]  (0.30, 0.65) circle (0.05 cm);
\draw (0.35,0.7)-- ++(0.45,0.45);
\filldraw[fill=pink, draw=black] (0.50,0.85)--(0.80,1.15)--(1.10,0.85) -- cycle;
\draw[text= black, font=\tiny] (0.80,1.0) node[rotate=-90]{+};
\draw (0.80,1.15)-- ++(0.45,-0.45);
\filldraw[fill=gray, draw=blue, line width =1]  (0.90,0) rectangle ++(0.70,0.70);
\draw[text = white, font=\tiny] (1.25,0.35) node[rotate=-90] {2N-2};
\draw[text= black, font=\tiny] (1.25,-0.2) node[rotate=-90]{C};
\draw[text = black, font=\tiny]  (1.6,0.35)--++(0.2,0) ++(0.2,0)node[rotate=-90]{$M_C$};
\draw (0.80,1.15) -- (0.80,1.45);
\draw[text=black,font=\tiny] (0.90,1.30) node[rotate=-90]{$\hat\mu$};
\filldraw[fill=white, draw=black]  (0.80,1.80) circle (0.35 cm);
\draw[text = black, font=\tiny] (0.80,1.80) node[rotate=-90] {2N};
\draw (0.80,2.15) --(0.80, 2.45);
\draw[text=black,font=\tiny] (0.90,2.30) node[rotate=-90]{$\hat{\tilde\mu}$};
\draw (0.80,2.45) -- ++(-0.45,0.45);
\draw(0.80,2.45) -- ++(0.45,0.45);
\filldraw[fill=cyan, draw=black] (0.80,2.45)--++ (0.30,0.30)--++(-0.60,0) -- cycle;
\draw[text= black, font=\tiny] (0.80,2.60) node[rotate=-90]{-};
\filldraw[fill=gray, draw=red, line width =1]  (0,2.9) rectangle ++(0.70,0.70);
\draw[text = white, font=\tiny] (0.35,3.25) node[rotate=-90] {2N-2};
\draw[text= black, font=\tiny] (0.35,3.8) node[rotate=-90]{B};
\draw[text = black, font=\tiny]  (0,3.25)--++(-0.2,0) ++(-0.2,-0.05) node[rotate=-90]{$M_B$};
\filldraw[fill=gray, draw=blue, line width =1]  (1.30, 2.95) circle (0.05 cm);
\end{tikzpicture}
\end{turn}
\caption{Intriligator-Seiberg dual: $\CU^{SO}_{c1}$}
\label{fig:soduality2}
\end{subfigure}

\begin{subfigure}[b]{2.8 in}
\centering
\begin{turn}{90}
\begin{tikzpicture}[scale=1.3, every node/.style={transform shape}]
\filldraw[fill=gray, draw=blue, line width =1]  (0,0) rectangle (0.70,0.70);
\draw[text = white, font=\tiny] (0.35,0.35) node[rotate=-90] {2N-2};
\draw[text= black, font=\tiny] (0.35,-0.2) node[rotate=-90] {D};
\draw[text = black, font=\tiny]  (0,0.35)--++(-0.2,0) ++(-0.15,-0.0) node[rotate=-90] {$\langle M_D \rangle$};
\draw (0.35,0.7)-- ++(0.45,0.45);

\filldraw[fill=pink, draw=black] (0.50,0.85)--(0.80,1.15)--(1.10,0.85) -- cycle;
\draw[text= black, font=\tiny] (0.80,1.0) node[rotate=-90]{+};

\draw (0.80,1.15)-- ++(0.45,-0.45);
\filldraw[fill=gray, draw=blue, line width =1]  (0.90,0) rectangle ++(0.70,0.70);
\draw[text = white, font=\tiny] (1.25,0.35) node[rotate=-90] {2N-2};
\draw[text= black, font=\tiny] (1.25,-0.2) node[rotate=-90]{C};
\draw[text = black, font=\tiny]  (1.6,0.35)--++(0.2,0) ++(0.15,0)node[rotate=-90]{$M_C$};
\draw (0.80,1.15) -- (0.80,1.45);
\draw[text=black,font=\tiny] (0.90,1.30) node[rotate=-90]{$\hat\mu$};
\filldraw[fill=white, draw=black]  (0.80,1.80) circle (0.35 cm);
\draw[text = black, font=\tiny] (0.80,1.80) node[rotate=-90] {2N};
\draw (0.80,2.15) --(0.80, 2.45);
\draw[text=black,font=\tiny] (0.90,2.30) node[rotate=-90]{$\hat{\tilde\mu}$};
\draw (0.80,2.45) -- ++(-0.45,0.45);
\draw(0.80,2.45) -- ++(0.45,0.45);
\filldraw[fill=cyan, draw=black] (0.80,2.45)--++ (0.30,0.30)--++(-0.60,0) -- cycle;
\draw[text= black, font=\tiny] (0.80,2.60) node[rotate=-90]{-};
\filldraw[fill=gray, draw=red, line width =1]  (0,2.9) rectangle ++(0.70,0.70);
\draw[text = white, font=\tiny] (0.35,3.25) node[rotate=-90] {2N-2};
\draw[text= black, font=\tiny] (0.35,3.8) node[rotate=-90]{B};
\draw[text = black, font=\tiny]  (0,3.25)--++(-0.2,0) ++(-0.2,-0.05) node[rotate=-90]{$M_B$};
\filldraw[fill=gray, draw=red, line width =1]  (0.9,2.9) rectangle ++(0.70,0.70);
\draw[text = white, font=\tiny] (1.25,3.25) node[rotate=-90] {2N-2};
\draw[text= black, font=\tiny] (1.25,3.8) node[rotate=-90]{A};
\draw[text = black, font=\tiny]  (1.6,3.25)--++(0.2,0) ++(0.15,0)node[rotate=-90]{$\langle M_A \rangle$};
\end{tikzpicture}
\end{turn}
\caption{Swapped dual : $\CU^{SO}_s$}
\label{fig:soduality3}
\end{subfigure}
\begin{subfigure}[b]{2.8in}
\begin{turn}{90}
\begin{tikzpicture}[scale=1.3, every node/.style={transform shape}]
\filldraw[fill=gray, draw=red, line width =1]  (0,0) rectangle (0.70,0.70);
\draw[text = white, font=\tiny] (0.35,0.35) node[rotate=-90] {2N-2};
\draw[text= black, font=\tiny] (0.35,-0.2) node[rotate=-90]{B};
\draw (0.35,0.7)-- ++(0.45,0.45);

\filldraw[fill=pink, draw=black] (0.50,0.85)--(0.80,1.15)--(1.10,0.85) -- cycle;
\draw[text= black, font=\tiny] (0.80,1.0)  node[rotate =-90]{+};

\draw (0.80,1.15)-- ++(0.45,-0.45);
\filldraw[fill=gray, draw=blue, line width =1]  (0.90,0) rectangle ++(0.70,0.70);
\draw[text = white, font=\tiny] (1.25,0.35)  node[rotate =-90] {2N-2};
\draw[text= black, font=\tiny] (1.25,-0.2)  node[rotate =-90]{C};
\draw[text = black, font=\tiny]  (1.6,0.35)--++(0.2,0) ++(0.11,-0.1) node[rotate =-90]{$M_C$};
\draw (0.80,1.15) -- (0.80,1.45);
\draw[text=black,font=\tiny] (0.90,1.30)  node[rotate =-90]{$\hat\mu$};
\filldraw[fill=white, draw=black]  (0.80,1.80) circle (0.35 cm);
\draw[text = black, font=\tiny] (0.80,1.80)  node[rotate =-90] {2N};
\draw (0.80,2.15) --(0.80, 2.45);
\draw[text=black,font=\tiny] (0.90,2.30)  node[rotate =-90]{$\hat{\tilde\mu}$};
\draw (0.80,2.45) -- ++(-0.45,0.45);
\draw(0.80,2.45) -- ++(0.45,0.45);
\filldraw[fill=cyan, draw=black] (0.80,2.45)--++ (0.30,0.30)--++(-0.60,0) -- cycle;
\draw[text= black, font=\tiny] (0.80,2.60)  node[rotate =-90]{-};
\filldraw[fill=gray, draw=red, line width =1]  (0,2.9) rectangle ++(0.70,0.70);
\draw[text = white, font=\tiny] (0.35,3.25)  node[rotate =-90]{2N-2};
\draw[text= black, font=\tiny] (0.35,3.8)  node[rotate =-90]{A};
\draw[text = black, font=\tiny]  (0,3.25)--++(-0.2,0) ++(-0.11,-0.07)  node[rotate =-90]{$\langle M_A \rangle$};
\filldraw[fill=gray, draw=blue, line width =1]  (1.30, 2.95) circle (0.05 cm);
\end{tikzpicture}
\end{turn}
\caption{Argyres-Seiberg type dual: $\CU^{SO}_{as}$ }
\label{fig:soduality4}
\end{subfigure}

\begin{subfigure}[b]{2.8in}
\centering
\begin{turn}{90}
\begin{tikzpicture}[scale=1.3, every node/.style={transform shape}]
\filldraw[fill=gray, draw=red, line width =1]  (0,0) rectangle (0.70,0.70);
\draw[text = white, font=\tiny] (0.35,0.35) node[rotate =-90] {2N-2};
\draw[text= black, font=\tiny] (0.35,-0.2) node[rotate =-90]{A};
\draw[text = black, font=\tiny]  (0,0.35)--++(-0.2,0) ++(-0.11,-0.1) node[rotate =-90]{$\langle M_A \rangle$};
\draw (0.35,0.7)-- ++(0.45,0.45);

\filldraw[fill=cyan, draw=black] (0.50,0.85)--(0.80,1.15)--(1.10,0.85) -- cycle;
\draw[text= black, font=\tiny] (0.80,1.0) node[rotate =-90]{-};

\draw (0.80,1.15)-- ++(0.45,-0.45);
\filldraw[fill=gray, draw=blue, line width =1]  (0.90,0) rectangle ++(0.70,0.70);
\draw[text = white, font=\tiny] (1.25,0.35) node[rotate =-90]{2N-2};
\draw[text= black, font=\tiny] (1.25,-0.2) node[rotate =-90]{C};
\draw (0.80,1.15) -- (0.80,1.45);
\draw[text=black,font=\tiny] (0.90,1.30) node[rotate =-90]{$\hat{\tilde\mu}$};
\filldraw[fill=white, draw=black]  (0.80,1.80) circle (0.35 cm);
\draw[text = black, font=\tiny] (0.80,1.80) node[rotate =-90] {2N};
\draw (0.80,2.15) --(0.80, 2.45);
\draw[text=black,font=\tiny] (0.90,2.30) node[rotate =-90]{$\hat{\mu}$};
\draw (0.80,2.45) -- ++(-0.45,0.45);
\draw(0.80,2.45) -- ++(0.45,0.45);
\filldraw[fill=pink, draw=black] (0.80,2.45)--++ (0.30,0.30)--++(-0.60,0) -- cycle;
\draw[text= black, font=\tiny] (0.80,2.60) node[rotate =-90]{+};
\filldraw[fill=gray, draw=red, line width =1]  (0,2.9) rectangle ++(0.70,0.70);
\draw[text = white, font=\tiny] (0.35,3.25) node[rotate =-90] {2N-2};
\draw[text= black, font=\tiny] (0.35,3.8) node[rotate =-90]{B};
\filldraw[fill=gray, draw=blue, line width =1]  (0.9,2.9) rectangle ++(0.70,0.70);
\draw[text = white, font=\tiny] (1.25,3.25)node[rotate =-90] {2N-2};
\draw[text= black, font=\tiny] (1.25,3.8) node[rotate =-90]{D};
\draw[text = black, font=\tiny]  (1.6,3.25)--++(0.2,0) ++(0.1,-0.1)node[rotate =-90]{$\langle M_D \rangle$};
\end{tikzpicture}
\end{turn}
\caption{Crossing type dual: $\CU^{SO}_{c2}$}
\label{fig:soduality5}
\end{subfigure}
\caption{Dual frames of $SO(2N)$ SQCD}
\label{fig:soduality}
\end{figure}

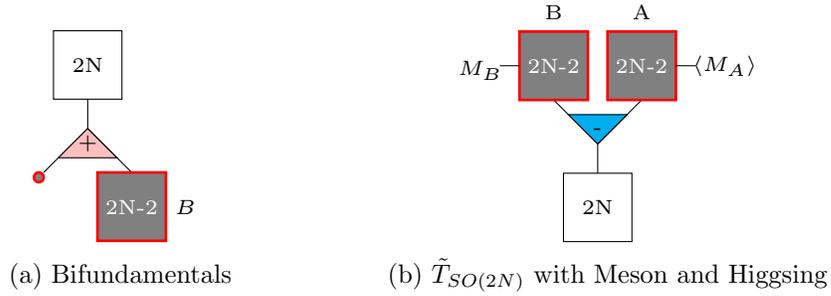
\begin{figure}[h]
\centering
\begin{subfigure}[b]{2.5in}
\centering
\begin{tikzpicture}[scale=1.3, every node/.style={transform shape}]
\filldraw[fill=gray, draw=red, line width =1]  (0.30, 0.65) circle (0.05 cm);
\draw (0.35,0.7)-- ++(0.45,0.45);

\filldraw[fill=pink, draw=black] (0.50,0.85)--(0.80,1.15)--(1.10,0.85) -- cycle;
\draw[text= black, font=\tiny] (0.80,1.0) node{+};

\draw (0.80,1.15)-- ++(0.45,-0.45);
\filldraw[fill=gray, draw=red, line width =1]  (0.90,0) rectangle ++(0.70,0.70);
\draw[text = white, font=\tiny] (1.25,0.35) node {2N-2};
\draw[text = black, font=\tiny]  (1.6,0.35) ++(0.2,0)node{$B$};
\draw (0.80,1.15) -- (0.80,1.45);
\filldraw[fill=white, draw=black]  (0.45,1.45) rectangle ++(0.70,0.70);
\draw[text = black, font=\tiny] (0.80,1.80) node {2N};
\end{tikzpicture}
\caption{Bifundamentals}
\end{subfigure}
\begin{subfigure}[b]{2.5in}
\centering
\begin{tikzpicture}[scale=1.3, every node/.style={transform shape}]
\filldraw[fill=white, draw=black]  (0.45,1.45) rectangle ++(0.70,0.70);
\draw[text = black, font=\tiny] (0.80,1.80) node {2N};
\draw (0.80,2.15) --(0.80, 2.45);
\draw (0.80,2.45) -- ++(-0.45,0.45);
\draw(0.80,2.45) -- ++(0.45,0.45);
\filldraw[fill=cyan, draw=black] (0.80,2.45)--++ (0.30,0.30)--++(-0.60,0) -- cycle;
\draw[text= black, font=\tiny] (0.80,2.60) node{-};
\filldraw[fill=gray, draw=red, line width =1]  (0,2.9) rectangle ++(0.70,0.70);
\draw[text = white, font=\tiny] (0.35,3.25) node {2N-2};
\draw[text= black, font=\tiny] (0.35,3.8) node{B};
\draw[text = black, font=\tiny]  (0,3.25)--++(-0.2,0) ++(-0.2,-0.05) node{$M_B$};
\filldraw[fill=gray, draw=red, line width =1]  (0.9,2.9) rectangle ++(0.70,0.70);
\draw[text = white, font=\tiny] (1.25,3.25) node {2N-2};
\draw[text= black, font=\tiny] (1.25,3.8) node{A};
\draw[text = black, font=\tiny]  (1.6,3.25)--++(0.2,0) ++(0.3,0)node{$\langle M_A \rangle$};
\end{tikzpicture}
\caption{$\tilde{T}_{SO(2N)}$ with Meson and Higgsing}
\end{subfigure}
\caption{Building blocks used to construct the electric and the swapped frames}
\label{fig:tsp+mm}
\end{figure}

The various duality frames obtained after Higgsing some of the $USp(2N-2)$ punctures are shown in figure \ref{fig:soduality}. 
The theories $\CU^{SO}$ and $\CU^{SO}_{c1}$ are related by Intriligator-Seiberg duality and their anomalies match in the usual manner.  For the purpose of matching the anomalies between $\CU^{SO}$ and $\CU^{SO}_s $, we observe that we only have to match the anomalies of the $SO(2N) \times USp(2N-2)$ bifundamental to the anomalies of the $\tilde{T}_{SO(2N)}$ block appropriately coupled to mesons (figure \ref{fig:tsp+mm}). 
For the  bifundamental we have 
\begin{equation}
\tr R \big|_{\textrm{bifund}}= \left(-\frac{1}{2} \right) (2N)(2N-2) = -N(2N-2) \ . 
\end{equation}
Note that on the dual side, after giving a vev to the meson $M_A$, the R-charge gets shifted: $R \rightarrow R-\rho(\sigma^3)$. This will not affect the contribution of the $\tilde{T}_{SO(2N)}$ block, since its $\tr \rho(\sigma^3) =0$. For the mesons, we will only consider the contributions of $(M_A)_{j ,-j}$ since the others decouple. This implies 
\begin{equation}
\begin{split}
\tr R \big|_{\langle M_A \rangle} &= \sum_j j
=\sum_{n=1}^{N-1} (2n-1) 
= (N-1)^2 \ . 
\end{split}
\end{equation} 
The meson $M_B$ does not contribute to the $R$-anomalies since its $R$-charge is not shifted and is equal to 1. Putting these together, we find that in this frame 
\begin{equation}
\tr R = \tr R \big|_{\tilde{T}_{SO(2N)}} + \tr R \big|_{\vev{M_A}} 
=-N(2N-2) \ , 
\end{equation} 
which matches with the corresponding anomaly of the bifundamental. 

Moving on, we now compare the $\tr R^3$ anomalies on the two sides and find 
\begin{equation}
\tr R^3 \big|_{\textrm{bifund}}=\left(-\frac{1}{2} \right)^3(2N)(2N-2) = -\frac{1}{2}N(N-1) \ . 
\end{equation}
On the dual side, since $R= R_0 - \rho(\sigma^3)$, where $R_0 = \frac{1}{2} R_{\mathcal{N}=2} + I_3$, therefore 
\begin{equation}
\tr R^3= \tr R_0^3 + 3 \tr R\rho^2 \ . 
\label{eq:r3an}
\end{equation}
Also $3 \tr R \rho^2 \delta^{ab}= \frac{3 \mathcal{I}}{2} \tr R_{\mathcal{N}=2} T^a_AT^b_A$, where $\mathcal{I}$ is the $SU(2)$ embedding index. Since our embedding takes $2N-2$ dimensional representation of $USp(2N-2)$ to the $2N-2$ dimensional representation of $SU(2)$, therefore 
\begin{equation}
\begin{split}
\mathcal{I} &= 2\sum_{j_z=1/2}^{N-3/2} j_z^2
=\frac{1}{6}(N-1)(4 N^2-8N+3) \ . 
\end{split}
\label{eq:emi}
\end{equation}
Thus, due to the shift in $R$-charges the $\tilde{T}_{SO(2N)}$ now contributes
\begin{equation}
\begin{split}
\tr R^3&= \tr R_0^3 + 3 \tr R\rho^2\\
&= -1 + \frac{13}{2} N -\frac{23}{2} N^2 + 8 N^3 - 2N^4 \ . 
\end{split}
\label{eq:embedding}
\end{equation}
Also 
\begin{equation}
\begin{split}
\tr R^3 \big|_{\langle M_A \rangle} = \sum_j j^3 
=\sum_{n=1}^{N-1} (2n-1)^3 
= 1-6N+11N^2-8N^3+2N^4 \ . 
\end{split}
\label{eq:mesons}
\end{equation}
Adding the contributions of the $\tilde{T}_{SO(2N)}$ block and the mesons we find that the $\tr R^3$ anomalies match with those of the bifundamental. The $\tr R T^a_B T^b_B$ and $\tr \mathcal{F} T^a_B T^b_B$ anomalies for the bifundamental are given by $(-\frac{1}{2} )(2N)$ and $(-1)(2N)$ respectively. On the dual side these have the same values as in the scenario before Higgsing. This is because  $\tr \rho T^a_B T^b_B = 0$ for the $\tilde{T}_{SO(2N)}$ block. We therefore conclude that these anomalies have the same value in the electric and the swapped theory.  

Similarly, we can match the anomaly coefficients of $\CU^{SO}$ and $\CU^{SO}_{as}$.  In $\CU^{SO}_{as}$ we have (up to the gaugino-contributions)
\begin{align}
\tr R &= \tr R|_{\tilde{T}_{SO(2N)}} + \tr R|_{\vev{M_A}} + 2N \sum_{m}\bigg(-\half -m \bigg) = -2N(2N-2)\\
\tr R^3 &= \tr R^3|_{\tilde{T}_{SO(2N)}} + \tr R^3|_{\vev{M_A}} + 2N \sum_{m}\bigg(-\half -m \bigg)^3 = -N(N-1)
\end{align}
which match with those in the electric theory. The coefficients of $\tr R T^a_B T^b_B$, $\tr R T^a_C T^b_C$, $\tr \CF T^a_B T^b_B$ and  $\tr \CF T^a_C T^b_C$ are not affected by Higgsing and therefore match with their electric counterparts.

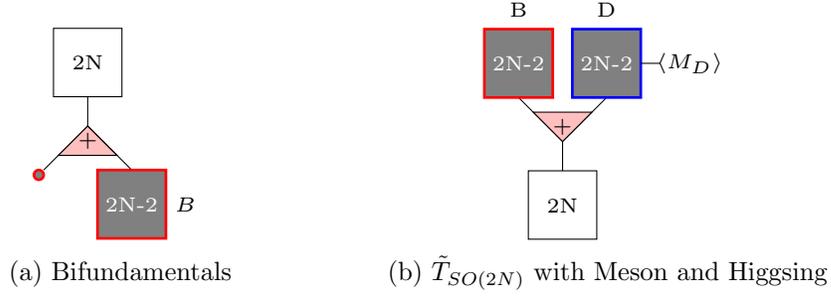
\begin{figure}[h]
\centering
\begin{subfigure}[b]{2.5in}
\centering
\begin{tikzpicture}[scale=1.3, every node/.style={transform shape}]
\filldraw[fill=gray, draw=red, line width =1]  (0.30, 0.65) circle (0.05 cm);
\draw (0.35,0.7)-- ++(0.45,0.45);

\filldraw[fill=pink, draw=black] (0.50,0.85)--(0.80,1.15)--(1.10,0.85) -- cycle;
\draw[text= black, font=\tiny] (0.80,1.0) node{+};

\draw (0.80,1.15)-- ++(0.45,-0.45);
\filldraw[fill=gray, draw=red, line width =1]  (0.90,0) rectangle ++(0.70,0.70);
\draw[text = white, font=\tiny] (1.25,0.35) node {2N-2};
\draw[text = black, font=\tiny]  (1.6,0.35) ++(0.2,0)node{$B$};
\draw (0.80,1.15) -- (0.80,1.45);
\filldraw[fill=white, draw=black]  (0.45,1.45) rectangle ++(0.70,0.70);
\draw[text = black, font=\tiny] (0.80,1.80) node {2N};
\end{tikzpicture}
\caption{Bifundamentals}
\end{subfigure}
\begin{subfigure}[b]{2.5in}
\centering
\begin{tikzpicture}[scale=1.3, every node/.style={transform shape}]
\filldraw[fill=white, draw=black]  (0.45,1.45) rectangle ++(0.70,0.70);
\draw[text = black, font=\tiny] (0.80,1.80) node {2N};
\draw (0.80,2.15) --(0.80, 2.45);
\draw (0.80,2.45) -- ++(-0.45,0.45);
\draw(0.80,2.45) -- ++(0.45,0.45);
\filldraw[fill=pink, draw=black] (0.80,2.45)--++ (0.30,0.30)--++(-0.60,0) -- cycle;
\draw[text= black, font=\tiny] (0.80,2.60) node{+};
\filldraw[fill=gray, draw=red, line width =1]  (0,2.9) rectangle ++(0.70,0.70);
\draw[text = white, font=\tiny] (0.35,3.25) node {2N-2};
\draw[text= black, font=\tiny] (0.35,3.8) node{B};
\filldraw[fill=gray, draw=blue, line width =1]  (0.9,2.9) rectangle ++(0.70,0.70);
\draw[text = white, font=\tiny] (1.25,3.25) node {2N-2};
\draw[text= black, font=\tiny] (1.25,3.8) node{D};
\draw[text = black, font=\tiny]  (1.6,3.25)--++(0.2,0) ++(0.3,0)node{$\langle M_D \rangle$};
\end{tikzpicture}
\caption{$\tilde{T}_{SO(2N)}$ with Meson and Higgsing}
\end{subfigure}
\caption{Building blocks used to construct the electric and the crossing frames}
\label{fig:tspmd}
\end{figure}

The anomaly coefficients in $\CU^{SO}_{c2}$ can also be matched to those in the other duality frames. This follows from the matching between the anomalies of the bifundamental and the $\tilde{T}_{SO(2N)}$ block (with mesons) shown in figure \ref{fig:tspmd}.

\section{Dualities for $USp(2N-2)$ gauge theory} \label{sec:Sp}
We now repeat the same procedure as the previous section for $USp(2N-2)$ gauge theory with $4N$ fundamentals. 

\subsection{Dualities for $USp(2N-2)$-coupled $\tilde{T}_{SO(2N)}$ theories}
\begin{figure}[h]
\centering
\begin{subfigure}[b]{2.8in}
\centering
\begin{turn}{90}
\begin{tikzpicture}[scale=1.25, every node/.style={transform shape}]
\filldraw[fill=gray, draw=red, line width=1]  (0,0) rectangle (0.70,0.70);
\draw[text = white, font=\tiny] (0.35,0.35) node[rotate=-90] {2N-2};
\draw[text= black, font=\tiny] (0.35,-0.2) node[rotate=-90]{A};
\draw (0.35,0.7)-- ++(0.45,0.45);

\filldraw[fill=pink, draw=black] (0.50,0.85)--(0.80,1.15)--(1.10,0.85) -- cycle;
\draw[text= black, font=\tiny] (0.80,1.0) node[rotate=-90]{+};

\draw (0.80,1.15)-- ++(0.45,-0.45);
\filldraw[fill=white, draw=red, line width=1]  (0.90,0) rectangle ++(0.70,0.70);
\draw[text = black, font=\tiny] (1.25,0.35) node[rotate=-90] {2N};
\draw[text= black, font=\tiny] (1.25,-0.2) node[rotate=-90]{B};
\draw (0.80,1.15) -- (0.80,1.45);
\draw[text=black,font=\tiny] (0.90,1.30) node[rotate=-90]{$\mu$};
\filldraw[fill=gray, draw=black]  (0.80,1.80) circle (0.35 cm);
\draw[text = white, font=\tiny] (0.80,1.80) node[rotate=-90] {2N-2};
\draw (0.80,2.15) --(0.80, 2.45);
\draw[text=black,font=\tiny] (0.90,2.30) node[rotate=-90]{$\tilde\mu$};
\draw (0.80,2.45) -- ++(-0.45,0.45);
\draw(0.80,2.45) -- ++(0.45,0.45);
\filldraw[fill=cyan, draw=black] (0.80,2.45)--++ (0.30,0.30)--++(-0.60,0) -- cycle;
\draw[text= black, font=\tiny] (0.80,2.60) node[rotate=-90]{-};
\filldraw[fill=white, draw=blue, line width=1]  (0,2.9) rectangle ++(0.70,0.70);
\draw[text = black, font=\tiny] (0.35,3.25) node[rotate=-90] {2N};
\draw[text= black, font=\tiny] (0.35,3.8) node[rotate=-90]{C};
\filldraw[fill=gray, draw=blue,line width=1]  (0.9,2.9) rectangle ++(0.70,0.70);
\draw[text = white, font=\tiny] (1.25,3.25) node[rotate=-90] {2N-2};
\draw[text= black, font=\tiny] (1.25,3.8) node[rotate=-90]{D};
\draw[text= black, font=\scriptsize] (-0.7,1.7) node[rotate=-90]{$W=c \tr \mu \tilde\mu$};
\end{tikzpicture}
\end{turn}
\caption{Electric theory: $\CT^{Sp}$.}
\label{fig:Ssp}
\end{subfigure}
\begin{subfigure}[b]{2.8in}
\centering
\begin{turn}{90}
\begin{tikzpicture}[scale=1.25, every node/.style={transform shape}]
\filldraw[fill=gray, draw=red, line width=1]  (0,0) rectangle (0.70,0.70);
\draw[text = white, font=\tiny] (0.35,0.35) node[rotate=-90] {2N-2};
\draw[text= black, font=\tiny] (0.35,-0.2) node[rotate=-90]{A};
\draw (0.35,0.7)-- ++(0.45,0.45);

\filldraw[fill=pink, draw=black] (0.50,0.85)--(0.80,1.15)--(1.10,0.85) -- cycle;
\draw[text= black, font=\tiny] (0.80,1.0) node[rotate=-90]{+};

\draw (0.80,1.15)-- ++(0.45,-0.45);
\filldraw[fill=white, draw=blue, line width=1]  (0.90,0) rectangle ++(0.70,0.70);
\draw[text = black, font=\tiny] (1.25,0.35) node[rotate=-90] {2N};
\draw[text= black, font=\tiny] (1.25,-0.2) node[rotate=-90]{C};
\draw[text = black, font=\tiny]  (1.6,0.35)--++(0.2,0) ++(0.2,0)node[rotate=-90]{$M_C$};
\draw (0.80,1.15) -- (0.80,1.45);
\draw[text=black,font=\tiny] (0.90,1.30) node[rotate=-90]{$\hat\mu$};
\filldraw[fill=gray, draw=black]  (0.80,1.80) circle (0.35 cm);
\draw[text = white, font=\tiny] (0.80,1.80) node[rotate=-90] {2N-2};
\draw (0.80,2.15) --(0.80, 2.45);
\draw[text=black,font=\tiny] (0.90,2.30) node[rotate=-90]{$\hat{\tilde\mu}$};
\draw (0.80,2.45) -- ++(-0.45,0.45);
\draw(0.80,2.45) -- ++(0.45,0.45);
\filldraw[fill=cyan, draw=black] (0.80,2.45)--++ (0.30,0.30)--++(-0.60,0) -- cycle;
\draw[text= black, font=\tiny] (0.80,2.60) node[rotate=-90]{-};
\filldraw[fill=white, draw=red,line width=1]  (0,2.9) rectangle ++(0.70,0.70);
\draw[text = black, font=\tiny] (0.35,3.25) node[rotate=-90] {2N};
\draw[text= black, font=\tiny] (0.35,3.8) node[rotate=-90]{B};
\draw[text = black, font=\tiny]  (0,3.25)--++(-0.2,0) ++(-0.2,-0.05) node[rotate=-90]{$M_B$};
\filldraw[fill=gray, draw=blue, line width=1]  (0.9,2.9) rectangle ++(0.70,0.70);
\draw[text = white, font=\tiny] (1.25,3.25) node[rotate=-90] {2N-2};
\draw[text= black, font=\tiny] (1.25,3.8) node[rotate=-90]{D};
\draw[text= black, font=\scriptsize] (-0.7,1.7) node[rotate=-90]{$W=\hat c \tr \hat\mu\hat{\tilde\mu} + \tr M_C\hat\mu_C + \tr M_B\hat\mu_B$};
\end{tikzpicture}
\end{turn}
\caption{Crossing frame 1: $\CT^{Sp}_{c1}$.}
\label{fig:Sspcc}
\end{subfigure}
\begin{subfigure}[b]{2.8 in}
\centering
\begin{turn}{90}
\begin{tikzpicture}[scale=1.25, every node/.style={transform shape}]
\filldraw[fill=gray, draw=blue,line width=1]  (0,0) rectangle (0.70,0.70);
\draw[text = white, font=\tiny] (0.35,0.35) node[rotate=-90] {2N-2};
\draw[text= black, font=\tiny] (0.35,-0.2) node[rotate=-90]{D};
\draw[text = black, font=\tiny]  (0,0.35)--++(-0.2,0) ++(-0.2,0) node[rotate=-90]{$ M_D $};
\draw (0.35,0.7)-- ++(0.45,0.45);

\filldraw[fill=pink, draw=black] (0.50,0.85)--(0.80,1.15)--(1.10,0.85) -- cycle;
\draw[text= black, font=\tiny] (0.80,1.0) node[rotate=-90]{+};

\draw (0.80,1.15)-- ++(0.45,-0.45);
\filldraw[fill=white, draw=red,line width=1]  (0.90,0) rectangle ++(0.70,0.70);
\draw[text = black, font=\tiny] (1.25,0.35) node[rotate=-90] {2N};
\draw[text= black, font=\tiny] (1.25,-0.2) node[rotate=-90]{B};
\draw (0.80,1.15) -- (0.80,1.45);
\draw[text=black,font=\tiny] (0.90,1.30) node[rotate=-90]{$\hat\mu$};
\filldraw[fill=gray, draw=black]  (0.80,1.80) circle (0.35 cm);
\draw[text = white, font=\tiny] (0.80,1.80) node[rotate=-90]{2N-2};
\draw (0.80,2.15) --(0.80, 2.45);
\draw[text=black,font=\tiny] (0.90,2.30) node[rotate=-90]{$\hat{\tilde\mu}$};
\draw (0.80,2.45) -- ++(-0.45,0.45);
\draw(0.80,2.45) -- ++(0.45,0.45);
\filldraw[fill=cyan, draw=black] (0.80,2.45)--++ (0.30,0.30)--++(-0.60,0) -- cycle;
\draw[text= black, font=\tiny] (0.80,2.60) node[rotate=-90]{-};
\filldraw[fill=white, draw=blue,line width=1]  (0,2.9) rectangle ++(0.70,0.70);
\draw[text = black, font=\tiny] (0.35,3.25) node[rotate=-90]{2N};
\draw[text= black, font=\tiny] (0.35,3.8) node[rotate=-90]{C};
\filldraw[fill=gray, draw=red,line width=1]  (0.9,2.9) rectangle ++(0.70,0.70);
\draw[text = white, font=\tiny] (1.25,3.25) node[rotate=-90]{2N-2};
\draw[text= black, font=\tiny] (1.25,3.8) node[rotate=-90]{A};
\draw[text = black, font=\tiny]  (1.6,3.25)--++(0.2,0) ++(0.2,0) node[rotate=-90]{$ M_A $};
\draw[text= black, font=\scriptsize] (-0.7,1.8) node[rotate=-90]{$W=\hat c \tr \hat\mu\hat{\tilde\mu} + \tr M_A\hat\mu_A\ + \tr M_D\hat\mu_D$};
\end{tikzpicture}
\end{turn}
\caption{ Crossing frame 2: $\CT^{Sp}_{c2}$}
\label{fig:spcr}
\end{subfigure}
\begin{subfigure}[b]{2.8 in}
\centering
\begin{turn}{90}
\begin{tikzpicture}[scale=1.25, every node/.style={transform shape}]
\filldraw[fill=gray, draw=red,line width=1]  (0,0) rectangle (0.70,0.70);
\draw[text = white, font=\tiny] (0.35,0.35) node[rotate=-90] {2N-2};
\draw[text= black, font=\tiny] (0.35,-0.2) node[rotate=-90]{A};
\draw (0.35,0.7)-- ++(0.45,0.45);
\filldraw[fill=pink, draw=black] (0.50,0.85)--(0.80,1.15)--(1.10,0.85) -- cycle;
\draw[text= black, font=\tiny] (0.80,1.0) node{+};
\draw (0.80,1.15)-- ++(0.45,-0.45);
\filldraw[fill=gray, draw=blue,line width=1]  (0.90,0) rectangle ++(0.70,0.70);
\draw[text = white, font=\tiny] (1.25,0.35) node[rotate=-90] {2N-2};
\draw[text= black, font=\tiny] (1.25,-0.2) node[rotate=-90]{D};
\draw (0.80,1.15) -- (0.80,1.45);
\draw[text=black,font=\tiny] (0.90,1.30) node[rotate=-90]{$\hat\mu$};
\filldraw[fill=white, draw=black]  (0.80,1.80) circle (0.35 cm);
\draw[text = black, font=\tiny] (0.80,1.80) node[rotate=-90] {2N};
\draw (0.80,2.15) --(0.80, 2.45);
\draw[text=black,font=\tiny] (0.90,2.30) node[rotate=-90]{$\hat{\tilde\mu}$};
\draw (0.80,2.45) -- ++(-0.45,0.45);
\draw(0.80,2.45) -- ++(0.45,0.45);
\filldraw[fill=cyan, draw=black] (0.80,2.45)--++ (0.30,0.30)--++(-0.60,0) -- cycle;
\draw[text= black, font=\tiny] (0.80,2.60) node[rotate=-90]{-};
\filldraw[fill=white, draw=blue,line width=1]  (0,2.9) rectangle ++(0.70,0.70);
\draw[text = black, font=\tiny] (0.35,3.25) node[rotate=-90] {2N};
\draw[text= black, font=\tiny] (0.35,3.8) node[rotate=-90]{C};
\filldraw[fill=white, draw=red,line width=1]  (0.9,2.9) rectangle ++(0.70,0.70);
\draw[text = black, font=\tiny] (1.25,3.25) node[rotate=-90] {2N};
\draw[text= black, font=\tiny] (1.25,3.8) node[rotate=-90]{B};
\draw[text= black, font=\scriptsize] (-0.7,1.7) node[rotate=-90]{$W=\hat c \tr \hat{\mu}\hat{\tilde\mu} + \tr M_B\hat{\mu}_B + \tr M_D \hat{\mu}_D$};
\draw[text = black, font=\tiny]  (1.6,3.25)--++(0.2,0) ++(0.2,0)node[rotate=-90]{$M_B$};
\draw[text = black, font=\tiny]  (1.6,0.35)--++(0.2,0) ++(0.2,0)node[rotate=-90]{$M_D$};
\end{tikzpicture}
\end{turn}
\caption{Crossing frame 3: $\CT^{Sp}_{c3}$}
\label{fig:Sso}
\end{subfigure}
\begin{subfigure}[b]{3 in}
\centering
\begin{turn}{90}
\begin{tikzpicture}[scale=1.25 , every node/.style={transform shape}]
\filldraw[fill=gray, draw=blue, line width=1]  (0,0) rectangle (0.70,0.70);
\draw[text = white, font=\tiny] (0.35,0.35) node[rotate=-90] {2N-2};
\draw[text= black, font=\tiny] (0.35,-0.2) node[rotate=-90]{D};
\draw[text = black, font=\tiny]  (0,0.35)--++(-0.2,0) ++(-0.1,-0.07) node[rotate=-90]{$M_D$};
\draw (0.35,0.7)-- ++(0.45,0.45);

\filldraw[fill=pink, draw=black] (0.50,0.85)--(0.80,1.15)--(1.10,0.85) -- cycle;
\draw[text= black, font=\tiny] (0.80,1.0) node[rotate=-90]{+};

\draw (0.80,1.15)-- ++(0.45,-0.45);
\filldraw[fill=white, draw=blue,line width=1]  (0.90,0) rectangle ++(0.70,0.70);
\draw[text = black, font=\tiny] (1.25,0.35) node[rotate=-90] {2N};
\draw[text= black, font=\tiny] (1.25,-0.2) node[rotate=-90]{C};
\draw[text = black, font=\tiny]  (1.6,0.35)--++(0.2,0) ++(0.1,-0.02)node[rotate=-90]{$M_C$};
\draw (0.80,1.15) -- (0.80,1.45);
\draw[text=black,font=\tiny] (0.90,1.30) node[rotate=-90]{$\hat\mu$};
\filldraw[fill=gray, draw=black]  (0.80,1.80) circle (0.35 cm);
\draw[text = white, font=\tiny] (0.80,1.80) node[rotate=-90] {2N-2};
\draw (0.80,2.15) --(0.80, 2.45);
\draw[text=black,font=\tiny] (0.90,2.30) node[rotate=-90]{$\hat{\tilde\mu}$};
\draw (0.80,2.45) -- ++(-0.45,0.45);
\draw(0.80,2.45) -- ++(0.45,0.45);
\filldraw[fill=cyan, draw=black] (0.80,2.45)--++ (0.30,0.30)--++(-0.60,0) -- cycle;
\draw[text= black, font=\tiny] (0.80,2.60) node[rotate=-90]{-};
\filldraw[fill=white, draw=red,line width=1]  (0,2.9) rectangle ++(0.70,0.70);
\draw[text = black, font=\tiny] (0.35,3.25) node[rotate=-90] {2N};
\draw[text= black, font=\tiny] (0.35,3.8) node[rotate=-90]{B};
\draw[text = black, font=\tiny]  (0,3.25)--++(-0.2,0) ++(-0.1,-0.07) node[rotate=-90]{$M_B$};
\filldraw[fill=gray, draw=red, line width=1]  (0.9,2.9) rectangle ++(0.70,0.70);
\draw[text = white, font=\tiny] (1.25,3.25) node[rotate=-90] {2N-2};
\draw[text= black, font=\tiny] (1.25,3.8) node[rotate=-90]{A};
\draw[text = black, font=\tiny]  (1.6,3.25)--++(0.2,0) ++(0.1,-0.02)node[rotate=-90]{$M_A$};
\draw[text= black, font=\scriptsize] (-0.7,1.3) node[rotate=-90]{$W=\hat c \tr \hat\mu\hat{\tilde\mu} + \tr M_A\hat\mu_A + \tr M_B\hat\mu_B + \tr M_C\hat\mu_C + \tr M_D\hat\mu_D$};
\end{tikzpicture}
\end{turn}
\caption{Swapped theory: $\CT^{Sp}_s$.}
\label{fig:Swsps}
\end{subfigure}
\caption{The $\CT^{Sp}$ theory, obtained by coupling two $\tilde{T}_{SO(2N)}$ blocks along a $USp(2N-2)$ puncture with an $\CN=1$ vector multiplet, and its duals obtained by moving the punctures around. Here we omit the anti-symmetric forms in the superpotential. The red/blue color means $\s = \pm$. }
\label{fig:spsp}
\end{figure}
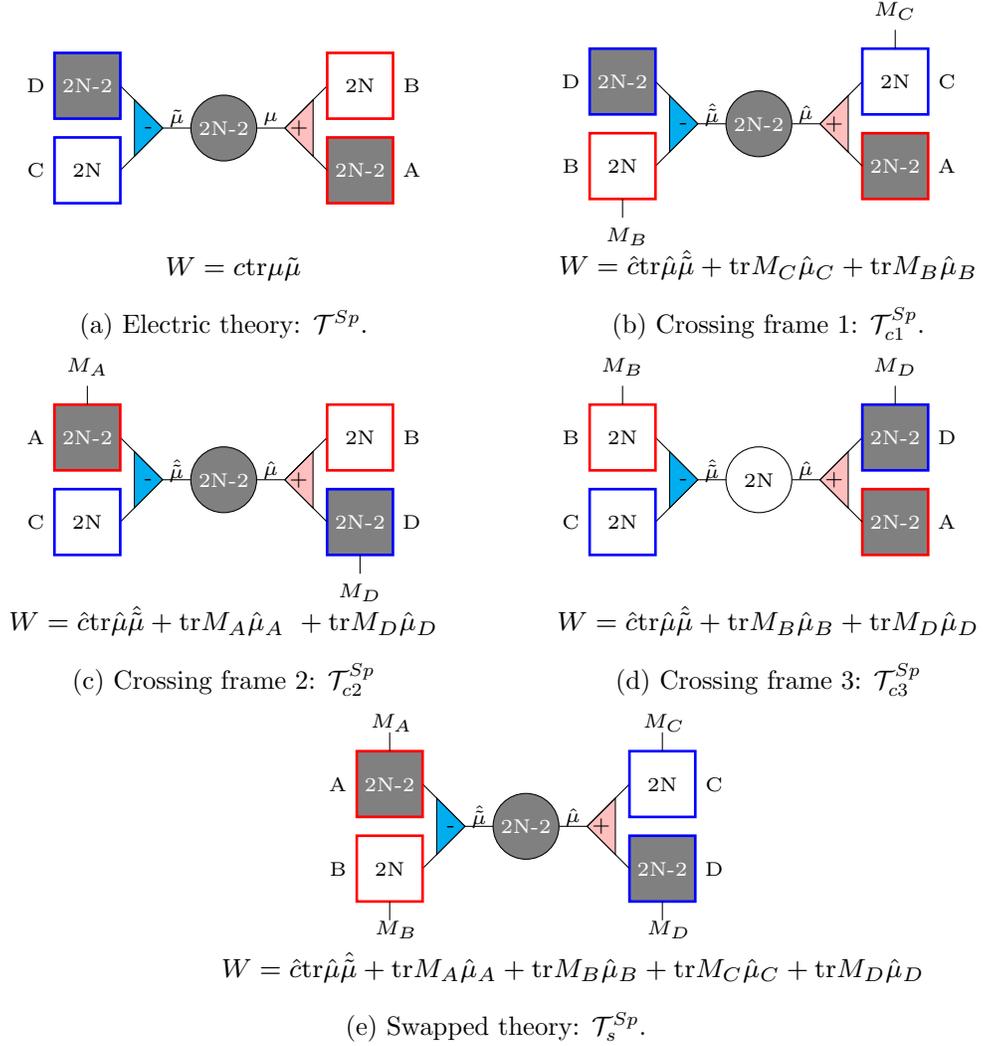

 We begin by considering two $\tilde{T}_{SO(2N)}$ blocks coupled to each other at a $USp(2N-2)$ puncture via an $\CN=1$ vector multiplet, giving the electric theory of figure \ref{fig:Ssp}. The superpotential for this theory is 
\be
 W=c \tr \mu \Omega \tilde\mu \Omega
\ee
We will henceforth denote this theory by $\CT^{Sp}$.  The frames dual to $\CT^{Sp}$ can be obtained by using the rules of section \ref{sec:N1M5} to move the punctures around. This gives us the set of theories shown in figure \ref{fig:spsp}.

We will call the theory in figure \ref{fig:Sspcc} as `crossing frame 1' and denote it by $\CT^{Sp}_{c1}$. It is obtained by exchanging punctures $B$ and $C$. Since these punctures will no longer have the same color as their pants, we will therefore have to integrate in mesons $M_B$ and $M_C$ transforming as the adjoints of the respective symmetries. The superpotential of the theory becomes
\be
W=\hat c \tr \hat\mu \Omega \hat{\tilde\mu} \Omega + \tr M_C\hat\mu_C + \tr M_B\hat\mu_B \ ,
\ee

Similarly when we exchange the puncture $A$ and $D$, we end up with the theory in `crossing frame 2'  (figure \ref{fig:spcr}) which is denoted by $\CT^{Sp}_{c2}$.  Once again we will have to couple mesons $M_A$ and $M_D$ via the superpotential 
\be
W=\hat c \tr \hat\mu \Omega \hat{\tilde\mu} \Omega + \tr M_A \Omega \hat\mu_A \Omega + \tr M_D \Omega \hat\mu_D \Omega \ ,
\ee

The theory in `crossing frame 3' (figure \ref{fig:Sso}) is obtained by exchanging punctures $B$ and $D$. This will correspond to a pair of pants decomposition where one of the pants has no outer automorphism twists. In other words it consists of an $\tilde{T}_{SO(2N)}$ block coupled to a $T_{SO(2N)}$ block at its $SO(2N)$ puncture. To compensate for the mismatch in the color of the punctures their respective pants, we will have to integrate in mesons $M_B$ and $M_D$ with the superpotential being
\be
W=\hat c \tr \hat\mu \hat{\tilde\mu} + \tr M_B\hat\mu_B + \tr M_D \Omega \hat\mu_D \Omega \  ,
\ee
Interestingly this gives us a duality between an $\CN=1$ theory with a $USp(2N-2)$ gauge group and a theory with $SO(2N)$ gauge group. We will denote the theory in this duality frame by $\CT^{Sp}_{c3}$.

The theory in figure \ref{fig:Swsps} will be called the `swapped' theory and we will denote it by $\CT^{Sp}_s$. It is obtained by moving the 4 punctures around such that none of them have the same color as the pants in which they reside. This will require us to integrate in mesons at each puncture. The superpotential will now become
\be
W=\hat c \tr \hat\mu\Omega\hat{\tilde\mu}\Omega + \tr M_A\Omega\hat\mu_A\Omega + \tr M_B\hat\mu_B + \tr M_C\hat\mu_C + \tr M_D\Omega\hat\mu_D\Omega \  .
\ee
\subsection{Dualities for $USp(2N-2)$ SQCD}
Now, let us consider the dual theories of SQCD. 

\subsubsection*{The Intriligator-Pouliot Duality}
\begin{figure}[h]
\centering
\begin{subfigure}[b]{2.8in}
\centering
\begin{tikzpicture}[scale=1.3, every node/.style={transform shape}]
\filldraw[fill=gray, draw=red, line width=1]  (0.30, 0.65) circle (0.05 cm);
\draw (0.35,0.7)-- ++(0.45,0.45);

\filldraw[fill=pink, draw=black] (0.50,0.85)--(0.80,1.15)--(1.10,0.85) -- cycle;
\draw[text= black, font=\tiny] (0.80,1.0) node{+};

\draw (0.80,1.15)-- ++(0.45,-0.45);
\filldraw[fill=white, draw=red, line width=1]  (0.90,0) rectangle ++(0.70,0.70);
\draw[text = black, font=\tiny] (1.25,0.35) node {2N};
\draw[text= black, font=\tiny] (1.25,-0.2) node{B};
\draw (0.80,1.15) -- (0.80,1.45);
\draw[text=black,font=\tiny] (0.90,1.30) node{$\mu$};
\filldraw[fill=gray, draw=black]  (0.80,1.80) circle (0.35 cm);
\draw[text = white, font=\tiny] (0.80,1.80) node {2N-2};
\draw (0.80,2.15) --(0.80, 2.45);
\draw[text=black,font=\tiny] (0.90,2.30) node{$\tilde\mu$};
\draw (0.80,2.45) -- ++(-0.45,0.45);
\draw(0.80,2.45) -- ++(0.45,0.45);
\filldraw[fill=cyan, draw=black] (0.80,2.45)--++ (0.30,0.30)--++(-0.60,0) -- cycle;
\draw[text= black, font=\tiny] (0.80,2.60) node{-};
\filldraw[fill=white, draw=blue,line width=1]  (0,2.9) rectangle ++(0.70,0.70);
\draw[text = black, font=\tiny] (0.35,3.25) node {2N};
\draw[text= black, font=\tiny] (0.35,3.8) node{C};
\filldraw[fill=gray, draw=blue, line width=1]  (1.30, 2.95) circle (0.05 cm);
\draw[text= black, font=\scriptsize] (0.7,-0.7) node{$W=c \tr \mu\tilde\mu$};
\end{tikzpicture}
\caption{Electric theory $\CU^{Sp}$}
\label{fig:spe}
\end{subfigure}
\quad
\begin{subfigure}[b]{2.8in}
\centering
\begin{tikzpicture}[scale=1.3, every node/.style={transform shape}]
\filldraw[fill=gray, draw=red,line width=1]  (0.30, 0.65) circle (0.05 cm);
\draw (0.35,0.7)-- ++(0.45,0.45);

\filldraw[fill=pink, draw=black] (0.50,0.85)--(0.80,1.15)--(1.10,0.85) -- cycle;
\draw[text= black, font=\tiny] (0.80,1.0) node{+};

\draw (0.80,1.15)-- ++(0.45,-0.45);
\filldraw[fill=white, draw=blue, line width=1]  (0.90,0) rectangle ++(0.70,0.70);
\draw[text = black, font=\tiny] (1.25,0.35) node {2N};
\draw[text= black, font=\tiny] (1.25,-0.2) node{C};
\draw[text = black, font=\tiny]  (1.6,0.35)--++(0.2,0) ++(0.2,0)node{$M_C$};
\draw (0.80,1.15) -- (0.80,1.45);
\draw[text=black,font=\tiny] (0.90,1.30) node{$\hat\mu$};
\filldraw[fill=gray, draw=black]  (0.80,1.80) circle (0.35 cm);
\draw[text = white, font=\tiny] (0.80,1.80) node {2N-2};
\draw (0.80,2.15) --(0.80, 2.45);
\draw[text=black,font=\tiny] (0.90,2.30) node{$\hat{\tilde\mu}$};
\draw (0.80,2.45) -- ++(-0.45,0.45);
\draw(0.80,2.45) -- ++(0.45,0.45);
\filldraw[fill=cyan, draw=black] (0.80,2.45)--++ (0.30,0.30)--++(-0.60,0) -- cycle;
\draw[text= black, font=\tiny] (0.80,2.60) node{-};
\filldraw[fill=white, draw=red,line width=1]  (0,2.9) rectangle ++(0.70,0.70);
\draw[text = black, font=\tiny] (0.35,3.25) node {2N};
\draw[text= black, font=\tiny] (0.35,3.8) node{B};
\draw[text = black, font=\tiny]  (0,3.25)--++(-0.2,0) ++(-0.2,-0.05) node{$M_B$};
\filldraw[fill=gray, draw=blue,line width=1]  (1.30, 2.95) circle (0.05 cm);
\draw[text= black, font=\scriptsize] (0.7,-0.7) node{$W=\hat c \tr \hat{\mu}\hat{\tilde\mu} + \tr M_C\hat\mu_C + \tr M_B\hat\mu_B$};
\end{tikzpicture}
\caption{Magnetic theory $\CU^{Sp}_{c1}$}
\label{fig:spm}
\end{subfigure}
\caption{Intriligator-Pouliot duality}
\end{figure}
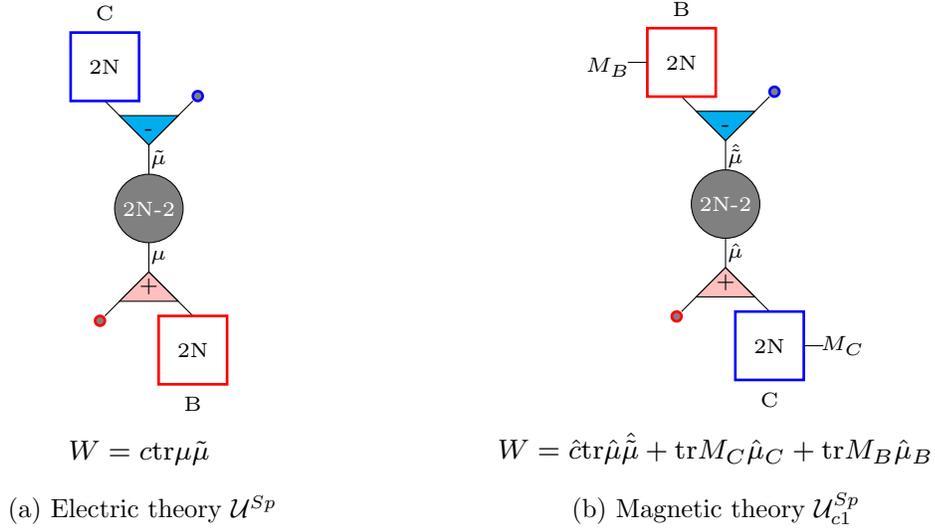
By Higgsing punctures $A$ and $D$ of $\CT^{Sp}$ and $\CT^{Sp}_{c1}$ (figure \ref{fig:Ssp} and \ref{fig:Sspcc}), with a vev to their adjoint representation operators, we obtain the usual pair of Intriligator-Pouliot dual theories \cite{Intriligator:1995ne}. The electric theory is given by figure \ref{fig:spe}. We will use the short-hand notation $\CU^{Sp}$ to denote this theory. It is a $USp(2N-2)$ gauge theory with $4N$ fundamental quarks. It has an $SO(2N)_B \times SO(2N)_C \subset SU(4N)$ global symmetry. Its superpotential is given by 
\begin{equation}
W= c \tr \mu\Omega\tilde\mu\Omega \ , 
\end{equation}
where now $\mu_{ij} = (Q_{\alpha i} Q_{\alpha j})_B$ and  $\tilde\mu_{ij} =( Q_{\alpha i} Q_{\alpha j})_C$. Here $(Q_B)_{\alpha i}$ is the quark transforming as the bifundamental of $SO(2N)_B \times USp(2N-2)$ while $(Q_C)_{\alpha i}$ is the bifundamental of $SO(2N)_C \times USp(2N-2)$.

Applying Intriligator-Pouliot duality to the above electric theory, we get a theory with $4N$ quarks $\hat{Q}$ transforming under a $USp(2N-2)$ gauge group. In the absence of any superpotential this theory will enjoy $SU(4N)$ global symmetry. The spectrum of the theory will also include mesons transforming in the anti-symmetric representation of $SU(4N)$. In terms of the $SO(2N)_B \times SO(2N)_C$ subgroup of the flavor symmetry the quarks split into bifundamentals of $SO(2N)_B \times USp(2N-2)$ and $SO(2N)_C \times USp(2N-2)$ while the meson splits into the following irreducible representations.
\begin{enumerate}
\item  anti-symmetric tensor of $SO(2N)_B$ : $M_{B \alpha \beta}$
\item  anti-symmetric tensor of $SO(2N)_C$ : $M_{C \alpha \beta}$
\item  bifundamental of $SO(2N)_B \times SO(2N)_C$ : $M_{\alpha \beta}$
\end{enumerate}
Note that $M_{\alpha\beta}$ is dual to the meson of the electric theory formed by $(Q_B)_{\alpha i} \Omega^{ij}( Q_C)_{\beta j}$. The dual superpotential becomes 
\begin{equation}
W_m = c \tr M M + \tr M_{B}\hat{Q}_{B}\Omega \hat{Q}_{B}+ \tr M_{C}\hat{Q}_{C}\Omega \hat{Q}_{C} + \tr M\hat{Q}_{B}\Omega \hat{Q}_{C}
\end{equation}
Integrating out the massive mesons $M_{\alpha\beta}$ then gives us the theory of figure \ref{fig:spm}. We will use $\CU^{Sp}_{c1}$ to denote this theory.


\subsubsection*{Non-Lagrangian dual 1: Swap} 
\begin{figure}[h]
\centering
\begin{turn}{90}
\begin{tikzpicture}[scale=1.3, every node/.style={transform shape}]
\filldraw[fill=gray, draw=blue, line width=1]  (0,0) rectangle (0.70,0.70);
\draw[text = white, font=\tiny] (0.35,0.35) node[rotate=-90] {2N-2};
\draw[text= black, font=\tiny] (0.35,-0.2) node[rotate=-90]{D};
\draw[text = black, font=\tiny]  (0,0.35)--++(-0.2,0) ++(-0.1,-0.07) node[rotate=-90]{$\vev{M_D}$};
\draw (0.35,0.7)-- ++(0.45,0.45);

\filldraw[fill=pink, draw=black] (0.50,0.85)--(0.80,1.15)--(1.10,0.85) -- cycle;
\draw[text= black, font=\tiny] (0.80,1.0) node[rotate=-90]{+};

\draw (0.80,1.15)-- ++(0.45,-0.45);
\filldraw[fill=white, draw=blue,line width=1]  (0.90,0) rectangle ++(0.70,0.70);
\draw[text = black, font=\tiny] (1.25,0.35) node[rotate=-90] {2N};
\draw[text= black, font=\tiny] (1.25,-0.2) node[rotate=-90]{C};
\draw[text = black, font=\tiny]  (1.6,0.35)--++(0.2,0) ++(0.08,-0.1)node[rotate=-90]{$M_C$};
\draw (0.80,1.15) -- (0.80,1.45);
\draw[text=black,font=\tiny] (0.90,1.30) node[rotate=-90]{$\hat\mu$};
\filldraw[fill=gray, draw=black]  (0.80,1.80) circle (0.35 cm);
\draw[text = white, font=\tiny] (0.80,1.80) node[rotate=-90] {2N-2};
\draw (0.80,2.15) --(0.80, 2.45);
\draw[text=black,font=\tiny] (0.90,2.30) node[rotate=-90]{$\hat{\tilde\mu}$};
\draw (0.80,2.45) -- ++(-0.45,0.45);
\draw(0.80,2.45) -- ++(0.45,0.45);
\filldraw[fill=cyan, draw=black] (0.80,2.45)--++ (0.30,0.30)--++(-0.60,0) -- cycle;
\draw[text= black, font=\tiny] (0.80,2.60) node[rotate=-90]{-};
\filldraw[fill=white, draw=red,line width=1]  (0,2.9) rectangle ++(0.70,0.70);
\draw[text = black, font=\tiny] (0.35,3.25) node [rotate=-90]{2N};
\draw[text= black, font=\tiny] (0.35,3.8) node[rotate=-90]{B};
\draw[text = black, font=\tiny]  (0,3.25)--++(-0.2,0) ++(-0.1,-0.07) node[rotate=-90]{$M_B$};
\filldraw[fill=gray, draw=red, line width=1]  (0.9,2.9) rectangle ++(0.70,0.70);
\draw[text = white, font=\tiny] (1.25,3.25) node[rotate=-90] {2N-2};
\draw[text= black, font=\tiny] (1.25,3.8) node[rotate=-90]{A};
\draw[text = black, font=\tiny]  (1.6,3.25)--++(0.2,0) ++(0.1,-0.1)node[rotate=-90]{$\vev{M_A}$};
\end{tikzpicture}
\end{turn}
\caption{Non-Lagrangian dual $\CU^{Sp}_s$ of $USp(2N-2)$ SQCD.}
\label{fig:Ssps}
\end{figure}
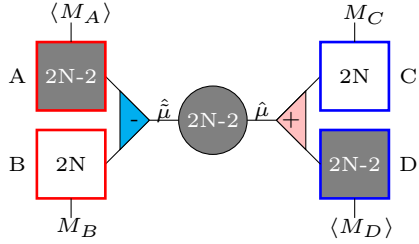
A non-Lagrangian dual (figure \ref{fig:Ssps}) of the electric theory $\CU^{Sp}$ is generated upon Higgsing the punctures $A$ and $D$ in $\CT^{Sp}_s$. This Higgsing is implemented by giving vev $\rho_{\varnothing}(\sigma^+)$ from eq.(\ref{eq:vevNull}) to the mesons $M_A$ and $M_D$. Upon considering the mesonic fluctuations around their vev and taking into account the breaking of flavor symmetries and the resulting non-conservation of their currents, we obtain the superpotential of our proposed non-Lagrangian dual:  
\begin{equation}
W=\hat c \tr \hat\mu\Omega\hat{\tilde\mu}\Omega + \tr M_C\hat\mu_C + \tr M_B\hat\mu_B +\sum_{j} (M_A)_{j,-j}(\hat{\mu}_A)_{j,j}+\sum_{j} (M_D)_{ j,-j}(\hat{\mu}_D)_{ j,j} \ .
\end{equation}
As usual the $R$- and $\CF$-charges get shifted to: 
\begin{equation}
\begin{aligned}
\mathcal{F} &= \mathcal{F}_0 + 2\rho^A(\sigma^3)- 2\rho^D(\sigma^3) \ , \\
R&=R_0 - \rho^A(\sigma^3)- \rho^D(\sigma^3) \ . 
\end{aligned}
\end{equation}
We will denote this theory by $\CU^{Sp}_{s}$. 

\subsubsection*{Non-Lagrangian dual 2: Argyres-Seiberg type dual}
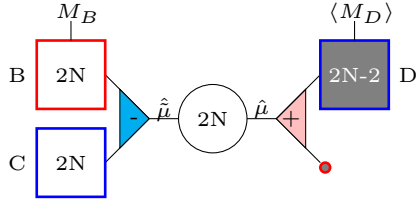
\begin{figure}[h]
\centering
\begin{turn}{90}
\begin{tikzpicture}[scale=1.3, every node/.style={transform shape}]
\filldraw[fill=gray, draw=red,line width=1]  (0.30, 0.65) circle (0.05 cm);
\draw (0.35,0.7)-- ++(0.45,0.45);

\filldraw[fill=pink, draw=black] (0.50,0.85)--(0.80,1.15)--(1.10,0.85) -- cycle;
\draw[text= black, font=\tiny] (0.80,1.0) node[rotate=-90]{+};

\draw (0.80,1.15)-- ++(0.45,-0.45);
\filldraw[fill=gray, draw=blue, line width=1]  (0.90,0) rectangle ++(0.70,0.70);
\draw[text = white, font=\tiny] (1.25,0.35) node[rotate=-90] {2N-2};
\draw[text= black, font=\tiny] (1.25,-0.2) node[rotate=-90]{D};
\draw (0.80,1.15) -- (0.80,1.45);
\draw[text=black,font=\tiny] (0.90,1.30) node[rotate=-90]{$\hat\mu$};
\filldraw[fill=white, draw=black]  (0.80,1.80) circle (0.35 cm);
\draw[text = black, font=\tiny] (0.80,1.80) node[rotate=-90] {2N};
\draw (0.80,2.15) --(0.80, 2.45);
\draw[text=black,font=\tiny] (0.90,2.30) node[rotate=-90]{$\hat{\tilde\mu}$};
\draw (0.80,2.45) -- ++(-0.45,0.45);
\draw(0.80,2.45) -- ++(0.45,0.45);
\filldraw[fill=cyan, draw=black] (0.80,2.45)--++ (0.30,0.30)--++(-0.60,0) -- cycle;
\draw[text= black, font=\tiny] (0.80,2.60) node[rotate=-90]{-};
\filldraw[fill=white, draw=blue, line width=1]  (0,2.9) rectangle ++(0.70,0.70);
\draw[text = black, font=\tiny] (0.35,3.25) node[rotate=-90] {2N};
\draw[text= black, font=\tiny] (0.35,3.8) node[rotate=-90]{C};
\filldraw[fill=white, draw=red, line width=1]  (0.9,2.9) rectangle ++(0.70,0.70);
\draw[text = black, font=\tiny] (1.25,3.25) node[rotate=-90] {2N};
\draw[text= black, font=\tiny] (1.25,3.8) node[rotate=-90]{B};
\draw[text = black, font=\tiny]  (1.6,3.25)--++(0.2,0) ++(0.08,-0.07)node[rotate=-90]{$M_B$};
\draw[text = black, font=\tiny]  (1.6,0.35)--++(0.2,0) ++(0.08,-0.07)node[rotate=-90]{$\langle M_D \rangle$};
\end{tikzpicture}
\end{turn}
\caption{Argyres-Seiberg type dual $\CU^{Sp}_{as}$ to $USp$ gauge theory.}
\label{fig:so}
\end{figure}

A more interesting non-Lagrangian dual is obtained if one considers Higgsing the $A$ and $D$ punctures of $\CT^{Sp}_{c3}$ (figure \ref{fig:Sso}). Closing the puncture for $USp(2N-2)_A$ reduces the corresponding $\tilde{T}_{SO(2N)}$ block to a bifundamental of $USp(2N-2) \times SO(2N)$. Giving  vev to $M_D$  then gives mass to one of the quarks. We therefore end up with a theory of $2N-3$ fundamentals of $SO(2N)$ coupled to a $T_{SO(2N)}$ block as shown in figure \ref{fig:so}. 
The dual superpotential now becomes
\begin{equation}
W=\hat c \tr \hat\mu\hat{\tilde\mu} + \tr M_B\hat\mu_B +\sum_{j} (M_D)_{ j,-j}(\hat{\mu}_D)_{ j,j} \ , 
\end{equation}
where $(\hat{\mu}_D)_{\alpha \beta} = (\hat{Q}_{m \alpha}\Omega^{ml}\hat{Q}_{l \beta})_D $ and the new $U(1)_{\mathcal{F}}$ and $U(1)_R$ charges are
\begin{equation}
\begin{aligned}
\mathcal{F} &= \mathcal{F}_0 - 2\rho^D(\sigma^3) \ ,\\
R&=R_0 - \rho^D(\sigma^3) \ . 
\end{aligned}
\end{equation}
We will use the short-hand notation $\CU^{Sp}_{as}$ for this theory.

\subsubsection*{Non-Lagrangian dual 3: Crossing type dual}

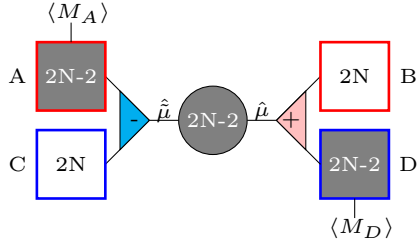
\begin{figure}[h]
\centering
\begin{turn}{90}
\begin{tikzpicture}[scale=1.3, every node/.style={transform shape}]
\filldraw[fill=gray, draw=blue,line width=1]  (0,0) rectangle (0.70,0.70);
\draw[text = white, font=\tiny] (0.35,0.35) node[rotate=-90] {2N-2};
\draw[text= black, font=\tiny] (0.35,-0.2) node[rotate=-90]{D};
\draw[text = black, font=\tiny]  (0,0.35)--++(-0.2,0) ++(-0.1,-0.07) node[rotate=-90]{$\langle M_D \rangle$};
\draw (0.35,0.7)-- ++(0.45,0.45);

\filldraw[fill=pink, draw=black] (0.50,0.85)--(0.80,1.15)--(1.10,0.85) -- cycle;
\draw[text= black, font=\tiny] (0.80,1.0) node[rotate=-90]{+};

\draw (0.80,1.15)-- ++(0.45,-0.45);
\filldraw[fill=white, draw=red,line width=1]  (0.90,0) rectangle ++(0.70,0.70);
\draw[text = black, font=\tiny] (1.25,0.35) node[rotate=-90] {2N};
\draw[text= black, font=\tiny] (1.25,-0.2) node[rotate=-90]{B};
\draw (0.80,1.15) -- (0.80,1.45);
\draw[text=black,font=\tiny] (0.90,1.30) node[rotate=-90]{$\hat\mu$};
\filldraw[fill=gray, draw=black]  (0.80,1.80) circle (0.35 cm);
\draw[text = white, font=\tiny] (0.80,1.80) node[rotate=-90] {2N-2};
\draw (0.80,2.15) --(0.80, 2.45);
\draw[text=black,font=\tiny] (0.90,2.30) node[rotate=-90]{$\hat{\tilde\mu}$};
\draw (0.80,2.45) -- ++(-0.45,0.45);
\draw(0.80,2.45) -- ++(0.45,0.45);
\filldraw[fill=cyan, draw=black] (0.80,2.45)--++ (0.30,0.30)--++(-0.60,0) -- cycle;
\draw[text= black, font=\tiny] (0.80,2.60) node[rotate=-90]{-};
\filldraw[fill=white, draw=blue,line width=1]  (0,2.9) rectangle ++(0.70,0.70);
\draw[text = black, font=\tiny] (0.35,3.25) node[rotate=-90] {2N};
\draw[text= black, font=\tiny] (0.35,3.8) node[rotate=-90]{C};
\filldraw[fill=gray, draw=red,line width=1]  (0.9,2.9) rectangle ++(0.70,0.70);
\draw[text = white, font=\tiny] (1.25,3.25) node[rotate=-90] {2N-2};
\draw[text= black, font=\tiny] (1.25,3.8) node[rotate=-90]{A};
\draw[text = black, font=\tiny]  (1.6,3.25)--++(0.2,0) ++(0.1,-0.07)node[rotate=-90]{$\langle M_A \rangle$};
\end{tikzpicture}
\end{turn}
\caption{The Crossing type dual $\CU^{Sp}_{c2}$ of $USp(2N-2)$ SQCD}
\label{fig:nlsp3}
\end{figure}

The crossing type dual of $\CU^{SO}$ can be obtained by exchanging its (closed) punctures $A$ and $D$. This will bring $A$ (and similarly $D$) into a pair pants whose color is opposite to that of $A$. The statement that these punctures are closed in $\CU^{Sp}$ is then equivalent to saying that the puntures are Higgsed by giving a vev to the mesons that we had to couple to the pants. We will denote this theory by $\CU^{Sp}_{c2}$. The quiver diagram for $\CU^{Sp}_{c2}$ is shown in figure \ref{fig:nlsp3}. Its superpotential is 
\be
 W= \hat c \tr \hat \mu \Omega \hat{\tilde{\mu}} \Omega+\sum_{j,m} (M_{A})_{j,-m} (\hat{\mu}_{B})_{ j,m} + \sum_{j,m} (M_{D})_{j,-m} (\hat{\mu}_{C})_{ j,m} \ ,
\ee
while $R$- and $\CF$-charges are 
\be
\begin{aligned}
\mathcal{F} &= \mathcal{F}_0 + 2\rho^B(\sigma^3) - 2\rho^C(\sigma^3)\ , \\
R&=R_0 - \rho^B(\sigma^3) - \rho^C(\sigma^3) \ , 
\end{aligned} 
\ee 
where $R_0$ and $\CF_0$ are the charges in the theory without a vev for the mesons i.e. $\CT^{Sp}_{c2}$ (see figure \ref{fig:spcr}).

\subsection{'t Hooft anomaly matching}
Let us go on to put the dualities to test. 

\subsubsection*{Non-Lagrangian duals}

It is a simple exercise to check that the $\tr R$ and $\tr R^3$ anomalies of the electric theory, the theories in the crossing frames 1 and 2, and the theory in the swapped frame match since the mesons have $R$-charge 1 and hence do not contribute to the R-anomalies. This also implies that the flavor central charge given by $K\delta^{ab} = -3 \tr RT^aT^b$ will also only get contributions from the $\tilde{T}_{SO(2N)}$ blocks and hence will match in all the these frames. 

It is instructive to match the  $\tr R$ and $\tr R^3$ anomalies of $\CT^{Sp}$ and $\CT^{Sp}_{c3}$. Thus in the electric frame these anomalies get contributions from the two $\tilde{T}_{SO(2N)}$ blocks and the gauginos in the $USp(2N-2)$, $\mathcal{N}=1$ vector multiplet. Each $\tilde{T}_{SO(2N)}$ block contributes 
\begin{equation}
\tr R \big{|}_{\tilde{T}_{SO(2N)}} = n_v-n_h = -2(N-1)^2 - N^2+1 \ , 
\end{equation} 
while the gauginos give 
\begin{equation}
\tr R \big{|}_{\textrm{gaugino}} = 1 (N-1)(2N-1) \ . 
\end{equation} 
This implies 
\begin{equation}
\begin{split}
\tr R \big{|}_{\CT^{Sp}} &= 2 \tr R \big{|}_{\tilde{T}_{SO(2N)}} + \tr R\big{|}_{\textrm{gaugino}}\\
&=-4N^2+5N-1 \ . 
\end{split}
\end{equation} 
Similarly, 
\begin{equation}
\begin{split}
\tr R^3 \big{|}_{\CT^{Sp}} 
=4N^3-10N^2+7N-1 \ . 
\end{split}
\end{equation} 
Let us calculate the above anomalies in $\CT^{Sp}_{c3}$. The mesons will not contribute since they have $R$-charge 1. Thus the contributions come from a $T_{SO(2N)}$ block, a $\tilde{T}_{SO(2N)}$ block and the $SO(2N)$ gauginos.   For the $T_{SO(2N)}$ block we find that 
\begin{equation}
\begin{split}
\tr R \big{|}_{T_{SO(2N)}} &=n_v-n_h
=(2-3N)N \ , 
\end{split}
\end{equation} 
and 
\begin{equation}
\begin{split}
\tr R^3|_{T_{SO(2N)}} &=n_v-\frac{1}{4}n_h =N(3-6N+2N^2) \ . 
\end{split}
\end{equation} 
The anomalies of $\CT^{Sp}_{c3}$ can now be computed:
\begin{equation}
\begin{split}
\tr R \big{|}_{\CT^{Sp}_{c3}} &= \tr R \big{|}_{T_{SO(2N)}} + \tr R \big{|}_{\tilde{T}_{SO(2N)}} + \tr R \big{|}_{\textrm{gaugino}} \\
&=-4N^2+5N-1 \ , 
\end{split}
\end{equation} 
and 
\begin{equation}
\begin{split}
\tr R^3 \big{|}_{\CT^{Sp}_{c3}} &= \tr R^3 \big{|}_{T_{SO(2N)}} + \tr R^3 \big{|}_{\tilde{T}_{SO(2N)}} + \tr R^3 \big{|}_{\textrm{gaugino}}\\
&=4N^3-10N^2+7N-1 \ . 
\end{split}
\end{equation} 
Thus we see the anomalies of the duality frames proposed here match perfectly. 

Let us now consider the matching of $\tr \mathcal{F}T^aT^b$ across the various duality frames. For the $\tilde{T}_{SO(2N)}$ block with $U(1)_\mathcal{F}$-charge $\sigma$, we have
\begin{equation}
\tr \mathcal{F}T^aT^b= \sigma \tr R_{\mathcal{N}=2}T^aT^b =- \frac{\sigma}{2} k_{\mathfrak{g}} \ . 
\end{equation}
To begin with, consider the anomaly coefficient for $T^a \in \mathfrak{sp}(2N-2)_A$. Note that $k_{\mathfrak{sp}(2N-2)}$ for $\tilde{T}_{SO(2N)}$ is $4N$ as can be checked by comparing the dual theories of figure \ref{fig:dTSp}. In the electric frame $\CT^{Sp}$, we find 
\begin{equation}
\tr \mathcal{F}T^a_AT^b_A= -2N\delta^{ab}
\end{equation}

This matches trivially to the anomaly coefficient of $\CT^{Sp}_{c1}$ and $\CT^{Sp}_{c3}$ .  It is much more interesting to compare this with the anomaly coefficient of $\CT^{Sp}_{c2}$ and $\CT^{Sp}_s$. After taking the contributions of the meson $M_A$ into account, the anomaly evaluates to 
\begin{equation}
\begin{split}
\tr\mathcal{F}T^a_AT^b_A &= 2N\delta^{ab} -2\tr_{\mathrm{adj}}T_A^aT_A^b. \\
&=2N\delta^{ab}-2(2N)\delta^{ab}\\
&=-2N\delta^{ab}
\end{split}
\end{equation}
This agrees with the original theory. We can analogously match the anomaly coefficient when $T^a \in \mathfrak{sp}(2N-2)_D$.  

 We now consider the case when $T^a \in \mathfrak{so}(2N)_B$. This time, by comparing the dual theories of figure \ref{fig:dTSOSP},  we find that the contribution of $\tilde{T}_{SO(2N)}$ to $k_{ \mathfrak{so}(2N)}$ is $4N-4$ and hence in $\CT^{Sp}$, the requisite coefficient is 
\begin{equation}
\tr \mathcal{F}T^a_BT^b_B= -(2N-2)\delta^{ab} \ . 
\end{equation}
After adding the contributions of the meson $M_B$ in the theories corresponding to  $\CT^{Sp}_{c1}$ and $\CT^{Sp}_s$ respectively , it is simple to check that their coefficients match the original theory. The above discussion also applies when comparing the anomaly coefficients with  $T^a \in \mathfrak{so}(2N)_C$ or $T^a \in \mathfrak{sp}(2N-2)_D$. The matching of these coefficient between $\CT^{Sp}$ and $\CT^{Sp}_{c2}$ is trivial.

It is much more non-trivial and interesting to match the anomalies of $SO(2N)_B$ and $SO(2N)_C$ in $\CT^{Sp}$ and $\CT^{Sp}_{c3}$. Let us start by comparing the $SO(2N)_C$ anomalies. In the electric theory we find that 
\begin{equation}
\begin{split}
\tr\mathcal{F}T^a_CT^b_C &=  - \tr R_{\mathcal{N}=2}T^a_CT^b_C\\
&=(2N-2)\delta^{ab}
\end{split}
\end{equation}
Using the linear quiver to evaluate $ \tr R_{\mathcal{N}=2}T^a_CT^b_C$ in the $T_{SO(2N)}$ block we find that the anomaly in the magnetic theory matches that in the electric theory.  We can then immediately see that the  anomalies of $SO(2N)_B$ will match in the electric and magnetic theory after including the contributions of the mesons, $M_B$.

\subsubsection*{Dual theories of $USp(2N-2)$ SQCD}

\begin{figure}[h]
\centering
\begin{subfigure}[b]{2.8in}
\centering
\begin{turn}{90}
\begin{tikzpicture}[scale=1.3, every node/.style={transform shape}]
\filldraw[fill=gray, draw=red,line width=1]  (0.30, 0.65) circle (0.05 cm);
\draw (0.35,0.7)-- ++(0.45,0.45);

\filldraw[fill=pink, draw=black] (0.50,0.85)--(0.80,1.15)--(1.10,0.85) -- cycle;
\draw[text= black, font=\tiny] (0.80,1.0) node[rotate=-90]{+};

\draw (0.80,1.15)-- ++(0.45,-0.45);
\filldraw[fill=white, draw=red,line width=1]  (0.90,0) rectangle ++(0.70,0.70);
\draw[text = black, font=\tiny] (1.25,0.35) node[rotate=-90] {2N};
\draw[text= black, font=\tiny] (1.25,-0.2) node[rotate=-90]{B};
\draw (0.80,1.15) -- (0.80,1.45);
\draw[text=black,font=\tiny] (0.90,1.30) node[rotate=-90]{$\mu$};
\filldraw[fill=gray, draw=black]  (0.80,1.80) circle (0.35 cm);
\draw[text = white, font=\tiny] (0.80,1.80) node[rotate=-90] {2N-2};
\draw (0.80,2.15) --(0.80, 2.45);
\draw[text=black,font=\tiny] (0.90,2.30) node[rotate=-90]{$\tilde\mu$};
\draw (0.80,2.45) -- ++(-0.45,0.45);
\draw(0.80,2.45) -- ++(0.45,0.45);
\filldraw[fill=cyan, draw=black] (0.80,2.45)--++ (0.30,0.30)--++(-0.60,0) -- cycle;
\draw[text= black, font=\tiny] (0.80,2.60) node[rotate=-90]{-};
\filldraw[fill=white, draw=blue,line width=1]  (0,2.9) rectangle ++(0.70,0.70);
\draw[text = black, font=\tiny] (0.35,3.25) node[rotate=-90] {2N};
\draw[text= black, font=\tiny] (0.35,3.8) node[rotate=-90]{C};
\filldraw[fill=gray, draw=blue,line width=1]  (1.30, 2.95) circle (0.05 cm);
\end{tikzpicture}
\end{turn}
\caption{$USp$ gauge theory: $\CU^{Sp}$}
\label{fig:duality1}
\end{subfigure}
\begin{subfigure}[b]{2.8in}
\centering
\begin{turn}{90}
\begin{tikzpicture}[scale=1.3, every node/.style={transform shape}]
\filldraw[fill=gray, draw=red,line width=1]  (0.30, 0.65) circle (0.05 cm);
\draw (0.35,0.7)-- ++(0.45,0.45);

\filldraw[fill=pink, draw=black] (0.50,0.85)--(0.80,1.15)--(1.10,0.85) -- cycle;
\draw[text= black, font=\tiny] (0.80,1.0) node[rotate=-90]{+};

\draw (0.80,1.15)-- ++(0.45,-0.45);
\filldraw[fill=white, draw=blue,line width=1]  (0.90,0) rectangle ++(0.70,0.70);
\draw[text = black, font=\tiny] (1.25,0.35) node[rotate=-90] {2N};
\draw[text= black, font=\tiny] (1.25,-0.2) node[rotate=-90]{C};
\draw[text = black, font=\tiny]  (1.6,0.35)--++(0.2,0) ++(0.15,0)node[rotate=-90]{$M_C$};
\draw (0.80,1.15) -- (0.80,1.45);
\draw[text=black,font=\tiny] (0.90,1.30) node[rotate=-90]{$\hat\mu$};
\filldraw[fill=gray, draw=black]  (0.80,1.80) circle (0.35 cm);
\draw[text = white, font=\tiny] (0.80,1.80) node[rotate=-90] {2N-2};
\draw (0.80,2.15) --(0.80, 2.45);
\draw[text=black,font=\tiny] (0.90,2.30) node[rotate=-90]{$\hat{\tilde\mu}$};
\draw (0.80,2.45) -- ++(-0.45,0.45);
\draw(0.80,2.45) -- ++(0.45,0.45);
\filldraw[fill=cyan, draw=black] (0.80,2.45)--++ (0.30,0.30)--++(-0.60,0) -- cycle;
\draw[text= black, font=\tiny] (0.80,2.60) node[rotate=-90]{-};
\filldraw[fill=white, draw=blue,line width=1]  (0,2.9) rectangle ++(0.70,0.70);
\draw[text = black, font=\tiny] (0.35,3.25) node[rotate=-90] {2N};
\draw[text= black, font=\tiny] (0.35,3.8) node[rotate=-90]{B};
\draw[text = black, font=\tiny]  (0,3.25)--++(-0.2,0) ++(-0.15,-0.05) node[rotate=-90]{$M_B$};
\filldraw[fill=gray, draw=blue,line width=1]  (1.30, 2.95) circle (0.05 cm);
\end{tikzpicture}
\end{turn}
\caption{Intriligator-Pouliot dual: $\CU^{Sp}_{c1}$}
\label{fig:duality2}
\end{subfigure}
\begin{subfigure}[b]{2.8in}
\centering
\begin{turn}{90}
\begin{tikzpicture}[scale=1.3, every node/.style={transform shape}]
\filldraw[fill=gray, draw=blue,line width=1]  (0,0) rectangle (0.70,0.70);
\draw[text = white, font=\tiny] (0.35,0.35) node[rotate=-90] {2N-2};
\draw[text= black, font=\tiny] (0.35,-0.2) node[rotate=-90]{D};
\draw[text = black, font=\tiny]  (0,0.35)--++(-0.2,0) ++(-0.15,-0.0) node[rotate=-90]{$\langle M_D \rangle$};
\draw (0.35,0.7)-- ++(0.45,0.45);

\filldraw[fill=pink, draw=black] (0.50,0.85)--(0.80,1.15)--(1.10,0.85) -- cycle;
\draw[text= black, font=\tiny] (0.80,1.0) node[rotate=-90]{+};

\draw (0.80,1.15)-- ++(0.45,-0.45);
\filldraw[fill=white, draw=blue,line width=1]  (0.90,0) rectangle ++(0.70,0.70);
\draw[text = black, font=\tiny] (1.25,0.35) node[rotate=-90] {2N};
\draw[text= black, font=\tiny] (1.25,-0.2) node[rotate=-90]{C};
\draw[text = black, font=\tiny]  (1.6,0.35)--++(0.2,0) ++(0.15,0)node[rotate=-90]{$M_C$};
\draw (0.80,1.15) -- (0.80,1.45);
\draw[text=black,font=\tiny] (0.90,1.30) node[rotate=-90]{$\hat\mu$};
\filldraw[fill=gray, draw=black]  (0.80,1.80) circle (0.35 cm);
\draw[text = white, font=\tiny] (0.80,1.80) node[rotate=-90] {2N-2};
\draw (0.80,2.15) --(0.80, 2.45);
\draw[text=black,font=\tiny] (0.90,2.30) node[rotate=-90]{$\hat{\tilde\mu}$};
\draw (0.80,2.45) -- ++(-0.45,0.45);
\draw(0.80,2.45) -- ++(0.45,0.45);
\filldraw[fill=cyan, draw=black] (0.80,2.45)--++ (0.30,0.30)--++(-0.60,0) -- cycle;
\draw[text= black, font=\tiny] (0.80,2.60) node[rotate=-90]{-};
\filldraw[fill=white, draw=red,line width=1]  (0,2.9) rectangle ++(0.70,0.70);
\draw[text = black, font=\tiny] (0.35,3.25) node[rotate=-90] {2N};
\draw[text= black, font=\tiny] (0.35,3.8) node[rotate=-90]{B};
\draw[text = black, font=\tiny]  (0,3.25)--++(-0.2,0) ++(-0.15,-0.05) node[rotate=-90]{$M_B$};
\filldraw[fill=gray, draw=red,line width=1]  (0.9,2.9) rectangle ++(0.70,0.70);
\draw[text = white, font=\tiny] (1.25,3.25) node[rotate=-90] {2N-2};
\draw[text= black, font=\tiny] (1.25,3.8) node[rotate=-90]{A};
\draw[text = black, font=\tiny]  (1.6,3.25)--++(0.2,0) ++(0.15,0)node[rotate=-90]{$\langle M_A \rangle$};
\end{tikzpicture}
\end{turn}
\caption{Swapped dual: $\CU^{Sp}_{s}$}
\label{fig:duality3}
\end{subfigure}
\begin{subfigure}[b]{2.8in}
\centering
\begin{turn}{90}
\begin{tikzpicture}[scale=1.3, every node/.style={transform shape}]
\filldraw[fill=gray, draw=red,line width=1]  (0.30, 0.65) circle (0.05 cm);
\draw (0.35,0.7)-- ++(0.45,0.45);

\filldraw[fill=pink, draw=black] (0.50,0.85)--(0.80,1.15)--(1.10,0.85) -- cycle;
\draw[text= black, font=\tiny] (0.80,1.0) node[rotate=-90]{+};

\draw (0.80,1.15)-- ++(0.45,-0.45);
\filldraw[fill=gray, draw=blue,line width=1]  (0.90,0) rectangle ++(0.70,0.70);
\draw[text = white, font=\tiny] (1.25,0.35) node[rotate=-90] {2N-2};
\draw[text= black, font=\tiny] (1.25,-0.2) node[rotate=-90]{D};
\draw (0.80,1.15) -- (0.80,1.45);
\draw[text=black,font=\tiny] (0.90,1.30) node[rotate=-90]{$\hat\mu$};
\filldraw[fill=white, draw=black]  (0.80,1.80) circle (0.35 cm);
\draw[text = black, font=\tiny] (0.80,1.80) node[rotate=-90] {2N};
\draw (0.80,2.15) --(0.80, 2.45);
\draw[text=black,font=\tiny] (0.90,2.30) node[rotate=-90]{$\hat{\tilde\mu}$};
\draw (0.80,2.45) -- ++(-0.45,0.45);
\draw(0.80,2.45) -- ++(0.45,0.45);
\filldraw[fill=cyan, draw=black] (0.80,2.45)--++ (0.30,0.30)--++(-0.60,0) -- cycle;
\draw[text= black, font=\tiny] (0.80,2.60) node[rotate=-90]{-};
\filldraw[fill=white, draw=blue,line width=1]  (0,2.9) rectangle ++(0.70,0.70);
\draw[text = black, font=\tiny] (0.35,3.25) node[rotate=-90] {2N};
\draw[text= black, font=\tiny] (0.35,3.8) node[rotate=-90]{C};
\filldraw[fill=white, draw=red,line width=1]  (0.9,2.9) rectangle ++(0.70,0.70);
\draw[text = black, font=\tiny] (1.25,3.25) node[rotate=-90] {2N};
\draw[text= black, font=\tiny] (1.25,3.8) node[rotate=-90]{B};
\draw[text = black, font=\tiny]  (1.6,3.25)--++(0.2,0) ++(0.15,0)node[rotate=-90]{$M_B$};
\draw[text = black, font=\tiny]  (1.6,0.35)--++(0.2,0) ++(0.15,0)node[rotate=-90]{$\langle M_D \rangle$};
\end{tikzpicture}
\end{turn}
\caption{Argyres-Seiberg type dual: $\CU^{Sp}_{as}$}
\label{fig:duality4}
\end{subfigure}
\begin{subfigure}[b]{2.8 in}
\centering
\begin{turn}{90}
\begin{tikzpicture}[scale=1.3, every node/.style={transform shape}]
\filldraw[fill=gray, draw=blue,line width=1]  (0,0) rectangle (0.70,0.70);
\draw[text = white, font=\tiny] (0.35,0.35) node[rotate=-90] {2N-2};
\draw[text= black, font=\tiny] (0.35,-0.2) node[rotate=-90]{D};
\draw[text = black, font=\tiny]  (0,0.35)--++(-0.2,0) ++(-0.1,-0.07) node[rotate=-90]{$\langle M_D \rangle$};
\draw (0.35,0.7)-- ++(0.45,0.45);

\filldraw[fill=pink, draw=black] (0.50,0.85)--(0.80,1.15)--(1.10,0.85) -- cycle;
\draw[text= black, font=\tiny] (0.80,1.0) node[rotate=-90]{+};

\draw (0.80,1.15)-- ++(0.45,-0.45);
\filldraw[fill=white, draw=red,line width=1]  (0.90,0) rectangle ++(0.70,0.70);
\draw[text = black, font=\tiny] (1.25,0.35) node[rotate=-90] {2N};
\draw[text= black, font=\tiny] (1.25,-0.2) node[rotate=-90]{B};
\draw (0.80,1.15) -- (0.80,1.45);
\draw[text=black,font=\tiny] (0.90,1.30) node[rotate=-90]{$\hat\mu$};
\filldraw[fill=gray, draw=black]  (0.80,1.80) circle (0.35 cm);
\draw[text = white, font=\tiny] (0.80,1.80) node[rotate=-90] {2N-2};
\draw (0.80,2.15) --(0.80, 2.45);
\draw[text=black,font=\tiny] (0.90,2.30) node[rotate=-90]{$\hat{\tilde\mu}$};
\draw (0.80,2.45) -- ++(-0.45,0.45);
\draw(0.80,2.45) -- ++(0.45,0.45);
\filldraw[fill=cyan, draw=black] (0.80,2.45)--++ (0.30,0.30)--++(-0.60,0) -- cycle;
\draw[text= black, font=\tiny] (0.80,2.60) node[rotate=-90]{-};
\filldraw[fill=white, draw=blue,line width=1]  (0,2.9) rectangle ++(0.70,0.70);
\draw[text = black, font=\tiny] (0.35,3.25) node[rotate=-90] {2N};
\draw[text= black, font=\tiny] (0.35,3.8) node[rotate=-90]{C};
\filldraw[fill=gray, draw=red,line width=1]  (0.9,2.9) rectangle ++(0.70,0.70);
\draw[text = white, font=\tiny] (1.25,3.25) node[rotate=-90] {2N-2};
\draw[text= black, font=\tiny] (1.25,3.8) node[rotate=-90]{A};
\draw[text = black, font=\tiny]  (1.6,3.25)--++(0.2,0) ++(0.1,-0.07)node[rotate=-90]{$\langle M_A \rangle$};
\end{tikzpicture}
\end{turn}
\caption{The crossing type dual: $\CU^{Sp}_{c2}$}
\label{fig:cros3}
\end{subfigure}
\caption{Dual frames of $USp$ SQCD}
\label{fig:duality}
\end{figure}
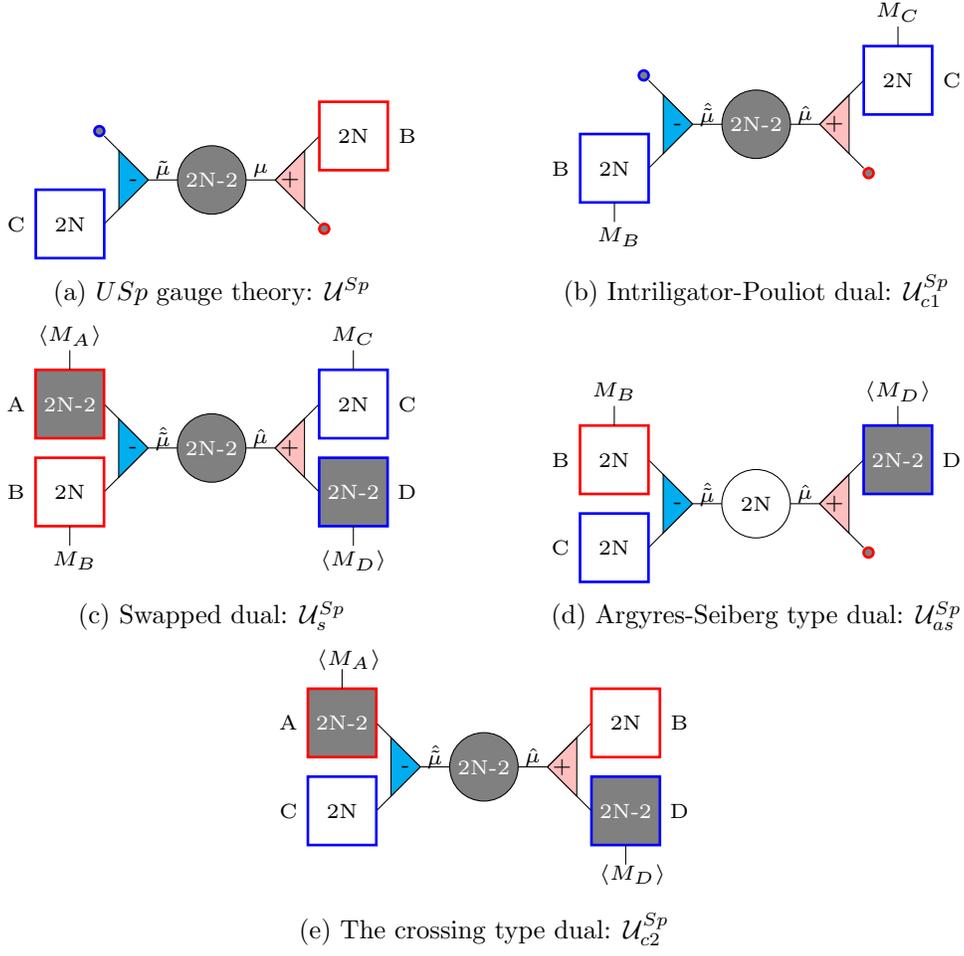

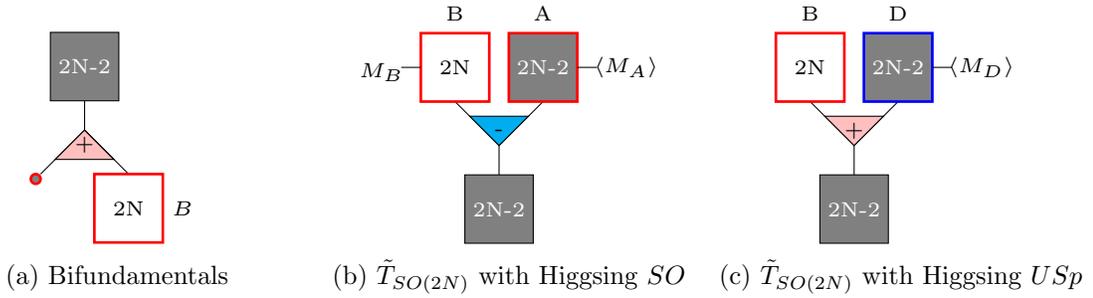
\begin{figure}[h]
\centering
\begin{subfigure}[b]{2in}
\centering
\begin{tikzpicture}[scale=1.3, every node/.style={transform shape}]
\filldraw[fill=gray, draw=red,line width=1]  (0.30, 0.65) circle (0.05 cm);
\draw (0.35,0.7)-- ++(0.45,0.45);

\filldraw[fill=pink, draw=black] (0.50,0.85)--(0.80,1.15)--(1.10,0.85) -- cycle;
\draw[text= black, font=\tiny] (0.80,1.0) node{+};

\draw (0.80,1.15)-- ++(0.45,-0.45);
\filldraw[fill=white, draw=red,line width=1]  (0.90,0) rectangle ++(0.70,0.70);
\draw[text = black, font=\tiny] (1.25,0.35) node {2N};
\draw[text = black, font=\tiny]  (1.6,0.35) ++(0.2,0)node{$B$};
\draw (0.80,1.15) -- (0.80,1.45);
\filldraw[fill=gray, draw=black]  (0.45,1.45) rectangle ++(0.70,0.70);
\draw[text = white, font=\tiny] (0.80,1.80) node {2N-2};
\end{tikzpicture}
\caption{Bifundamentals}
\label{fig:bb1}
\end{subfigure}
\begin{subfigure}[b]{2.0in}
\centering
\begin{tikzpicture}[scale=1.3, every node/.style={transform shape}]
\filldraw[fill=gray, draw=black]  (0.45,1.45) rectangle ++(0.70,0.70);
\draw[text = white, font=\tiny] (0.80,1.80) node {2N-2};
\draw (0.80,2.15) --(0.80, 2.45);
\draw (0.80,2.45) -- ++(-0.45,0.45);
\draw(0.80,2.45) -- ++(0.45,0.45);
\filldraw[fill=cyan, draw=black] (0.80,2.45)--++ (0.30,0.30)--++(-0.60,0) -- cycle;
\draw[text= black, font=\tiny] (0.80,2.60) node{-};
\filldraw[fill=white, draw=red,line width=1]  (0,2.9) rectangle ++(0.70,0.70);
\draw[text = black, font=\tiny] (0.35,3.25) node {2N};
\draw[text= black, font=\tiny] (0.35,3.8) node{B};
\draw[text = black, font=\tiny]  (0,3.25)--++(-0.2,0) ++(-0.2,-0.05) node{$M_B$};
\filldraw[fill=gray, draw=red,line width=1]  (0.9,2.9) rectangle ++(0.70,0.70);
\draw[text = white, font=\tiny] (1.25,3.25) node {2N-2};
\draw[text= black, font=\tiny] (1.25,3.8) node{A};
\draw[text = black, font=\tiny]  (1.6,3.25)--++(0.2,0) ++(0.3,0)node{$\langle M_A \rangle$};
\end{tikzpicture}
\caption{$\tilde{T}_{SO(2N)}$ with Higgsing $SO$}
\label{fig:bb2}
\end{subfigure}
\begin{subfigure}[b]{2.0in}
\centering
\begin{tikzpicture}[scale=1.3, every node/.style={transform shape}]
\filldraw[fill=gray, draw=black]  (0.45,1.45) rectangle ++(0.70,0.70);
\draw[text = white, font=\tiny] (0.80,1.80) node {2N-2};
\draw (0.80,2.15) --(0.80, 2.45);
\draw (0.80,2.45) -- ++(-0.45,0.45);
\draw(0.80,2.45) -- ++(0.45,0.45);
\filldraw[fill=pink, draw=black] (0.80,2.45)--++ (0.30,0.30)--++(-0.60,0) -- cycle;
\draw[text= black, font=\tiny] (0.80,2.60) node{+};
\filldraw[fill=white, draw=red,line width=1]  (0,2.9) rectangle ++(0.70,0.70);
\draw[text = black, font=\tiny] (0.35,3.25) node {2N};
\draw[text= black, font=\tiny] (0.35,3.8) node{B};
\filldraw[fill=gray, draw=blue, line width=1]  (0.9,2.9) rectangle ++(0.70,0.70);
\draw[text = white, font=\tiny] (1.25,3.25) node {2N-2};
\draw[text= black, font=\tiny] (1.25,3.8) node{D};
\draw[text = black, font=\tiny]  (1.6,3.25)--++(0.2,0) ++(0.3,0)node{$\langle M_D \rangle$};
\end{tikzpicture}
\caption{$\tilde{T}_{SO(2N)}$ with Higgsing $USp$}
\label{fig:bb3}
\end{subfigure}
\caption{The building blocks of $\CU^{Sp}$, $\CU^{Sp}_s$ and $\CU^{Sp}_{c2}$}
\label{fig:tsp+m}
\end{figure}

The various duality frames obtained after Higgsing some of the $USp(2N-2)$ punctures are summarized in figure \ref{fig:duality}. 
Since $\CU^{Sp}$ and $\CU^{Sp}_{c1}$ are related by Intriligator-Pouliot duality and their anomalies match without much ado.  For the purpose of matching the anomalies between $\CU^{Sp}$ and $\CU^{Sp}_s$, we observe that we only have to match the anomalies of the $SO(2N) \times USp(2N-2)$ bifundamental to the anomalies of the $\tilde{T}_{SO(2N)}$ block appropriately coupled to mesons (figure \ref{fig:tsp+m}). 
For the  bifundamental we have 
\begin{equation}
\tr R \big|_{\textrm{bifund}}=(-\frac{1}{2})(2N)(2N-2) = -N(2N-2) \ . 
\end{equation}
On the dual side, after giving a vev to the mesons, $M_A$, the R-charge gets shifted: $R \rightarrow R-\rho(\sigma^3)$. This will not affect the contribution of the $\tilde{T}_{SO(2N)}$ block, since its $\tr \rho(\sigma^3) =0$.  However for the mesons, we will only consider the contributions of $M_{A,j ,-j}$ since the rest decouple. This implies 
\begin{equation}
\begin{split}
\tr R \big|_{\langle M_A \rangle} &= \sum_j j
=\sum_{n=1}^{N-1} (2n-1) 
= (N-1)^2 \ . 
\end{split}
\end{equation} 
Also $M_B$ does not contribute to the $R$-anomalies since their $R$-charge is not shifted and is equal to 1. Putting these together, we find that in this frame 
\begin{equation}
\begin{split}
\tr R &= \tr R|_{\tilde{T}_{SO(2N)}} + \tr R|_{M_A}
=-N(2N-2) \ , 
\end{split}
\end{equation} 
which is identical to the corresponding anomaly of the bifundamental. 

Moving on, we now compare the $\tr R^3$ anomalies on the two sides and find 
\begin{equation}
\tr R^3|_{\textrm{bifund}}=\left( -\frac{1}{2} \right)^3(2N)(2N-2) = -\frac{1}{2}N(N-1) \ . 
\end{equation}
On the dual side, since $R= R_0 - \rho(\sigma^3)$, where $R_0 = \frac{1}{2} R_{\mathcal{N}=2} + I_3$, therefore 
\begin{equation}
\tr R^3= \tr R_0^3 + 3 \tr R\rho^2 \ . 
\end{equation}
Adding the contributions of $\tilde{T}_{SO(2N)}$ and the mesons using \eqref{eq:embedding} and \eqref{eq:mesons}, we find that the $\tr R^3$ anomalies match with those of the bifundamental. The $\tr R T^a_B T^b_B$ and $\tr \mathcal{F} T^a_B T^b_B$ anomalies for the bifundamental are given by $(-\frac{1}{2} )(2N-2)$ and $(-1)(2N-2)$ respectively. On the dual side these have the same values as in the scenario before Higgsing. This is because  $\tr \rho T^a_B T^b_B = 0$ for the $\tilde{T}_{SO(2N)}$block. We therefore conclude that these anomalies have the same value in $\CU^{Sp}$ and $\CU^{Sp}_s$. The anomalies of $\CU^{Sp}$ and $\CU^{Sp}_{c2}$ can also be matched in a similar manner by comparing the contributions made by their building blocks shown in figure \ref{fig:bb1} and \ref{fig:bb3}.
  
We now compare the anomalies of the Argyres-Seiberg type dual, $\CU^{Sp}_{as}$. Note that in $\CU^{Sp}$
\begin{equation}
\begin{split}
\tr R
&=2 \left(-\frac{1}{2} \right)(2N)(2N-2) + (N-1)(2N-1)\\
&= -2N^2+N+1 \ , 
\end{split}
\end{equation}
and 
\begin{equation}
\begin{split}
\tr R^3&=2 \left(-\frac{1}{2} \right)^3(2N)(2N-2) + (N-1)(2N-1)\\
&= (N-1)^2 \ .
\end{split}
\end{equation}
In the $\CU^{Sp}_{as}$, the R-charges are shifted to $R=R_0 - \rho(\sigma^3)$.  Also, the meson, $M_B$, does not get a vev. It therefore  has an $R$-charge 1 and hence does not contribute. In the $T_{SO(2N)}$ block, $\tr \rho(\sigma^3) =0$, which implies
\begin{equation}
\begin{split}
 \tr R \big|_{T_{SO(2N)}} &= \tr R_0 \big|_{T_{SO(2N)}} 
= (2-3N)N \ . 
\end{split}
\end{equation} 
The contribution from those components of $M_D$ which continue to stay coupled to the theory after giving a vev is 
\begin{equation}
\begin{split}
\tr R|_{\langle M_D \rangle} &= \sum_j j
=\sum_{n=1}^{N-1} (2n-1) 
= (N-1)^2 \ . 
\end{split}
\end{equation}
For the purpose of anomaly matching we can consider the $2N-3$ fundamentals coupled to the $T_{SO(2N)}$ block as a bifundamental of $SO(2N) \times USp(2N-2)$ with shifted R-charges. As usual the shift will correspond to the embedding of $SU(2)$ in $USp(2N-2)$. The shift in the R-charge of the bifundamental does not change its contribution to $tr R$, since $\tr \rho =0$ for the bifundamental. Thus we find that $\tr R$ in $\CU^{Sp}_{as}$ is given by 
\begin{equation}
\begin{split}
\tr R &= \tr R \big|_{T_{SO(2N)}} + \tr R \big|_{M_D} + \tr R\big|_{\textrm{bifund}}+\tr R\big|_{\textrm{gaugino}}\\
&=(2-3N)N+  (N-1)^2-N(2N-2) + N(2N-1)\\
&=-2N^2+N+1 \ . 
\end{split}
\end{equation} 
This shows perfect agreement with the corresponding anomaly in $\CU^{Sp}$.  Similarly, we find  
\begin{equation}
\begin{split}
\tr R^3|_{T_{SO(2N)}} &= \tr R_0^3|_{T_{SO(2N)}} \\
&= N(3-6N+2N^2) \ . 
\end{split}
\end{equation}
As was mentioned before, the meson, $M_B$ will contribute trivially while the contribution from those modes of $M_D$ that are still coupled to the theory becomes  
\begin{equation}
\begin{split}
\tr R^3 \big|_{\langle M_D \rangle} &= \sum_j j^3 
= \sum_{n=1}^{N-1} (2n-1) 
= 1-6N+11N^2-8N^3+2N^4 \ . 
\end{split}
\end{equation}
The contribution of the bifundamental is given by 
\begin{equation}
\begin{split}
\tr R^3 \big|_{\textrm{bifund}}&=\tr R^3_0 \big|_{\textrm{bifund}} + 3 \tr R\rho^2 \\
& = -\frac{1}{2}N(N-1) + \frac{1}{2}(N-1)(4N^2-8N+3)(-N)\\
&=-N(N-1)(2N^2-4N+2) \ . 
\end{split}
\end{equation}
Combining all these contributions we find 
\begin{equation}
\begin{split}
\tr R^3 &= \tr R^3 \big|_{T_{SO(2N)}} + \tr R^3 \big|_{M_D} + \tr R^3 \big|_{\textrm{bifund}} + \tr R^3 \big|_{\textrm{gaugino}}
=(N-1)^2 \ , 
\end{split}
\end{equation} 
hence providing a nontrivial check of our proposal. It can also be checked, via a pretty direct calculation, that the $\tr RT^aT^b$ and $\tr \mathcal{F}T^aT^b$ anomalies also match in these theories.

\section{Dualities for the $G_2$ gauge theory} \label{sec:G2}
In this section, we study a $G_2$ gauge theory and its dual frames. The $G_2$ gauge group can be obtained from $\Gamma = D_4$ theory with $\IZ_3$ outer-automorphism twist. Since the $D_4$ theory allows both $\IZ_2$-twisting $\s_2$ and $\IZ_3$-twisting $\s_3$. We should take the twist lines with slightly more care to go to various different dual frames.

We study the $G_2$ gauge theory with $8$ fundamental quarks in the $7$ dimensional representation of $G_2$. A dual theory for the $G_2$ gauge theory was first proposed in \cite{Pouliot:1995zc} where the dual theory is given by $SU$ gauge group with anti-symmetric tensors. We find new dual descriptions for the $G_2$ gauge theory flowing to the same fixed point in the IR. We test the duality via anomaly matching and comparison of superconformal indices. 

\subsection{$G_2$ gauge theory and its dual from coupled $E_7$ blocks}
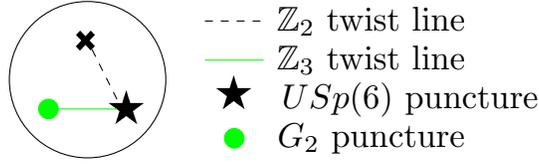
\begin{figure}[h]
\centering
\begin{tikzpicture}[scale=1.3, every node/.style={transform shape}]
\draw[ draw=black]  (0,0) circle (0.8 cm);
\filldraw[fill= green, draw = green] (-0.4,-0.3) circle (0.10 cm);
\draw[text = black ] (0.4 ,-0.3) node {\FiveStar};
\draw[text = black,font=\tiny ] (-0.01 ,0.4) node {\XSolidBold};
\draw[draw = green] (-0.4,-0.3) -- (0.3,-0.3);
\draw[dashed] (0.3,-0.2) -- (0,0.4);
\draw[dashed] (1.2,0.6) -- (1.8,0.6);
\draw[text = black, font=\small] (2.9 ,0.6) node {$\mathbb{Z}_2$ twist line};
\draw[draw=green] (1.2,0.2) -- (1.8,0.2);
\draw[text = black, font=\small] (2.9 ,0.2) node {$\mathbb{Z}_3$ twist line};
\draw[text = black ] (1.5 ,-0.15) node {\FiveStar};
\draw[text = black,font=\small] (3.3, -0.2) node {$USp(6)$ puncture};
\filldraw[fill= green, draw = green] (1.5,-0.6) circle (0.10 cm);
\draw[text = black,font=\small] (2.9, -0.6) node {$G_2$ puncture};
\end{tikzpicture}
\caption{Three punctured sphere with $USp(6)$ and $G_2$ punctures. }
\label{fig:E7}
\end{figure}

To obtain the $G_2$-dual we propose the following procedure: start with the strongly coupled block of \cite{Tachikawa:2010vg} given by $D_4$ theory on a three punctured sphere with a twisted null puncture, a $USp(6)$ puncture and a $G_2$ puncture as in figure \ref{fig:E7}. Even though the $E_7$ flavor symmetry is not manifest, the theory exhibits enhanced $E_7$ symmetry which is the theory of Minahan-Nemeschansky \cite{Minahan:1996cj}. We will demonstrate in section \ref{sec:index} that the superconformal index of the theory of figure \ref{fig:E7} agrees with the $E_7$ theory. 

\begin{figure}[h]
\centering
\begin{subfigure}[b]{1.8in}
\centering
\begin{tikzpicture}[scale=1.3, every node/.style={transform shape}]
\draw (0,0) arc (-60:240:0.6cm);
\draw (0.1,0.3) .. controls (-0.07,0) and (-0.07,-0.5) .. (0.1,-0.8);
\draw (-0.7,0.3) .. controls (-0.53,0) and (-0.53,-0.5) .. (-0.7,-0.8);
\draw (0,-0.5) arc (60:-240:0.6cm);
\draw[text = black,font=\small ] (0.05 ,0.7) node {\FiveStar};
\draw[text = black,font=\tiny ] (-0.15 ,0.85) node {$A$};
\draw[text = black,font=\small ] (0.05 ,-1.2) node {\FiveStar};
\draw[text = black,font=\tiny ] (-0.2 ,-1.3) node {$B$};
\draw[draw=green] (0.04 ,0.64)  .. controls (-0.25,0) and (-0.25,-0.5) .. (0.03 ,-1.15);
\draw[text = black,font=\tiny ] (-0.65 ,0.5) node {\XSolidBold};
\draw[text = black,font=\tiny ] (-0.55 ,0.75) node {$D$};
\draw[text = black,font=\tiny ] (-0.65 ,-1) node {\XSolidBold};
\draw[text = black,font=\tiny ] (-0.55 ,-1.25) node {$C$};
\draw[dashed] (-0.65,0.5) -- (0.05,0.7);
\draw[dashed] (-0.65,-1) -- (0.05,-1.2);
\end{tikzpicture}
\caption{$G_2$ frame}
\label{fig:N2G2gauge}
\end{subfigure}
\begin{subfigure}[b]{1.8in}
\centering
\begin{tikzpicture}[scale=1.3, every node/.style={transform shape}]
\draw (0,0) arc (-60:240:0.6cm);
\draw (0.1,0.3) .. controls (-0.07,0) and (-0.07,-0.5) .. (0.1,-0.8);
\draw (-0.7,0.3) .. controls (-0.53,0) and (-0.53,-0.5) .. (-0.7,-0.8);
\draw (0,-0.5) arc (60:-240:0.6cm);
\draw[text = black,font=\small ] (0.05 ,0.7) node {\FiveStar};
\draw[text = black,font=\tiny ] (-0.15 ,0.85) node {$B$};
\draw[text = black,font=\small ] (0.05 ,-1.2) node {\FiveStar};
\draw[text = black,font=\tiny ] (-0.2 ,-1.3) node {$A$};
\draw[draw=green] (-0.3, -0.25) ellipse (0.35 cm and 1 mm);
\draw[text = black,font=\tiny ] (-0.65 ,0.5) node {\XSolidBold};
\draw[text = black,font=\tiny ] (-0.55 ,0.75) node {$D$};
\draw[text = black,font=\tiny ] (-0.65 ,-1) node {\XSolidBold};
\draw[text = black,font=\tiny ] (-0.55 ,-1.25) node {$C$};
\draw[dashed] (-0.65,0.5) -- (0.05,0.7);
\draw[dashed] (-0.65,-1) -- (0.05,-1.2);
\end{tikzpicture}
\caption{$Spin(8)$ frame}
\label{fig:N2Z3SO8}
\end{subfigure}
\begin{subfigure}[b]{1.8in}
\centering
\begin{tikzpicture}[scale=1.3, every node/.style={transform shape}]
\draw (0,0) arc (-60:240:0.6cm);
\draw (0.1,0.3) .. controls (-0.07,0) and (-0.07,-0.5) .. (0.1,-0.8);
\draw (-0.7,0.3) .. controls (-0.53,0) and (-0.53,-0.5) .. (-0.7,-0.8);
\draw (0,-0.5) arc (60:-240:0.6cm);
\draw[text = black,font=\tiny ] (0.05 ,0.7) node {\XSolidBold};
\draw[text = black,font=\tiny ] (-0.15 ,0.85) node {$C$};
\draw[text = black,font=\small ] (0.05 ,-1.2) node {\FiveStar};
\draw[text = black,font=\tiny ] (-0.2 ,-1.3) node {$A$};
\draw[draw=green] (0.04 ,0.64)  .. controls (-0.25,0) and (-0.25,-0.5) .. (0.03 ,-1.15);
\draw[text = black,font=\tiny ] (-0.65 ,0.5) node {\XSolidBold};
\draw[text = black,font=\tiny ] (-0.55 ,0.75) node {$D$};
\draw[text = black,font=\small ] (-0.65 ,-1) node {\FiveStar};
\draw[text = black,font=\tiny ] (-0.55 ,-1.25) node {$B$};
\draw[dashed] (-0.65,0.5) -- (0.05,0.7);
\draw[dashed] (-0.65,-1) -- (0.05,-1.2);
\end{tikzpicture}
\caption{Argyres-Seiberg like frame}
\label{fig:N2G2asdual}
\end{subfigure}
\caption{S-duality for the $G_2$-coupled two $E_7$ theories. }
\label{fig:z3twist}
\end{figure}
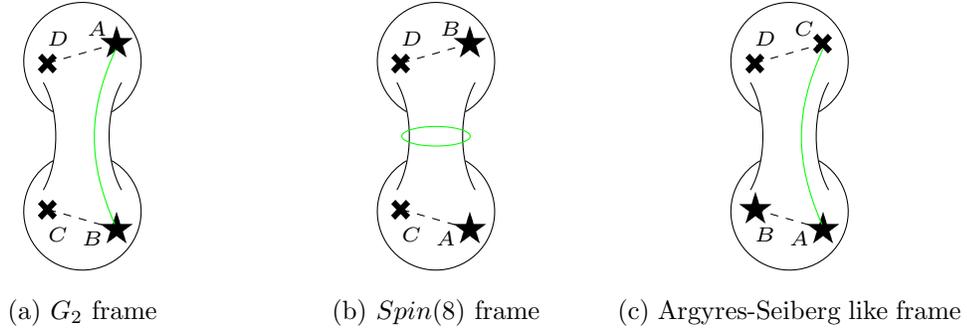

Now prepare two copies of this theory. By gauging the $G_2$ symmetry common to the two blocks we obtain an $\mathcal{N}=2$ SCFT with a $G_2$ gauge group which can be represented by figure \ref{fig:N2G2gauge}. We can obtain its S-dual by exchanging the punctures. One of its S-dual can be obtained by exchanging two null punctures. It is given by an $\mathcal{N}=2$ SCFT with $Spin(8)$ gauge symmetry along with three hypermultiplets in $\mathbf{8}_V$ and three hypermultiplets in $\mathbf{8}_S$ representations which can be represented as in the figure \ref{fig:N2Z3SO8}. This duality was first found in \cite{Argyres:2007tq}.  Another frame can be found by colliding two null punctures and two $USp(6)$ punctures. This is similar to the Argyres-Seiberg duality, where in this case we partially gauge the theory with $USp(6)^2 \times G_2$ flavor symmetry. 

\subsubsection*{$\CN=1$ duality from $E_7$ blocks}
Let us go to the $\CN=1$ construction. It can be done by giving colors to the punctures and the pair of pants. In figure \ref{fig:N2G2gauge}, let's color the two punctures on the bottom to be red, and the other two punctures to be blue. Also color the pair of pants on the bottom to be red and the other to be blue. Since the color of the punctures and the pants are the same, we can identify the `matter content' to be the same two $E_7$ blocks as before. Then we glue two $G_2$ punctures by $\CN=1$ vector multiplet with the superpotential
\begin{equation}
W=c \tr \mu\tilde\mu \ , 
\end{equation}
where $\mu$ and $\tilde\mu$ transform in the adjoint representation of $G_2$.  

A dual frame is described by a $Spin(8)$ gauge theory, with quarks in the $\mathbf{8}_V \times \mathbf{6}$ of $Spin(8) \times USp(6)_A$ and another in the $\mathbf{8}_S \times \mathbf{6}$ representations of $Spin(8) \times USp(6)_B$. There are also mesons transforming in the adjoint representations of $USp(6)_A$ and $USp(6)_B$ respectively. One can also prove the duality starting from $\CN=2$ construction and then giving mass to the chiral adjoint in the vector multiplet if we assume the chiral ring relation
\begin{equation}
\tr {\mu_{G_2}}^2 = \tr (\mu_{USp(6)}\Omega)^2 \ ,  
\end{equation}
and then following the procedure of \cite{Gadde:2013fma}. 
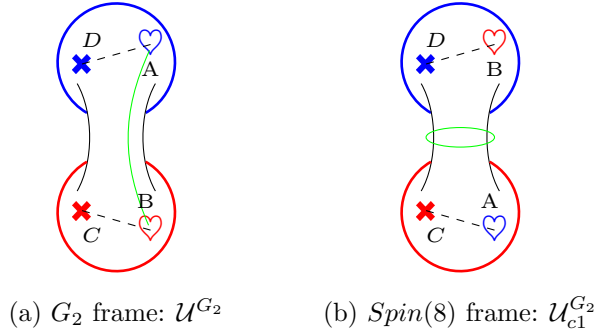
\begin{figure}[h]
\centering
\begin{subfigure}[b]{1.5in}
\centering
\begin{tikzpicture}[scale=1.3, every node/.style={transform shape}]
\draw[color=blue, line width=1] (0,0) arc (-60:240:0.6cm);
\draw (0.1,0.3) .. controls (-0.07,0) and (-0.07,-0.5) .. (0.1,-0.8);
\draw (-0.7,0.3) .. controls (-0.53,0) and (-0.53,-0.5) .. (-0.7,-0.8);
\draw[color=red, line width=1] (0,-0.5) arc (60:-240:0.6cm);
\draw[text = blue,font=\small ] (0.05 ,0.7) node {$\heartsuit$};
\draw[text = red,font=\small] (0.05 ,-1.2) node {$\heartsuit$};
\draw[draw=green] (0.04 ,0.64)  .. controls (-0.25,0) and (-0.25,-0.5) .. (0.03 ,-1.15);
\draw[text = blue,font=\tiny ] (-0.65 ,0.5) node {\XSolidBold};
\draw[text = black,font=\tiny ] (-0.55 ,-1.25) node {$C$};
\draw[text = red,font=\tiny ] (-0.65 ,-1) node {\XSolidBold};
\draw[text = black,font=\tiny ] (-0.55 ,0.75) node {$D$};
\draw[dashed] (-0.65,0.5) -- (0.05,0.7);
\draw[dashed] (-0.65,-1) -- (0.05,-1.2);
\draw[text = black,font=\tiny ] (0.05 ,0.45) node {A};
\draw[text = black,font=\tiny ] (0.0 ,-0.90) node {B};
\end{tikzpicture}
\caption{$G_2$ frame: $\CU^{G_2}$}
\label{fig:G2th}
\end{subfigure}
\begin{subfigure}[b]{2in}
\centering
\begin{tikzpicture}[scale=1.3, every node/.style={transform shape}]
\draw[color=blue, line width=1] (0,0) arc (-60:240:0.6cm);
\draw (0.1,0.3) .. controls (-0.07,0) and (-0.07,-0.5) .. (0.1,-0.8);
\draw (-0.7,0.3) .. controls (-0.53,0) and (-0.53,-0.5) .. (-0.7,-0.8);
\draw[color=red, line width=1] (0,-0.5) arc (60:-240:0.6cm);
\draw[text = black,font=\small, color=red ] (0.05 ,0.7) node {$\heartsuit$};
\draw[text = black,font=\tiny ] (0.05 ,0.45) node {B};
\draw[text = black,font=\small, color=blue ] (0.05 ,-1.2) node {$\heartsuit$};
\draw[text = black,font=\tiny ] (0.0 ,-0.90) node {A};
\draw[draw=green] (-0.3, -0.25) ellipse (0.35 cm and 1 mm);
\draw[text = black,font=\tiny, color=blue ] (-0.65 ,0.5) node {\XSolidBold};
\draw[text = black,font=\tiny ] (-0.55 ,-1.25) node {$C$};
\draw[text = black,font=\tiny, color=red ] (-0.65 ,-1) node {\XSolidBold};
\draw[text = black,font=\tiny ] (-0.55 ,0.75) node {$D$};
\draw[dashed] (-0.65,0.5) -- (0.05,0.7);
\draw[dashed] (-0.65,-1) -- (0.05,-1.2);
\end{tikzpicture}
\caption{$Spin(8)$ frame: $\CU^{G_2}_{c1}$}
\label{fig:G2spin8}
\end{subfigure}
\label{fig:G2lag}
\caption{Lagrangian duals to the $G_2$ gauge theory with $8$ fundamentals}
\end{figure}
The dual superpotential is given by
\be
 W = \hat{c} \tr \hat{\mu}\hat{\tilde\mu} +\tr M_A \hat{\mu}_A + \tr M_B \hat{\mu}_B \ , 
\ee
where $\hat{\mu}_A = Q_A Q_A$ and $\hat{\mu}_B = Q_B Q_B$.

Upon Higgsing the $USp(6)$ flavor symmetries, in the electric frame, down to $USp(4)$, we obtain two copies of bifundamentals of $G_2 \times USp(4)$ with the $G_2$ being gauged. Higgsing is achieved by giving a vev to the adjoint of $USp(6)$ along the partition: $6=[2,1^4]$. We will use the short-hand notation $\CU^{G_2}$ to denote this theory. In the dual frame we will have to give the same vev to the mesons $M_A$ and $M_B$. This will generate a mass for the dual quarks with $SU(2)$ quantum numbers $(j=\frac{1}{2}, m=-\frac{1}{2})$. We integrate these out and obtain the low energy theory which is described by 5 vectors and 5 spinors of the $Spin(8)$ gauge group and transforming as $\mathbf{4} \oplus \mathbf{1}$ of their respective $USp(4)$ flavor symmetries. The low energy superpotential in the dual frame becomes
\be
W=\hat{c} \tr \hat{\mu}\hat{\tilde\mu}+\sum_{j} M_{A j,-j}\hat{\mu}_{A j,j} +\sum_{j} M_{B j,-j}\hat{\mu}_{B j,j} \ , 
\ee
with $\hat{\mu}_{A j,j}$ being quadratics $Spin(8)$ invariants. The $R$-charge in magnetic frame is shifted by $R\rightarrow R-	\rho^A(\sigma^3)-\rho^B(\sigma^3)$, where as usual $\rho$ specifies the $SU(2)$ embedding in $USp(6)$. The $U(1)_{\CF}$ gets shifted to  $\CF\rightarrow \CF - 2\rho^A(\sigma^3)+2\rho^B(\sigma^3)$.  Some of the mesons decouple.and we are left with the mesons $M_{j,j,k}$ coupled to the magnetic theory. This theory will be denoted by the symbol $\CU^{G_2}_{c1}$.

\subsubsection*{Non-Lagrangian duals}
\begin{figure}[h]
\centering
\begin{subfigure}[b]{2in}
\centering
\begin{tikzpicture}[scale=1.3, every node/.style={transform shape}]
\draw[color=blue, line width=1] (0,0) arc (-60:240:0.6cm);
\draw (0.1,0.3) .. controls (-0.07,0) and (-0.07,-0.5) .. (0.1,-0.8);
\draw (-0.7,0.3) .. controls (-0.53,0) and (-0.53,-0.5) .. (-0.7,-0.8);
\draw[color=red, line width=1] (0,-0.5) arc (60:-240:0.6cm);
\draw[text = black,font=\tiny, color=red ] (0.05 ,0.7) node {\XSolidBold};
\draw[text = black,font=\tiny ] (-0.15 ,0.85) node {$C$};
\draw[text = black,font=\small, color=blue ] (0.05 ,-1.2) node {$\heartsuit$};
\draw[text = black,font=\tiny ] (0.0 ,-0.90) node {A};
\draw[draw=green] (0.04 ,0.64)  .. controls (-0.25,0) and (-0.25,-0.5) .. (0.03 ,-1.15);
\draw[text = black,font=\tiny, color=blue ] (-0.65 ,0.5) node {\XSolidBold};
\draw[text = black,font=\tiny ] (-0.55 ,0.75) node {$D$};
\draw[text = black,font=\small, color=red ] (-0.65 ,-1) node {$\heartsuit$};
\draw[text = black,font=\tiny ] (-0.50 ,-0.75) node {B};
\draw[dashed] (-0.65,0.5) -- (0.05,0.7);
\draw[dashed] (-0.65,-1) -- (0.05,-1.2);
\end{tikzpicture}
\caption{AS frame: $\CU^{G_2}_{as}$}
\label{fig:G2as}
\end{subfigure}
\begin{subfigure}[b]{2in}
\centering
\begin{tikzpicture}[scale=1.3, every node/.style={transform shape}]
\draw[color=red, line width=1] (0,0) arc (-60:240:0.6cm);
\draw (0.1,0.3) .. controls (-0.07,0) and (-0.07,-0.5) .. (0.1,-0.8);
\draw (-0.7,0.3) .. controls (-0.53,0) and (-0.53,-0.5) .. (-0.7,-0.8);
\draw[color=blue, line width=1] (0,-0.5) arc (60:-240:0.6cm);
\draw[text = blue,font=\small ] (0.05 ,0.7) node {$\heartsuit$};
\draw[text = red,font=\small] (0.05 ,-1.2) node {$\heartsuit$};
\draw[draw=green] (0.04 ,0.64)  .. controls (-0.25,0) and (-0.25,-0.5) .. (0.03 ,-1.15);
\draw[text = blue,font=\tiny ] (-0.65 ,0.5) node {\XSolidBold};
\draw[text = black,font=\tiny ] (-0.55 ,0.75) node {$D$};
\draw[text = red,font=\tiny ] (-0.65 ,-1) node {\XSolidBold};
\draw[text = black,font=\tiny ] (-0.55 ,-1.25) node {$C$};
\draw[dashed] (-0.65,0.5) -- (0.05,0.7);
\draw[dashed] (-0.65,-1) -- (0.05,-1.2);
\draw[text = black,font=\tiny ] (0.05 ,0.45) node {A};
\draw[text = black,font=\tiny ] (0.0 ,-0.90) node {B};
\end{tikzpicture}
\caption{Swapped $G_2$: $\CU^{G_2}_s$}
\label{fig:G2swap}
\end{subfigure}
\begin{subfigure}[b]{2 in}
\centering
\begin{tikzpicture}[scale=1.3, every node/.style={transform shape}]
\draw[color=red, line width=1] (0,0) arc (-60:240:0.6cm);
\draw (0.1,0.3) .. controls (-0.07,0) and (-0.07,-0.5) .. (0.1,-0.8);
\draw (-0.7,0.3) .. controls (-0.53,0) and (-0.53,-0.5) .. (-0.7,-0.8);
\draw[color=blue, line width=1] (0,-0.5) arc (60:-240:0.6cm);
\draw[text = black,font=\small, color=red ] (0.05 ,0.7) node {$\heartsuit$};
\draw[text = black,font=\tiny ] (0.05 ,0.45) node {B};
\draw[text = black,font=\small, color=blue ] (0.05 ,-1.2) node {$\heartsuit$};
\draw[text = black,font=\tiny ] (0.0 ,-0.90) node {A};
\draw[draw=green] (-0.3, -0.25) ellipse (0.35 cm and 1 mm);
\draw[text = black,font=\tiny, color=blue ] (-0.65 ,0.5) node {\XSolidBold};
\draw[text = black,font=\tiny ] (-0.55 ,0.75) node {$D$};
\draw[text = black,font=\tiny, color=red ] (-0.65 ,-1) node {\XSolidBold};
\draw[text = black,font=\tiny ] (-0.55 ,-1.25) node {$C$};
\draw[dashed] (-0.65,0.5) -- (0.05,0.7);
\draw[dashed] (-0.65,-1) -- (0.05,-1.2);
\end{tikzpicture}
\caption{Crossing-type: $\CU^{G_2}_{c2}$}
\label{fig:G2crossing}
\end{subfigure}
\caption{Non-Lagrangian dual theories for the $\CN=1$ $G_2$ gauge theory with $8$ fundamentals}
\label{fig:G2nonLag}
\end{figure}
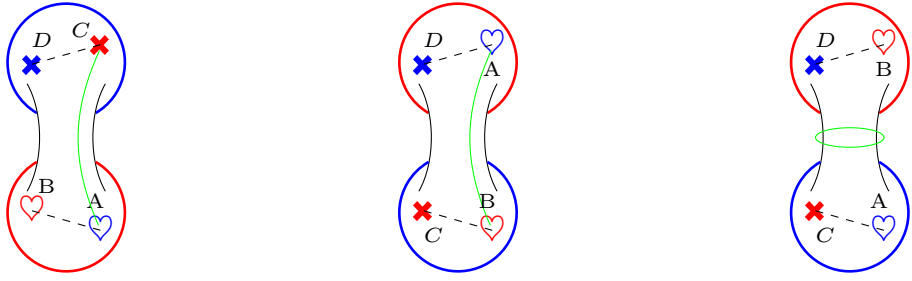

We can also get several non-Lagrangian duals to the $G_2$ theory using different colored pair-of-pants decompositions. See the figure \ref{fig:G2nonLag}. 

\subsubsection*{Non-Lagrangian dual 1: Argyres-Seiberg type}

The Argyres-Seiberg type dual of figure \ref{fig:G2as} is obtained by colliding the punctures $A$ and $B$ on the Riemann surface. This will land us upon a theory consisting of an $E_7$ block (with $G_2 \times USp(6) \subset E_7$ manifest) coupled to a $USp(6) \times USp(4) \times G_2$ block via an $\CN=1$, $G_2$ vector multiplet. We will also have to integrate in mesons (with appropriate vevs) to compensate for the mismatch between the colors of the punctures and the pair of pants. Its superpotential is 
\be
W = \hat{c} \tr \hat{\mu}\hat{\tilde\mu} +\tr \sum_{j} (M_A)_{ j,-j}(\hat{\mu}_A)_{ j,j} + \sum_{j} (M_C)_{ j,-j}(\hat{\mu}_C)_{ j,j}
\ee
with the shifted charges being 
\begin{align}
R &\rightarrow R-	\rho^A(\sigma^3)-\rho^C(\sigma^3) \ , \\
\CF &\rightarrow \CF - 2\rho^A(\sigma^3)+2\rho^C(\sigma^3) \ .
\end{align}
We will use the symbol $\CU^{G_2}_{as}$ to denote this theory.

\subsubsection*{Non-Lagrangian dual 2: Swapped $G_2$}

We arrive at the swapped $G_2$ frame by permuting all the four punctures such that we exchange $A$ with $B$ and $C$ with $D$.  This is equivalent to coupling two $USp(6)^2 \times G_2$ theories along their $G_2$ puncture.  We will have to integrate in 4 mesons $M_A$, $M_B$, $M_C$ and $M_D$.  We give vevs to these mesons such that $USp(6)_C$ and $USp(6)_D$ get completely Higgsed while $USp(6)_A$ and $USp(6)_B$ get Higgsed down to their respective $USp(4)$. This theory will henceforth be denoted by $\CU^{G_2}_s$. Its superpotential becomes 
\be
\begin{split}
W = & ~ \hat{c} \tr \hat{\mu}\hat{\tilde\mu}+\tr \sum_{j} (M_A)_{ j,-j}(\hat{\mu}_A)_{ j,j} + \sum_{j} (M_B)_{ j,-j}(\hat{\mu}_B)_{ j,j} + \\
 &\sum_{j} (M_C)_{ j,-j}(\hat{\mu}_C)_{ j,j} + \sum_{j} (M_D)_{ j,-j}(\hat{\mu}_D)_{ j,j}  \ ,
\end{split}
\ee
while the charges get shifted such that
\begin{align}
R &\rightarrow R-	\rho^A(\sigma^3)-\rho^B(\sigma^3) -	\rho^C(\sigma^3)-\rho^D(\sigma^3) \ , \\
\CF &\rightarrow \CF - 2\rho^A(\sigma^3)+2\rho^B(\sigma^3) + 2\rho^C(\sigma^3)-2\rho^D(\sigma^3) \ .
\end{align}

\subsubsection*{Non-Lagrangian dual 3: Crossing-type}

The crossing-type frame is shown in figure \ref{fig:G2crossing}. It consists of two blocks with $USp(6) \times USp(4) \times SO(8)$ flavor symmetry glued along their $SO(8)$ puncture. The spectrum of the theory also includes mesons $M_C$ and $M_D$ as the punctures $C$ and $D$ lie in pants that are colored oppositely to their own color. We will give a vev to the mesons such that the $USp(6)$ flavor symmetry of these punctures gets completely Higgsed. The superpotential then becomes 
\be
W = &\hat{c} \tr \hat{\mu}\hat{\tilde\mu} + \sum_{j} (M_C)_{ j,-j}(\hat{\mu}_C)_{ j,j} + \sum_{j} (M_D)_{ j,-j}(\hat{\mu}_D)_{ j,j}  \ ,
\ee
and the new charges are given by 
\begin{align}
R &\rightarrow R-\rho^C(\sigma^3)-\rho^D(\sigma^3) \ , \\
\CF &\rightarrow \CF + 2\rho^C(\sigma^3)-2\rho^D(\sigma^3) \ .
\end{align}
We will use the symbol $\CU^{G_2}_{c2}$ to represent this theory.

\subsection{Anomaly matching}

We now show that the anomalies of our proposed dual frames match.  

\subsubsection*{$\tr R$ and $\tr R^3$}

In the $G_2$ electric theory, we find that 
\begin{align}
\tr R &= 14 + \left(-\frac{1}{2} \right) (8 \times 7) = -14 \ , \\
\tr R^3 &= 14 +  \left(-\frac{1}{2} \right)^3 (8 \times 7) = 7 \ .  
\end{align}
After considering the shift in the charges, we find that in $\CU^{G_2}_{c1}$ frame 
\begin{equation}
\begin{aligned}
\tr R &= 28 + 2\sum_{j,m}\left( -\frac{1}{2} -m \right) \times 8 + 2 \sum_{j} j\\
&= 28-48+6 = -14 \ ,
\end{aligned}
\end{equation}
which is same as the result obtained for the electric theory. Similarly for the $\tr R^3$ anomaly in the $Spin(8)$ theory we obtain
\begin{equation}
\begin{aligned}
\tr R^3 &= 28 + 2\sum_{j,m}\left( -\frac{1}{2} -m \right)^3 \times 8 + 2 \sum_{j} j^3\\
& = 28 - 24+3 = 7 \ , \\
\end{aligned}
\end{equation}
which matches with the electric theory.

The effective number of hypers and vectors in a block with $USp(6) \times USp(4) \times SO(8)$ flavor symmetry is 102 and 72 respectively \cite{Chacaltana:2013oka}.  Using this result we find that the $\tr R$ anomaly in the $\CU^{G_2}_{c2}$ is
\be
\tr R = 2(72-102) + 28 + 2 \times 9 =-14 \ ,
\ee
here the first term on the RHS corresponds to the contribution of the non-Lagrangian blocks to $\tr R$, the second term is the contribution from $SO(8)$ gauginos while the last term is the contribution of the mesons used to Higgs the $USp(6)$ flavor symmetry of the blocks. Using (\ref{eq:r3an}) and  (\ref{eq:emi})  along with the fact that in $\tilde{T}_{SO(2N)}$, $k_{\mathfrak{sp}(2N-2)} = 4N$ we find that in $\CU^{G_2}_{c2}$
\be
\tr R^3 = -327+ 28+306 =7 \ .
\ee
As before the various terms on the RHS are obtained from the contribution of the non-Lagrangian blocks, the SO(8) gauginos and the mesons respectively.

The effective number of hypers and vectors in the block with $USp(6)^2 \times G_2$ symmetries can be obtained by comparing the $\CN=2$ theory obtained by gluing two such blocks along their $G_2$ puncture and its $S$-dual corresponding to two copies of the block with $USp(6)^2 \times SO(8)$ punctures glued along their $SO(8)$ puncture with a $\IZ_3$ twist around the cylinder. This will also provide us with the central charges of the various flavor symmetries. Following this procedure we find that in the  $USp(6)^2 \times G_2$ block, $n_v =86$ and $n_h=112$. Using this and including the contribution of the mesons that stay coupled to the theory (after Higgsing one of the $USp(6)$ down to $USp(4)$ and completely Higgsing the other $USp(6)$), we find that in $\CU^{G_2}_s$
\be
\tr R = 2\times(86-112)+ 14+ 2 \times 9+2 \times 3 = -14 \ .
\ee
If we now calculate the $\tr R^3$ anomaly in this theory, we find 
\be
\tr R^3 = -316 \times 2 + 14 +\frac{3}{2} \times 2 + 153 \times 2 =7  \ ,
\ee
 This is in agreement with our proposal. 

We can use our knowledge of the number of hypers and vectors and central charges in the $USp(6) \times USp(4) \times SO(8)$ block to evaluate this data for the $USp(6) \times USp(4) \times G_2$ block which are: $n_v=79$ and $n_h = 102$. The $\tr R$ anomaly can now be calculated in $\CU^{G_2}_{as}$ and is found to match with that in the other duality frames:
\be
\tr R|_{\CU^{G_2}_{as}} = (79-102) + (7-24) + 14 + 9 + 3 =-14 \ ,
\ee
here the first term on the RHS is the contribution from the $USp(6) \times USp(4) \times G_2$ block while second term is the contribution from the $E_7$ theory. The third term is the contribution of $G_2$ gauginos while the last two terms are the contributions of the mesonic excitations.  The coefficient of $\tr R^3$ in this theory is 
\be
\tr R^3= -209 + \frac{95}{2} + 14 + 153 +\frac{3}{2} = 7 \ .
\ee
This is consistent with our expectations.

\subsubsection*{$\tr R\mathcal{F}^2$}

In the $G_2$ theory, each block contributes 
\begin{equation}
\tr R \CF^2 =  \left(-\frac{1}{2} \right)(4 \times 7) =-14 \ . 
\end{equation}
In the $SO(8)$ theory, the $\mathcal{F}$ charges are shifted such that $\mathcal{F} \rightarrow \mathcal{F} - 2 \rho^A (\sigma^3)+2 \rho^B (\sigma^3)$. The contribution of the pants with color `$\sigma$' is therefore given by 
\begin{equation}
\begin{aligned}
\tr R \CF^2 &= \sum_{j,m}\bigg(-\frac{1}{2}-m\bigg)(\sigma+ 2 \sigma m)^2 \times 8 +  \sum_{j} j (-2 \sigma -2 \sigma j)^2 = -14 \ . 
\end{aligned}
\end{equation}
This shows a perfect match with the $G_2$ theory.

In the non-Lagrangian duals,  the shifted charges $R = R_0 - \rho (\sigma^3) $ and $\mathcal{F} = \mathcal{F} - 2 \sigma \rho(\sigma^3)$, (for pants with color `$\sigma$') give rise to the following expression for $\tr R \CF^2$:
\be 
\tr R \CF^2 = \tr (R_0 - \rho)(\mathcal{F}_0 - 2 \sigma \rho)^2 = \tr R_0 \CF_0^2 + 6 \CI \tr R_{\CN=2}T^a T^b
\label{eq:rf2an}
\ee 
where we have used the $SU(2)$ embedding index $\CI$ to evaluate $\tr R \rho^2$ and $\tr \CF \rho^2$. The final expression in (\ref{eq:rf2an}) is independent of the color of pants, as should be the case. Also, on each pair-of-pants $\tr R_0 \CF_0^2 = -n_h$. Using this and taking the contribution of mesons into account, it can be verified that each pair-of-pants in the decomposition of $\CU^{G_2}_{c2}$ and $\CU^{G_2}_s$, contributes a $-14$ to the anomaly, thereby establishing the match with the electric frame.  In $\CU^{G_2}_{as}$, since the pair-of-pants decomposition is not symmetric thus the pants contribute different amounts to the total anomaly. The pants with $USp(6) \times USp(4) \times SO(8)$ punctures contributes $-92$ while the other pant contributes $64$, thereby bringing the total to $-28$ which is same as in the electric theory.
 
\subsubsection*{$\tr RT^a T^b$ and $\tr \mathcal{F}T^aT^b$}

After Higgsing the $USp(6)$ punctures in  the $G_2$-frame of figure \ref{fig:N2G2gauge}, we are left with a $USp(4)_A \times USp(4)_B$ flavor symmetry which is enhanced to $USp(8)$ in the electric theory when there is no superpotential. We now match the 'tHooft anomalies of these flavor symmetries in the electric and the magnetic frames. In the $G_2$ theory we find that 
\begin{align}
\tr RT^a_AT^b_A &= 7 \times(-\frac{1}{2}) \times \tr_{\tiny\Square} T^a_A T^b_A = -\frac{7}{2} \delta^{ab} \ , \\
\tr \mathcal{F}T^a_AT^b_A &= 7 \times(-1) \times \tr_{\tiny\Square} T^a_A T^b_A  = -7 \delta^{ab} \ . 
\end{align}
It is straight forward to check that these match with those in the $SO(8)$ theory, once we use the shifted R and $\mathcal{F}$ charges. Thus in the SO(8) theory we have 
\begin{equation}
\begin{aligned}
\tr RT^a_A T^b_A &= 8 \times \tr \left( -\frac{1}{2} + \rho \right)T^a_A T^b_A + \sum_j  \tr j T^a_AT^b_A
= -\frac{7}{2}\delta^{ab} \ , 
\end{aligned}
\end{equation}
and 
\begin{equation}
\begin{aligned}
\tr \mathcal{F}T^a_A T^b_A &= 8 \times \tr (1 + 2 \rho)T^a_A T^b_A + \sum_j  \tr(-2-2j)T^a_AT^b_A
= -7\delta^{ab} \ , 
\end{aligned}
\end{equation}
which is same as the corresponding anomalies of the $G_2$ theory. The same discussion will also apply in the case of anomalies for the $USp(4)_B$ flavor symmetries.

In $\CU^{G_2}_{c2}$  the anomaly coefficients can be obtained  from the flavor central charges: $\tr RT^a_AT^b_A  = \frac{1}{2}\tr R_{\CN=2}T^a_AT^b_A $ and $\tr \mathcal{F}T^a_AT^b_A = \tr R_{\CN=2}T^a_AT^b_A$. Since $k_{\mathfrak{sp}(4)} = 7$, we find that the anomaly coefficients match those in the electric frame. The same holds for the anomalies of $USp(4)_B$.

The anomalies of in $\CU^{G_2}_s$ can be obtained from the embedding index of $USp(4)$ in $USp(6)$. Thus for the pair-of-pants containing the puncture $A$ we find
\be
\tr R_{\CN=2}T^a_AT^b_A = \CI \tr R_{\CN=2}T^a_{\mathfrak{sp}(6)}T^b_{\mathfrak{sp}(6)}\ .
\ee
Since the $\mathbf{6}$ of $USp(6)$ becomes $\mathbf{4} \oplus \mathbf{1} \oplus \mathbf{1}$ of $USp(4)$, therefore $\CI = 1$. We will also have to add the contribution of the mesons. Thus 
\be
\begin{split}
\tr RT^a_AT^b_A|_{\CU^{G_2}_s} & =  \frac{1}{2}\CI \tr R_{\CN=2}T^a_{\mathfrak{sp}(6)}T^b_{\mathfrak{sp}(6)} + \sum_{j} tr{j T^a_A T^b_A}\\
& =( -4 + \frac{1}{2} \times 1) \delta ^{ab}= -\frac{7}{2} \delta ^{ab} \ .
\end{split}
\ee
Similarly we can show that $\tr \CF T^a_AT^b_A|_{\CU^{G_2}_s}  = -7 \delta ^{ab} $. The anomalies of $USp(4)_B$ match those in the electric frame in an analogous manner.

The anomalies of $\CU^{G_2}_{as}$ can also be shown to match after using the fact that $k_{\mathfrak{sp}(4)_B} = 7$ and proceeding in the same way as in $\CU^{G_2}_s$ for the anomalies of $USp(4)_A$.


\section{Superconformal index} \label{sec:index}
In this section, we put our new dualities to test by comparing the superconformal indices for the dual theories. We first review superconformal indices for the $\CN=2$ theories of class $\CS$ studied in \cite{Gadde:2009kb, Gadde:2011ik, Gadde:2011uv, Gaiotto:2012xa} which was extended to the case of type $D$ by \cite{Lemos:2012ph}. In the process, we close some of the loose ends regarding the $\IZ_{2, 3}$-twisted punctures of $D_n$ theories. Then we compute the superconformal indices for the $\CN=1$ theories studied in section \ref{sec:SO}, \ref{sec:Sp}, \ref{sec:G2} using a similar formalism developed in \cite{Beem:2012yn, Gadde:2013fma}. 

\subsection{$\CN=2$ index}
The $\CN=2$ superconformal index is defined as
\be
 I = \Tr (-1)^F \left( \frac{t}{pq} \right)^r p^{j_2 + j_1} q^{j_2 - j_1} t^R \prod_{i} x_i^{f_i} \ , 
\ee
where $(j_1, j_2)$ are the Cartans of the Lorentz group $SU(2)_1 \times SU(2)_2$, $r$ and $R$ are the $U(1)_R$ and $SU(2)_R$ generators respectively. The $f_i$ denote the Cartans for the flavor symmetry group. For any class $\CS$ theories, the indices can be thought of as a correlation function for a topological field theory. It turns out that the indices for a class $\CS$ theory defined by a Riemann surface $\CC$ with genus $g$ and $n$ twisted or untwisted punctures labeled by $\rho_{1, \cdots, n}$ can be written as
\be \label{eq:tqftindex}
 I = \sum_{\lambda} \frac{\prod_{I=1}^n K_{\rho_I}(\vec{a}_I) P_\lambda (\vec{a}_{\rho_I} ) }{(K_\varnothing P_{\lambda} (t^\varnothing) )^{2g-2+n}} \ , 
\ee
where the summation is over the representations $\lambda$ of $\Gamma$. Let us explain the meaning of various symbols.

\begin{itemize}
\item
The function $P_\lambda$ is some special function defined by requiring the function $f_\lambda (\vec a) = K_{\textrm{full}}(\vec{a}) P_\lambda (\vec{a})$ to be orthonormal under the measure given by the vector multiplet index $I_V (\vec a)$:
\be \label{eq:idxortho}
 \oint [d \vec z] I_V (\vec z) f_\lambda (\vec a) f_{\lambda'} (\vec a) = \delta_{\lambda \lambda'} \ . 
\ee
The function $P_\lambda$ can be Schur function or Macdonald polynomial or related to the wave function of elliptic Ruijsenaars-Schneider model depending on the number of fugacities $(p, q, t)$ we want to keep. The $P_\lambda$ also depend on the choice of twisted/untwisted puncture.

\item
The $K$-factor $K_\rho$ is labeled by a embedding $\rho$ of $SU(2)$ into $G$, where $G = \Gamma$ for the untwisted puncture and $G$ is the group formed by folding the Dynkin diagram with the choice of outer-automorphism as in the table \ref{table:GammaG}. 
The embedding $\rho$ induces a decomposition of adjoint into the form $\oplus_j R_j \otimes V_j$ where $V_j$ is the spin-$j$ irrep of $SU(2)$ and $R_j$ are representations for the flavor symmetry group associated to the puncture. For the case of the Macdonald index ($p=0$), the $K$-factor can be written as \cite{Mekareeya:2012tn}
\be
 K_\Lambda (\vec a)= \textrm{PE} \left[\sum_j \frac{t^{j+1}}{1-q} \tr_{R_j} (\vec a) \right] \ , 
\ee
where PE stands for the plethystic exponential. For example, for the full puncture, it is simply given by
\be
 K_{\textrm{full}} (\vec a) = \textrm{PE} \left[ \frac{t}{1-q} \chi_{\textrm{adj}} (\vec{a})\right] \ .  
\ee
For the null puncture $\varnothing$, it is given by
\be
 K_{\varnothing} = \textrm{PE} \left[ \frac{t^{d_i}}{1-q} \right] = \prod_{i=1}^{\textrm{rank}(\Gamma)} (t^{d_i}; q)^{-1} \ , 
\ee
where $d_i$ are the degrees of invariants of $G$ and $(x; q) = \prod_{i=0}^\infty (1- x q^i)$ is the Pochhammer symbol. The general form of $K_\Lambda(\vec a; p,q,t)$ has been conjectured in \cite{Gadde:2013fma} to be
\be
 K_\Lambda (\vec a)= \textrm{PE} \left[\sum_j \frac{t^{j+1}-pqt^j}{(1-q)(1-p)} \tr_{R_j} (\vec a) \right] \ . 
 \label{eq:prefactor}
\ee 
\item
The argument $\vec{a}_{\rho_I}$ can be determined by looking at the embedding of $\rho(SU(2)) \times G_F$ into $G$ where $\rho(SU(2))$ is image under the map $\rho$ and $G_F$ is the flavor symmetry group associated to the puncture. The fundamental of $G$ can be decomposed in terms of spin-$j$ irreps of $SU(2)$ as $\textrm{fund}_G = \oplus_j R^F_j \otimes V_j$.  
One can match the fugacities by using characters. First write down the character for the fundamental of $G$. And then compare it with the characters of the representations of $SU(2) \times G_F$. By comparing the two, one can map the fugacities for the $G_F$ to the fugacities of $G$ appear in $P_\lambda (\vec a)$. See the section 4.2.1 of \cite{Lemos:2012ph} for more details. 

\end{itemize}
Now, let us focus on the examples of twisted $D_n$-type theories. We will restrict our discussion to the case of Macdonald index $p=0$. 

We implemented computation of Macdonald polynomials using the procedure outlined in appendix B of \cite{Mekareeya:2012tn} through direct Gram-Schmidt process using Mathematica and LieART \cite{Feger:2012bs}. There is more efficient method of computing Macdonald polynomials for $A, B, C, D, E_{6, 7}$ through determinantal construction \cite{determinantal2003}. We refer to appendix A of \cite{Lemos:2012ph} for a nice review on the construction of Macdonald and Hall-Littlewood polynomials.

\subsubsection*{$D_n$-type theories with $\IZ_2$-twist}
The function $P_\lambda$ in our case becomes the normalized Macdonald polynomial of type $G$ where $G$ is either $\Gamma = D_n$ or $G=C_{n-1}$ depending on the choice of untwisted and twisted puncture. 
\be
 P_\lambda (\vec{a}) = N_\lambda^{-1/2}  P_{M, G}^\lambda (\vec{a}; q, t) \ , 
\ee
where $P_{M, G}$ is the Macdonald polynomial given by the root-system of $G$.\footnote{In general, $P_M$ is labeled by an affine root system. There is many to one map between the affine root systems and the group $G$. In our case, only the Macdonald polynomial for $G$ appears. The other ones such as the dual root system $G^\vee$ and the non-reduced affine root system $(C^\vee_n, C_n)$ appear when we consider outer-automorphism twisted index. \cite{Mekareeya:2012tn}}  The $N_\lambda (q, t) $ is a normalization factor given by inner product of two Macdonald polynomials 
\be
 N_\lambda = \langle P_{M, G}^\lambda, P_{M, G}^{\lambda} \rangle = 
 \int [d \vec z]_G \textrm{PE} \left[ \frac{-q + t}{1-q} \chi_{\textrm{adj}} (\vec z) \right] P_{M, G}^\lambda (\vec z) P_{M, G}^\lambda (\vec z)  \ , 
\ee
where $[dx]_G$ stands for the Haar measure of the group $G$. 
For the $D_4$ case, we have two different choice of twisting, namely $\IZ_2$ and $\IZ_3$ which gives $C_3$ and $G_2$. We will treat this special case later in this section.

The superconformal index for the ${T}_{SO(2n)}$ theory is given by
\be
 I = \frac{K^{SO}_{\textrm{full}}(\vec{a}_1) K^{SO}_{\textrm{full}}(\vec{a}_2) K^{SO}_{\textrm{full}}(\vec{a}_3)}{K^{SO}_\varnothing}  \sum_{\lambda \in R_{SO(2n)}} \frac{ P^{SO}_\lambda (\vec{a}_1 ) P^{SO}_\lambda (\vec{a}_2 ) P^{SO}_\lambda (\vec{a}_3 ) }{P^{SO}_{\lambda} (t^\varnothing) } \ , 
\ee
where the $P_\lambda$ is given by the $SO(2n)$ Macdonald polynomial. One can start from this theory and then by partially closing or Higgsing the punctures, to obtain general theory corresponding to a 3 punctured sphere. In more extreme limit, one can consider completely closing the punctures. Then the index should be trivial, which completely fixes the factor in the denominator which is the structure constant of the TQFT.

More generally, when we have twisted punctures, the structure constant can be fixed by requiring it to become trivial when we close all the three punctures. Therefore we can write the index for the $\tilde{T}(SO(2n))$ theory as
\be
 I = \frac{K^{SO}_{\textrm{full}} (\vec a) K^{USp}_{\textrm{full}} (\vec b_1) K^{USp}_{\textrm{full}} (\vec b_2) }{K^{SO}_\varnothing } \sum_{\lambda \in R_{USp(2n-2)}} \frac{ P^{SO}_\lambda (\vec{a}) P^{USp}_\lambda (\vec{b}_1) P^{USp}_\lambda (\vec{b}_2) }{P^{SO}_\lambda (t^\varnothing) } \ , 
\ee
where the sum is over the representations of $USp(2n-2)$ not $SO(2n)$. For the $P^{SO(2n)}$, we restrict the sum to the case of outer-automorphism invariant representations. In terms of Dynkin labels, they are of the form $[\lambda_1, \lambda_2, \cdots, \lambda_{n-1}, \lambda_{n-1}]$. 

One can completely close one of the $USp$ puncture to obtain the free theory of $SO(2n)\times USp(2n-2)$ bifundamental half-hypermultiplets. It is given by
\be \label{eq:bifundidx}
 I_{\textrm{bifund}} = \frac{K^{SO}_{\textrm{full}} (\vec a) K^{USp}_{\textrm{full}} (\vec b) K^{USp}_{\varnothing} }{K^{SO}_\varnothing } \sum_{\lambda \in R_{USp(2n-2)}} \frac{ P^{SO}_\lambda (\vec{a}) P^{USp}_\lambda (\vec{b}) P^{USp}_\lambda (t^{\varnothing}) }{P^{SO}_\lambda (t^\varnothing) } \ .
\ee
We have checked this relation up to $n=5$ and to a few orders in $q$. 

When we glue three punctured spheres, we integrate with a vector multiplet measure. From the orthonormality condition \eqref{eq:idxortho}, we arrive at the same result of \eqref{eq:tqftindex}. One interesting aspect here is that whenever there is a twisted puncture, summation over the representations of $\Gamma$ reduces to that of $G$. 

\subsubsection*{$D_4$-type theories with $\IZ_3$-twist} 
The $\Gamma = D_4$ theory can be twisted in two different ways because the outer-automorphism group is generated by $\IZ_2$ and also $\IZ_3$. The $\IZ_2$ twisting gives $C_3 = USp(6)$ puncture and the $\IZ_3$ twisting gives $G_2$ puncture. Consider the three punctured sphere given by one $USp(6)$ puncture and one $G_2$ puncture with twisted null puncture as in the figure \ref{fig:E7}. From the TQFT structure, we can write its index as

\be \label{eq:E7index}
 I_{E_7} (\vec a, \vec b) = \frac{K^{G_2}_{\textrm{full}} (\vec a) K^{USp}_{\textrm{full}} (\vec b) K^{USp}_\varnothing}{K^{SO}_\varnothing} \sum_{\lambda \in R_{G_2}} \frac{P^{G_2}_\lambda (\vec a) P^{USp}_\lambda (\vec b) P^{USp}_\lambda (t^\varnothing) }{P^{SO}_\lambda (t^\varnothing)} \ . 
\ee
Here the sum is over the representations of $G_2$. For the $SO(8)$ and $USp(6)$ punctures, this means summing over the representations invariant under the $\IZ_3$ action. In terms of the Dynkin labels, they are $[\lambda_1, \lambda_2, \lambda_1, \lambda_1]$ and $[\lambda_1, \lambda_2, \lambda_1]$ for the $G_2$ representation $[\lambda_2, \lambda_1]$. 

The TQFT structure requires S-duality invariance of the index. In our case, it translates to the condition that the indices for the first two frames of $G_2$-coupled two $E_7$ theories as in figure \ref{fig:z3twist} being equal. We should have
\be
\oint [d \vec \omega] I_{\textrm{vec}}^{G_2} (\vec \omega) I_{E_7} (\vec \omega, \vec a) I_{E_7} (\vec \omega, \vec b) = \oint [d \vec z] I_{\textrm{vec}}^{SO(8)}(z) I_{\textrm{bifund}} (\vec z, \vec{a}) I_{\textrm{bifund}}(\vec{ \tilde{z}}, \vec{b}) \ , 
\label{eq:G2SO8ind}
\ee
where $I^G_{\textrm{vecl}}$ is the vector multiplet index for the gauge group $G$ and $I_{\textrm{bifund}}$ denotes the index of the $SO(8)\times USp(6)$ bifundamentals \eqref{eq:bifundidx}. We represent the $G_2$ fugacities with $\vec\omega$ while the $SO(8)$ fugacities are given by $\vec{z} = (z_1, z_2, z_3, z_4)$ and $\vec{\tilde{z}} = (z_4, z_2, z_3, z_1)$. The transformation of $SO(8)$ fugacities from $\vec{z}$ to $\vec{\tilde{z}}$ implements the $\IZ_3$ twist around the $SO(8)$ cylinder in figure \ref{fig:N2Z3SO8}. Orthogonality of the $SO(8)$ wave-functions upon integration with respect to the $SO(8)$ vector multiplets implies that only those representations that are of the $\IZ_3$ invariant form mentioned before, contribute to the RHS of \eqref{eq:G2SO8ind}. This is enough to show the identity of \eqref{eq:G2SO8ind}

As a remark, we find that the index for the $E_7$ theory can also be written as 
\be \label{eq:e7idxnop}
 I_{E_7} (\vec a, \vec b) = \frac{K^{G_2}_\varnothing K^{G_2}_{\textrm{full}} (\vec a) K^{USp}_{\textrm{full}} (\vec b) }{K^{USp}_\varnothing} \sum_{\lambda \in R_{G_2}} \frac{P^{G_2}_\lambda ( t^\varnothing) P^{G_2}_\lambda (\vec a) P^{USp}_\lambda (\vec b) }{ P^{USp}_\lambda (t^\varnothing) } \ . 
\ee
We can get this form from the identity 
\be \label{eq:g2tqft}
\left( P^{USp}_\lambda (t^\varnothing) \right)^2 = P^{SO}_\lambda (t^\varnothing) P^{G_2}_\lambda (t^\varnothing)  \ , 
\ee
where the representations $\lambda$ are now restricted to belong to the $\IZ_3$ invariant form discussed above. We do not have an analytic proof of the identity \eqref{eq:g2tqft}, but we were able to check this relation for several low-dimensional representations. 

From the form \eqref{eq:e7idxnop}, the index becomes $1$ upon closing all the punctures. For the case of UV curves without twisted punctures, we always get $1$ upon closing all the punctures. It is not clear whether it should be the case with twisted punctures, because even after closing a twisted puncture it still carries non-trivial information. Nevertheless, it turns out that the superconformal index is unity for the theory having a UV curve with only null punctures (with or without twist) of type $A_n, D_n$.  

\subsubsection*{Enhancement of Global symmetry $USp(6)\times G_2$ to $E_7$}
As we have discussed in section \ref{sec:G2}, the theory given by $USp(6)$ and $G_2$ punctures is expected to  have enhanced $E_7$ global symmetry \cite{Tachikawa:2010vg}. Here we check this explicitly through the computation of index. We find that the index of this theory computed by \eqref{eq:E7index} can be indeed written in terms of the characters of $E_7$. 

The product algebra $G_2 \times USp(6)$ is embedded into $E_7$ such that \cite{McKay:99021}
\be
 56 &\to& (7, 6) \oplus (1, 14) \ , \\ 
 133 &\to& (7, 14) \oplus (14, 1) \oplus (1, 21) \ ,\\
 7371 &\to& (27, 90) \oplus (14, 70) \oplus (64, 14) \oplus (7, 189) \oplus (77', 1)   \nn \\
 &{ }& \oplus (27, 14) \oplus(7, 70)  \oplus (14, 21) \oplus (1, 126') \oplus (1, 90)\\
 &{ }&  \oplus (7, 21) \oplus (7, 14)  \oplus (27, 1) \oplus (1, 14) \oplus (1, 1) \ . \nn 
\ee
We find that the index of the $USp(6) \times G_2$ theory can be written in terns of the $E_7$ characters. For example, the Schur index $(p=0, q=t)$ can be written as
\be
 I_{\text{Schur}} = 1 + \chi^{E_7}_{133} (\vec a, \vec b) q + (\chi^{E_7}_{7371} (\vec a, \vec b)+ \chi^{E_7}_{133} (\vec a, \vec b) + 1) q^2 + \cdots \ , 
\ee
where we used the above decompositions to write as $\chi^{E_7}_{133} (\vec a, \vec b) = \chi^{G_2}_7 (\vec a) \chi^{USp(6)}_{14} (\vec b) + \chi^{G_2}_{14}(\vec a) \cdot 1 + 1 \cdot \chi^{USp(6)}_{21} (\vec b)$ and so on. 

Especially, the Hall-Littlewood index $(p=0, q=0)$ is known to reproduce the Hilbert Series of the Higgs branch when the UV curve has genus 0 \cite{Gadde:2011uv}. The Higgs branch of Minahan-Nemeschansky $E_7$ theory is known to be the moduli space of $E_7$ instantons with instanton number 1. The Hilbert series of 1 instanton moduli space is entirely given in terms of the characters for the symmetric product of adjoint representations:
\be
 \textrm{Hilb}(\CM_{G, k=1}) = \sum_{n \ge 0} \chi_{Sym^n(\textrm{adj})} t^{n} \  .
\ee
This relation for the exceptional group was proven in \cite{VinbergPopov, Garfinkle} and studied in the physics literatures by \cite{Benvenuti:2010pq, Keller:2011ek, Keller:2012da}. We verified that the Hall-Littlewood index for the $USp(6)\times G_2$ theory is indeed written in terms of the characters of the adjoint representations of $E_7$
\be
 I_{HL} = \sum_{n \ge 0} \chi [n, 0, 0, 0, 0, 0, 0] t^n \ , 
\ee
where we used the Dynkin label here. 

\subsubsection*{Bifundamentals of $G_2 \times USp(4)$ through Higgsing the $E_7$ theory}
As we have discussed in section \ref{sec:G2}, we can obtain a free theory of $G_2 \times USp(4)$ bifundamentals by partially Higgsing the $USp(6)$ global symmetry down to $USp(4)$ of the $E_7$ theory. We obtain the $K$-factor from decomposing the adjoint of $USp(6)$ to the representations of $SU(2)\times USp(4)$ which is
\be
 K_{USp(4)} = \textrm{PE} \left[ \frac{1}{1-q} \left(\chi_{[2, 0]} t + \chi_{[1, 0]} t^{3/2} + \chi_{[0, 0]} t^2\right) \right] \ , 
\ee
where we used Dynkin labels to write the representation of $USp(4)$. The fugacities for the $USp(4)$ puncture is $(t, t^{1/2} b_1, t^{1/2} b_2)$ in the $\a$-basis meaning all the weights are given as a linear combination of the simple roots. The fugacities for the null puncture is $(t, t^{-1})$ for the $G_2$ and $(t^{5/2}, t^4, t^{9/2})$ for the $USp(6)$ in the $\a$-basis.


\subsection{$\CN=1$ index}
Now, let us move on to the discussion of the superconformal indices of $\CN=1$ class $\CS$ theories. 
The $\CN=1$ superconformal index is defined as 
\be
 I(z; p, q, \xi) = \Tr (-1)^F p^{j_1 - j_2 + R/2}  q^{j_1 + j_2 + R/2} \xi^{-\CF/2} z^Q \ , 
\ee
where $\CF$ is the $U(1)_\CF$ global symmetry preserved in the class $\CS$ theory.  

The $\CN=1$ index of theories constructed in the present paper can be obtained from the $\CN=2$ index of their building blocks. These building blocks can be classified into the colored $T_N^{\sigma}$ blocks ($\sigma= \pm$) and the $\CN=1$ and $\CN=2$ vector multiplets that couple them together. Their contribution to the  $\CN=1$ index is given by $I_{\CN=1}=I_{\CN=2}(p,q,t=\xi^{\sigma}\sqrt{pq})$, where $\xi^{\sigma}$ gives their charge with respect to the $U(1)_{\CF}$ flavor symmetry.\footnote{ Here for the sake of brevity, we have omitted the fugacities for all flavor symmetries of the three punctured spheres. Nevertheless they are there and will be important for matching the index across various duality frames.} As mentioned previously, the underlying TQFT structure implies that  the $\CN=2$ superconformal index of class $\CS$ theories can be written in terms of orthogonal functions $f_{\lambda}(\vec a; p,q,t)$. It is expected that in general $f_{\lambda}(\vec a; p,q,t)$ are related to the wave-functions of elliptic Ruijsenaars-Schneider model. There is some evidence  that for theories of type $A_N$ these functions satisfy the identity  \cite{Gaiotto:2012xa}
\be
f_{\lambda}(\vec a; p,q,t) = \text{PE} \left[ \frac{t-pq/t}{(1-p)(1-q)}\chi_{\textrm{adj}}({\vec{a}}) \right] f_{\lambda}(\vec a; p,q,\frac{pq}{t}) \ . 
\ee   
We will henceforth assume that this identity continues to hold for theories of type $D_N$ and their outer-automorphism twists. This identity implies that the functions $P_{\lambda}(p,q,t)$ are invariant under $t \leftrightarrow pq/t$. Upon reducing this to the case of $\CN=1$ index, it ensures the invariance of $P_{\lambda}(p,q,\xi^{\sigma}\sqrt{pq})$ under $\xi \leftrightarrow \xi^{-1}$. The superconformal index of two $T_N^{\sigma}$ blocks coupled by an $\CN=1$ vector multiplet can be written as 
\be
I(\vec a, \vec b; \vec c, \vec d) = \oint [d \vec z] I^{\CN=1}_{\textrm{vec}}(\vec z) I_{T_N^+}(\vec z, \vec a, \vec b) I_{T_N^-}(\vec z, \vec c, \vec d) \ , 
\label{eq:SCI1}
\ee
where $\vec a, \vec b,\vec c, \vec d$ are the fugacities for the flavor symmetries of the theory while $\vec z$ are the fugacities for the gauge group. $ I_{T_N^{\sigma}}$ is the $\CN=1$ index of $T_N^{\sigma}$ theory obtained from its $\CN=2$ index. Due to orthonormality of the wavefunctions, the index in (\ref{eq:SCI1}) formally simplifies  to 
\be
I(\vec a, \vec b; \vec c, \vec d) = \sum_{\lambda} \frac{f^+_{\lambda}(\vec a) f^+_{\lambda}(\vec b) f^-_{\lambda}(\vec c) f^-_{\lambda}(\vec d) }{f^+_{\lambda}(\varnothing) f^-_{\lambda}(\varnothing)}
\label{eq:Isimp} \ . 
\ee
Here $f^{\sigma}_{\lambda}(\vec a)$ is short-hand for $f_{\lambda}(\vec a;p,q, t=\xi^{\sigma}\sqrt{pq})$ and has to be chosen appropriately according to the flavor symmetry of puncture ``a". $f^\pm_{\lambda}(\varnothing)$ correspond to the structure constants in the $\CN=2$ index. The sum in (\ref{eq:Isimp}) is over the set of representations  whose Dynkin labels are of the form explained earlier in the paper. 

\subsubsection*{$SO$ dualities}
We first compare the superconformal index of the unHiggsed theories across the various duality frames. In the electric theory, $\CT^{SO}$, we find that the index can be written as 
\be
I_{\CT^{SO}}(\vec a, \vec b; \vec c, \vec d) =\frac{K^{USp}_+(\vec a) K^{USp}_+(\vec b) K^{USp}_-(\vec c) K^{USp}_-(\vec d)}{K^{SO}_{\varnothing ,+} K^{SO}_{\varnothing ,-}} \sum_{\lambda} \frac{P^{USp}_{\lambda}(\vec a) P^{USp}_{\lambda}(\vec b) P^{USp}_{\lambda}(\vec c) P^{USp}_{\lambda}(\vec d) }{P^{SO}_{\lambda}(t^\varnothing) P^{SO}_{\lambda}(t^\varnothing)} \ . \nn \\
\label{eq:Isoe}
\ee
 In the crossing frame, $\CT^{SO}_c$, the punctures $B$ and $C$ are exchanged with each other. Their $U(1)_{\CF}$ charges switch signs and we had to integrate in mesons $M_B$ and $M_C$ with $U(1)_{\CF}$ charges being $-2$ and $+2$  respectively. The index of $\CT^{SO}_c$ then becomes 
\be
I_{\CT^{SO}_c}(\vec a, \vec c; \vec b, \vec d)=M^+(\vec b)M^-(\vec c) I_{\CT^{SO}}(\vec a, \vec c; \vec b, \vec d) \ ,
\label{eq:Isomc}
\ee
where $M^{\sigma}(\vec x)$ is the contribution of the mesons having $\CF$-charge $-2\sigma$ and flavor fugacities $\vec x$
\be
M^{\sigma}(\vec x) = \textrm{PE} \left[ \frac{\sqrt{pq}(\xi^{\sigma}-\xi^{-\sigma})}{(1-p)(1-q)} \chi_{\textrm{adj}}(\vec x) \right] .
\ee
The equality of the indices in (\ref{eq:Isoe}) and (\ref{eq:Isomc}) then follows from the identity
\be
 M^{\sigma}(\vec x) K^{-\sigma}(\vec x) =K^{\sigma}(\vec x) \ . 
 \label{eq:mes}
\ee
We can repeat this exercise for the index of the theory $\CT^{SO}_s$ , in the swapped frame wherein we find  
\be
I_{\CT^{SO}_s}(\vec d, \vec c; \vec b, \vec a)=M^+(\vec a)M^+(\vec b)M^-(\vec c)M^-(\vec d) I_{\CT^{SO}}(\vec d, \vec c; \vec b, \vec a) \ . 
\label{eq:Isoms}
\ee
The identity in (\ref{eq:mes}) can now be used to match the indices in the various duality frames. 

The procedure of Higgsing the $USp(2N-2)$ punctures can be implemented in the index by transmuting the $USp(2N-2)$ fugacities into the fugacities of the partially closed puncture. As has been mentioned earlier this can be achieved by comparing the character of the $USp(2N-2)$ fundamental written in terms of the fugacities of the $USp(2N-2)$ symmetry, to the character written in terms of the $SU(2) \times G_F \subset USp(2N-2)$. The $SU(2)$  here is embedded into $USp(2N-2)$ through the vev we use to Higgs the puncture while $G_F$ is residual flavor symmetry left invariant by the vev. The fugacity for $SU(2)$ characters is required to be $\tau= (\xi^{\sigma}\sqrt{pq})^{1/2}$. The redundancy in the choice of fugacities corresponds to the Weyl symmetries of $USp(2N-2)$. The prefactor $K_\Lambda(\vec a; p,q,t=\xi^{\sigma}\sqrt{pq})$ is given by (\ref{eq:prefactor}).

Applying this to close the punctures $A$ and $D$ we find that the index for the electric theory $\CU^{SO}$ can be written as 
\be
I_{\CU^{SO}}(\varnothing, \vec b; \vec c, \varnothing) =\frac{K^{USp}_{\varnothing,+} K^{USp}_{\varnothing,-} K^{USp}_+(\vec b) K^{USp}_-(\vec c) }{K^{SO}_{\varnothing ,+} K^{SO}_{\varnothing ,-}} \sum_{\lambda} \frac{P^{USp}_{\lambda}((\xi\sqrt{pq})^{\varnothing}) P^{USp}_{\lambda}((\xi^{-1}\sqrt{pq})^{\varnothing}) P^{USp}_{\lambda}(\vec b) P^{USp}_{\lambda}(\vec c)  }{P^{SO}_{\lambda}((\xi\sqrt{pq})^{\varnothing}) P^{SO}_{\lambda}((\xi^{-1}\sqrt{pq})^{\varnothing})}\ . \nn \\
\label{eq:IHsoe}
\ee
In the Intriligator-Seiberg (magnetic) frame $\CU^{SO}_{c1}$, the superconformal index is 
\be
I_{\CU^{SO}_{c1}}(\varnothing, \vec c; \vec b, \varnothing)=M^+(\vec b)M^-(\vec c) I_{\CU^{SO}}(\varnothing, \vec c; \vec b, \varnothing)  \ , 
\label{eq:IHsomc}
\ee
which matches with the index of the electric theory upon using (\ref{eq:mes}). 

In the swapped frame it is the mesons that get a vev, leading to a shift in the $R$-and $\CF$-charges. The shift of the charges can be accommodated into the index by the following substitution: in the $\tilde{T}^{\sigma}_{N}$ block of the swapped theory, replace the fugacities for $USp(2N-2)$ with those for  $SU(2) \times G_F\subset USp(2N-2)$ using $\xi^{\sigma}/\sqrt{pq}$ as the fugacity for $SU(2)$.  The index of the swapped theory, $\CU^{SO}_s$, is therefore given by
\be
 I_{\CU^{SO}_s}= M^+_{\varnothing} M^-_{\varnothing}M^+(\vec b)M^-(\vec c) I_{\CT^{SO}}((\xi/\sqrt{pq})^{\varnothing} , \vec c ; \vec b, (\xi^{-1}/\sqrt{pq})^{\varnothing}) \ , 
\ee
where $M^{\sigma}_{\varnothing}$ is the contribution from the mesonic excitations $M_{Aj,-j}$ and $M_{Dj,-j}$ that stay coupled to the theory: 
\be
M^{\sigma}_{\varnothing} = \prod_j \textrm{PE} \left[ \frac{(\xi^{\sigma}\sqrt{pq})^{1+j}-pq/(\xi^{\sigma}\sqrt{pq})^{1+j}}{(1-p)(1-q)} \right] \ .  
\ee
Similarly the index for the theory, $\CU^{SO}_{as}$,  in the Argyres-Seiberg frame can be written as 
\be
 I_{\CU^{SO}_{as}}= M^+_{\varnothing} M^-(\vec c) I_{\CT^{SO}}(\vec c, \vec b ;  (\xi^{-1}/\sqrt{pq})^{\varnothing}, \varnothing) \ , 
\ee
while the index for the theory, $\CU_{c2}$, in the crossing frame is given by 
\be
 I_{\CU^{SO}_{c2}}= M^+_{\varnothing} M^-_{\varnothing} I_{\CT^{SO}}((\xi/\sqrt{pq})^{\varnothing} ,  \vec b ;  \vec c,  (\xi^{-1}/\sqrt{pq})^{\varnothing}) \ . 
\ee
The equality of the indices in the various duality frames can be established by using the identity
\be
 M^{\sigma}_{\varnothing} K^{USp}_{-\sigma}((\xi^{-\sigma}/\sqrt{pq})^{\varnothing})= K^{USp}_{\varnothing,\sigma} \ . 
 \label{eq:higgsedmes}
\ee
along with (\ref{eq:mes}) and the invariance of $P_{\lambda}^{USp}$ under the Weyl symmetries of $USp(2N-2)$.\footnote{More specifically we use the fact that $P_{\lambda}^{USp}(\vec a) = P_{\lambda}^{USp}(\vec a^{-1})$. }

\subsubsection*{$USp$ dualities}
Following a similar procedure as in the case of the SQCD with $SO(2N)$ gauge group, we can now write down the index of the various duality frames of SQCD with $USp(2N-2)$ gauge group. Before Higgsing some the punctures, we compare the indices of the unHiggsed theories in the various duality frames we obtain by moving the punctures around. The index for the electric theory, $\CT^{Sp}$ is
\bea
I_{\CT^{Sp}}(\vec a, \vec b; \vec c, \vec d) =\frac{K^{USp}_+(\vec a) K^{SO}_+(\vec b) K^{SO}_-(\vec c) K^{USp}_-(\vec d)}{K^{SO}_{\varnothing ,+} K^{SO}_{\varnothing ,-}} \sum_{\lambda} \frac{P^{USp}_{\lambda}(\vec a) P^{SO}_{\lambda}(\vec b) P^{SO}_{\lambda}(\vec c) P^{USp}_{\lambda}(\vec d) }{P^{SO}_{\lambda}(t^\varnothing) P^{SO}_{\lambda}(t^\varnothing)} \ , \nn \\ 
\label{eq:Ispe}
\eea
where the sum now is over the representations of $USp(2N-2)$, as was explained earlier. In the duality frame $\CT^{Sp}_{c1}$ obtained by exchanging punctures $B$ and $C$, we find 
\be
I_{\CT^{Sp}_{c1}}(\vec a, \vec c; \vec b, \vec d)=M^+(\vec b)M^-(\vec c) I_{\CT^{Sp}}(\vec a, \vec c; \vec b, \vec d) \ , 
\ee
Similarly the index of the crossing theory $\CT^{Sp}_{c2}$, obtained by exchanging punctures $A$ and $D$, is 
\be
I_{\CT^{Sp}_{c2}}(\vec d, \vec b; \vec c, \vec a)=M^+(\vec a)M^-(\vec d) I_{\CT^{Sp}}(\vec d, \vec b; \vec c, \vec a) \ . 
\ee
In the frame $\CT^{Sp}_{c3}$, obtained by exchanging puncture $B$ and $D$, the index becomes 
\be
I_{\CT^{Sp}_{c3}}(\vec a, \vec d; \vec c, \vec b)=M^+(\vec b)M^-(\vec d) I_{\CT^{Sp}}(\vec a, \vec d; \vec c, \vec b) \ . 
\ee
The index for the theory $\CT^{Sp}_{s}$ in the swapped frame is 
\be
I_{\CT^{Sp}_s}(\vec d, \vec c; \vec b, \vec a)=M^+(\vec a)M^+(\vec b)M^-(\vec c)M^-(\vec d) I_{\CT^{Sp}}(\vec d, \vec c; \vec b, \vec a)  \ .
\label{eq:Ispms}
\ee
Equality of the above indices follows from (\ref{eq:mes}).

Upon appropriately Higgsing the punctures $A$ and $D$ we find that the index in the electric theory $\CU^{Sp}$ can be written as 
\be
I_{\CU^{Sp}}(\varnothing, \vec b; \vec c, \varnothing) =\frac{K^{USp}_{\varnothing,+} K^{USp}_{\varnothing,-} K^{SO}_+(\vec b) K^{SO}_-(\vec c) }{K^{SO}_{\varnothing ,+} K^{SO}_{\varnothing ,-}} \sum_{\lambda} \frac{P^{USp}_{\lambda}((\xi\sqrt{pq})^{\varnothing}) P^{USp}_{\lambda}((\xi^{-1}\sqrt{pq})^{\varnothing}) P^{SO}_{\lambda}(\vec b) P^{SO}_{\lambda}(\vec c)  }{P^{SO}_{\lambda}((\xi\sqrt{pq})^{\varnothing}) P^{SO}_{\lambda}((\xi^{-1}\sqrt{pq})^{\varnothing})} \ . \nn \\
\label{eq:IHspe`}
\ee
The index of Intriligator-Pouliot theory $\CU^{Sp}_{c1}$ is 
\be
I_{\CU^{Sp}_{c1}}(\varnothing, \vec c; \vec b, \varnothing)=M^+(\vec b)M^-(\vec c) I_{\CU^{Sp}}(\varnothing, \vec c; \vec b, \varnothing) \ . 
\label{eq:IHspIP}
\ee
For the crossing theory $\CU^{Sp}_{c2}$, the index is given by 
\be
 I_{\CU^{Sp}_{c2}}= M^+_{\varnothing} M^-(\varnothing) I_{\CT^{Sp}}((\xi/\sqrt{pq})^{\varnothing} ,  \vec b ;  \vec c,  (\xi^{-1}/\sqrt{pq})^{\varnothing}) \ .
\ee
Similarly in the swapped frame $\CU^{Sp}_s$ and the Argyres-Seiberg dual frame $\CU^{Sp}_{as}$, the respective superconformal indices are:
\begin{align}
I_{\CU^{Sp}_s} &= M^+_{\varnothing} M^-_{\varnothing}M^+(\vec b)M^-(\vec c) I_{\CT^{Sp}}((\xi/\sqrt{pq})^{\varnothing} , \vec c ; \vec b, (\xi^{-1}/\sqrt{pq})^{\varnothing}) \ , \\
I_{\CU^{Sp}_{as}} & = M^+_{\varnothing} M^-(\vec c) I_{\CT^{Sp}}(\vec c, \vec b ;  (\xi^{-1}/\sqrt{pq})^{\varnothing}, \varnothing) \ .
\end{align}
The indices in the various duality frames match owing to the identities (\ref{eq:mes}) and (\ref{eq:higgsedmes}) and the Weyl invariance of $P^{USp}_{\lambda}$.

\subsubsection*{$G_2$ dualities}
The index of the theories involved in the $G_2$ dualities proposed by us can be written in terms of the $\CN=1$ index of the theory $\CT^{G_2}$ obtained by coupling two $\tilde{T}_{SO(8)}$ blocks with an $\CN=1$, $SO(8)$ vector multiplet and a $\textrm{Z}_3$ twist around the cylinder that couples two spheres. The superconformal index for this theory is 
\be
I_{\CT^{G2}}(\vec p, \vec q; \vec r, \vec s) =\frac{K^{USp}_+(\vec p) K^{USp}_+(\vec q) K^{USp}_-(\vec r) K^{USp}_-(\vec s)}{K^{SO}_{\varnothing ,+} K^{SO}_{\varnothing ,-}} \sum_{\lambda} \frac{P^{USp}_{\lambda}(\vec p) P^{USp}_{\lambda}(\vec q) P^{USp}_{\lambda}(\vec r) P^{USp}_{\lambda}(\vec s) }{P^{SO}_{\lambda}(t^\varnothing) P^{SO}_{\lambda}(t^\varnothing)} \ , \nn
\\ \label{eq:Ig2}
\ee
where the sum is over $G_2$ representations. The electric theory $\CU^{G_2}$ is built from bifundamentals of $G_2 \times USp(4)$ and its index is 
\be
I_{\CU^{G_2}}(\vec a; \vec b) = I_{\CT^{G2}}(\varnothing, \vec a (\xi \sqrt{pq})^{\heartsuit}; \vec b (\xi^{-1} \sqrt{pq})^{\heartsuit}, \varnothing) \ .
\ee
Here $\vec a$  and $\vec b$ are the fugacities for $USp(4)_A$ and $USp(4)_B$ respectively and $\heartsuit$ represents the embedding of $SU(2)$  in $USp(6)$ that reduces the flavor symmetry of the puncture down to $USp(4)$.

In the $Spin(8)$ frame, the superconformal index of the theory is given by 
\be
I_{\CU^{G_2}_{c1}} = M^+_{\heartsuit}(\vec a) M^-_{\heartsuit}(\vec b)  I_{\CT^{G2}}(\varnothing, \vec b (\xi/ \sqrt{pq})^{\heartsuit}; \vec a (\xi^{-1} /\sqrt{pq})^{\heartsuit}, \varnothing) \ , 
\ee
where $M^{\sigma}_{\heartsuit}(\vec a)$ are the mesons that remain in the theory after Higgsing the corresponding $USp(6)$puncture down to $USp(4)$ which is given by
\be
M^{\sigma}_{\heartsuit}(\vec a) &=& \textrm{PE} \left[ \frac{(\xi^{\sigma}\sqrt{pq})-pq/(\xi^{\sigma}\sqrt{pq})}{(1-p)(1-q)} \chi_{\textrm{adj}}(\vec a) \right]  \nn \\
&{ }& \times ~ \textrm{PE} \left[ \frac{(\xi^{\sigma}\sqrt{pq})^{\frac{3}{2}}-pq/(\xi^{\sigma}\sqrt{pq})^{\frac{3}{2}}}{(1-p)(1-q)} \chi_{\textrm{f}}(\vec a) \right] \\
&{ }& \times ~ \textrm{PE} \left[ \frac{(\xi^{\sigma}\sqrt{pq})^{2}-pq/(\xi^{\sigma}\sqrt{pq})^{2}}{(1-p)(1-q)} \right] \ . \nn 
\ee
In the crossing-type frame we find 
\be
I_{\CU^{G_2}_{c2}} = M^+_{\varnothing}M^-_{\varnothing}  I_{\CT^{G2}}( (\xi/ \sqrt{pq})^{\varnothing},  \vec a (\xi \sqrt{pq})^{\heartsuit}; \vec b (\xi^{-1} \sqrt{pq})^{\heartsuit} , (\xi^{-1} /\sqrt{pq})^{\varnothing}) \ .
\ee
The superconformal index for the Argyres-Seiberg type dual can be written as 
\be
I_{\CU^{G_2}_{as}} = M^+_{\heartsuit}(\vec a)M^-_{\varnothing}  I_{\CT^{G2}}( (\xi/ \sqrt{pq})^{\varnothing},\varnothing  ; \vec b (\xi^{-1} \sqrt{pq})^{\heartsuit} , \vec a (\xi^{-1} /\sqrt{pq})^{\heartsuit}) \ .
\ee
Similarly the index of the theory in the swapped $G_2$ frame is
\be
I_{\CU^{G_2}_s} = M^+_{\varnothing}M^-_{\varnothing}  M^+_{\heartsuit}(\vec a)M^-_{\heartsuit}(\vec b)  I_{\CT^{G2}}( (\xi/ \sqrt{pq})^{\varnothing},\vec b (\xi/\sqrt{pq})^{\heartsuit}  ;  (\xi^{-1}/ \sqrt{pq})^{\heartsuit} , \vec a (\xi^{-1} /\sqrt{pq})^{\heartsuit}) \ . ~~~~~~
\ee
The indices in all these frames match upon using the Weyl invariance of $P^{USp}_{\lambda}$ along with (\ref{eq:mes}) and the generalized form of (\ref{eq:higgsedmes}) given by 
\be
 M^{\sigma}_{\Lambda} K^{USp}_{-\sigma}((\xi^{-\sigma}/\sqrt{pq})^{\Lambda})= K^{USp}_{\Lambda,\sigma} \ . 
 \label{eq:higgsedmesgen}
\ee
Therefore we find the indices all agree on five dual frames of the $G_2$ gauge theory.

\acknowledgments
We would like to thank Ibrahima Bah, Kenneth Intriligator, Abhijit Gadde, Kazunobu Maruyoshi, John McGreevy, Yuji Tachikawa and Wenbin Yan for discussions and correspondence. 
We would also like to thank the Simons Center for Geometry and Physics and the organizers of the 2013 Summer Simons Workshop in Mathematics and Physics for their hospitality while this work was being conceived.
JS thanks Korea Institute for Advanced Study for the hospitality where this work is finalized. This work is supported by DOE grant DOE-FG03-97ER40546.

\appendix

\section{Chiral ring relations of $T_{SO(2N)}$ and $\tilde{T}_{SO(2N)}$ theories} \label{app:chiralring}
\subsection{$T_{SO(2N)}$}
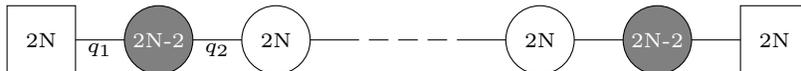
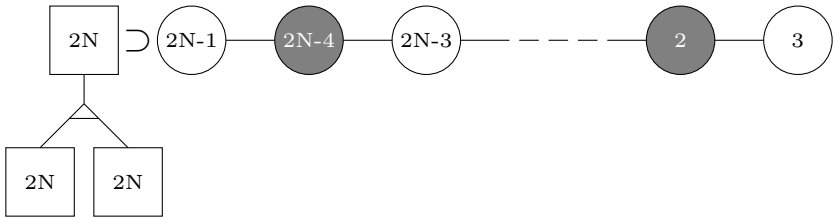
\begin{figure}[h]
\centering
\begin{subfigure}[b]{5in}
\centering
\begin{tikzpicture}[scale=1.3, every node/.style={transform shape}]
\draw (0,0) rectangle (0.70,0.70);
\draw[text = black, font=\tiny] (0.35,0.35) node {2N};
\draw (0.70,0.35)--(1.2,0.35);
\filldraw[fill=gray, draw=black] (1.55,0.35) circle (0.35 cm);
\draw[text = white, font=\tiny] (1.55,0.35) node {2N-2};
\draw (1.9,0.35)--(2.4,0.35);
\draw(2.75,0.35) circle (0.35 cm);
\draw[text = black, font=\tiny] (2.75,0.35) node {2N};
\draw (3.10,0.35) -- ++(0.5,0) ++(0.1,0) -- ++(0.2,0) ++(0.1,0) -- ++(0.2,0) ++(0.1,0) --++(0.2,0) ++(0.1,0) -- ++(0.5,0);
\draw (5.45,0.35) circle (0.35 cm);
\draw[text = black, font=\tiny] (5.45,0.35) node {2N};
\draw (5.8,0.35) -- ++(0.5,0);
\filldraw[fill=gray, draw=black] (6.65,0.35) circle (0.35);
\draw[text = white, font=\tiny] (6.65,0.35) node {2N-2};
\draw (7,0.35) -- ++(0.5,0);
\draw (7.5,0) rectangle (8.2,0.7);
\draw[text = black, font=\tiny] (7.85,0.35) node {2N};
\draw[font=\tiny] (0.95,0.25) node{$q_1$} (2.15,0.25) node{$q_2$};
\end{tikzpicture}
\caption{Linear quiver with $SO(2N)$ ends}
\end{subfigure}

\hfill

\begin{subfigure}[b]{5in}
\centering
\begin{tikzpicture}[scale=1.3, every node/.style={transform shape}]
\draw (0,0) rectangle (0.70,0.70);
\draw[text = black, font=\tiny] (0.35,0.35) node {2N};
\draw (0.35,0.7)-- ++(0.45,0.45);
\draw (0.65,1.0) -- (0.95,1.0);
\draw (0.80,1.15)-- ++(0.45,-0.45);
\draw (0.90,0) rectangle ++(0.70,0.70);
\draw[text = black, font=\tiny] (1.25,0.35) node {2N};
\draw (0.80,1.15) -- (0.80,1.45);
\draw (0.45,1.45) rectangle ++(0.70,0.70);
\draw[text = black, font=\tiny] (0.80,1.80) node {2N};
\draw[text = black] (1.35,1.80) node {$\supset$};
\draw(1.90,1.80) circle (0.35 cm);
\draw[text = black, font=\tiny] (1.90,1.80) node {2N-1};
\draw (2.25,1.80)-- ++(0.5,0);
\filldraw[fill=gray, draw=black] (3.1,1.80) circle (0.35 cm);
\draw[text = white, font=\tiny] (3.1,1.80) node {2N-4};
\draw (3.45,1.80)-- ++(0.5,0);
\draw(4.3,1.8) circle (0.35 cm);
\draw[text = black, font=\tiny] (4.3,1.80) node {2N-3};
\draw (4.65,1.80) -- ++(0.5,0) ++(0.1,0) -- ++(0.2,0) ++(0.1,0) -- ++(0.2,0) ++(0.1,0) --++(0.2,0) ++(0.1,0) -- ++(0.5,0);
\draw [fill=gray, draw=black] (6.9, 1.8) circle (0.35 cm);
\draw[text = white, font=\tiny] (6.9,1.8) node {2};
\draw (7.25,1.8) -- ++(0.5,0);
\draw(8.1,1.8) circle (0.35);
\draw[text = black, font=\tiny] (8.1,1.8) node {3};
\end{tikzpicture}
\caption{Dual frame with $T_{SO(2N)}$ block}
\end{subfigure}
\caption{The linear quiver dual to $T_{SO(2N)}$ coupled to a superconformal tail}
\label{fig:dTSO}
\end{figure}

Consider the $\mathcal{N}=2$ superconformal quiver gauge theory with the gauge groups $$ USp(2N-2) \times SO(2N) \times \cdots \times SO(2N) \times USp(2N-2) \ , $$ with a total of $2N-3$ gauge factors and also $N$ fundamentals at the two end of the quiver, from which we realize the $SO(2N)$ flavor symmetry at each ends. This is dual to a $T_N$ block with $SO(2N)^3$ flavor symmetry, coupled to a superconformal tail given by $$SO(2N-1) \times USp(2N-4) \times SO(2N-2) \times \cdots \times USp(2) \times SO(3) \ . $$ Pictorially we can represent the two dual theories by figure \ref{fig:dTSO}. 

Note that in the dual frame the $SO(2N-1)$ sub-group of one of the three $SO(2N)$ flavor symmetries of the $T_N$ block is gauged while the other two $SO(2N)$ flavor symmetries are in one to one correspondence with flavor symmetries at the ends of the linear quiver.  We thus expect the operator $\mu_{1\alpha \beta}$ transforming in the adjoint representation of $SO(2N)_1$ to be identified with $\Omega_{i j} q^{\phantom{1}i}_{1\alpha} q^{\phantom{1}j}_{1\beta}$ in the linear quiver. Here $\Omega$ is the invariant anti-symmetric form of the $USp(2N-2)$ group. Similarly we can also identify the operator that corresponds to the dual of $\mu_{2 \alpha \beta}$. We now want to establish the chiral ring relation
\begin{equation}
\tr \mu_1^2 = \tr \mu_2^2 \ . 
\end{equation}
To see this note that the $F$-term equation of motion of the linear quiver are given by 
\begin{align}
q_{1\alpha}^{\phantom{1} i} q_{1\alpha}^{\phantom{1} j}+q_{2\beta}^{\phantom{2} i} q_{2\beta}^{\phantom{2} j} &= 0 \ , \nn \\
\Omega_{i j}(q_{2\alpha}^{\phantom{1} i} q_{2\beta}^{\phantom{2} j}+q_{3\alpha}^{\phantom{3} i} q_{3\beta}^{\phantom{3} j}) &= 0 \ , \\
q_{3\alpha}^{\phantom{3} i} q_{3\alpha}^{\phantom{3} j}+q_{4\beta}^{\phantom{4} i} q_{4\beta}^{\phantom{4} j} &= 0 \ , \nn \\
\vdots \notag
\end{align} 
Using these relations we find that 
\begin{align}
\tr \mu_1^2 &=\mu_{1\alpha\beta}\mu_{1\beta\alpha} \nn \\
&= \Omega_{ij}\Omega_{lm}q_{1\alpha}^{\phantom{1} i} q_{1\beta}^{\phantom{2} j}q_{1\beta}^{\phantom{1} l} q_{1\alpha}^{\phantom{2} m} \nn \\
&= \Omega_{ij}\Omega_{lm}q_{1\alpha}^{\phantom{1} i}q_{1\alpha}^{\phantom{2} m}q_{1\beta}^{\phantom{2} j}q_{1\beta}^{\phantom{1} l} \nn \\
&=\Omega_{ij}\Omega_{lm}q_{2\alpha}^{\phantom{2} i} q_{2\alpha}^{\phantom{2} m} q_{2\beta}^{\phantom{2} j} q_{2\beta}^{\phantom{2} l} \nn \\
&=(\Omega_{ij}q_{2\alpha}^{\phantom{2} i} q_{2\beta}^{\phantom{2} j} )(\Omega_{lm}q_{2\beta}^{\phantom{2} l} q_{2\alpha}^{\phantom{2} m})\\
&=\Omega_{ij} \Omega_{lm}q_{3\alpha}^{\phantom{2} i} q_{3\beta}^{\phantom{2} j}q_{3\beta}^{\phantom{2} l} q_{3\alpha}^{\phantom{2} m} \nn \\
&=\Omega_{ij} \Omega_{lm}q_{4\alpha}^{\phantom{2} i} q_{4\beta}^{\phantom{2} j}q_{4\beta}^{\phantom{2} l} q_{4\alpha}^{\phantom{2} m}\nn \\
&= \tr \dot{\mu}^2 \ , \nn 
\end{align}
where $\dot{\mu}_{\alpha\beta}$ is the operator transforming in the adjoint of the $SO(2N)$ gauge group in the linear quiver. Propagating this relation across the quiver we then establish that $\tr \mu_1^2 = \tr \mu_2^2$. By symmetry we thus expect that in the strongly coupled $T_{SO(2N)}$ block the following chiral ring relation holds 
\begin{equation}
\tr \mu_1^2 = \tr \mu_2^2 =  \tr \mu_3^2 \ . 
\end{equation}

\subsection{$\tilde{T}_{SO(2N)}$}

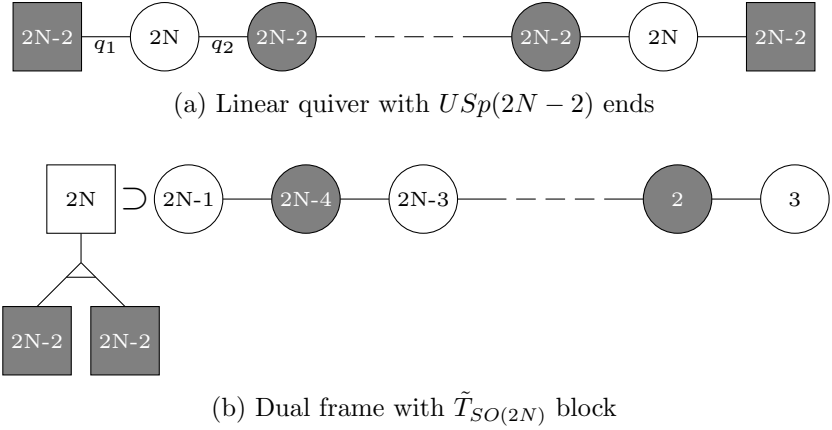
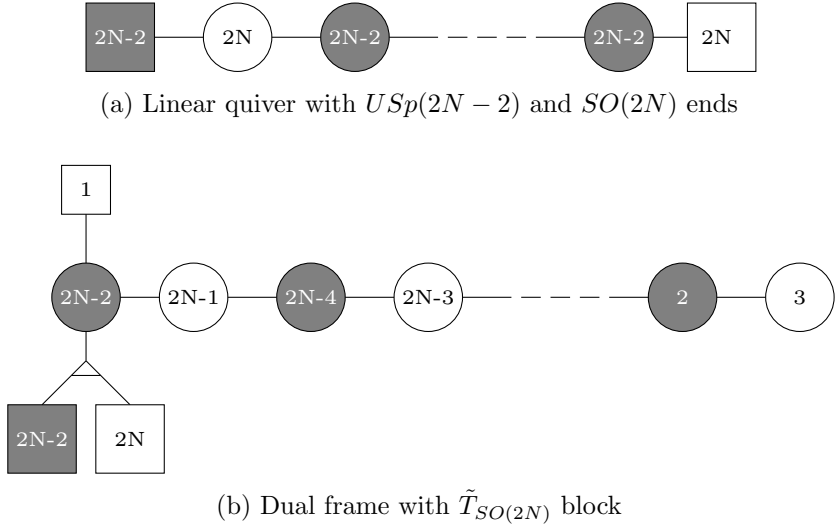
\begin{figure}
\centering
\begin{subfigure}[b]{5in}
\centering
\begin{tikzpicture}[scale=1.3, every node/.style={transform shape}]
\filldraw[fill=gray, draw=black]  (0,0) rectangle (0.70,0.70);
\draw[text = white, font=\tiny] (0.35,0.35) node {2N-2};
\draw (0.70,0.35)--(1.2,0.35);
\filldraw[fill=white, draw=black] (1.55,0.35) circle (0.35 cm);
\draw[text = black, font=\tiny] (1.55,0.35) node {2N};
\draw (1.9,0.35)--(2.4,0.35);
\filldraw[fill=gray, draw=black] (2.75,0.35) circle (0.35 cm);
\draw[text = white, font=\tiny] (2.75,0.35) node {2N-2};
\draw (3.10,0.35) -- ++(0.5,0) ++(0.1,0) -- ++(0.2,0) ++(0.1,0) -- ++(0.2,0) ++(0.1,0) --++(0.2,0) ++(0.1,0) -- ++(0.5,0);
\filldraw[fill=gray, draw=black]  (5.45,0.35) circle (0.35 cm);
\draw[text = white, font=\tiny] (5.45,0.35) node {2N-2};
\draw (5.8,0.35) -- ++(0.5,0);
\filldraw[fill=white, draw=black] (6.65,0.35) circle (0.35);
\draw[text = black, font=\tiny] (6.65,0.35) node {2N};
\draw (7,0.35) -- ++(0.5,0);
\filldraw[fill=gray, draw=black] (7.5,0) rectangle (8.2,0.7);
\draw[text = white, font=\tiny] (7.85,0.35) node {2N-2};
\draw[font=\tiny] (0.95,0.25) node{$q_1$} (2.15,0.25) node{$q_2$};
\end{tikzpicture}
\caption{Linear quiver with $USp(2N-2)$ ends} 
\end{subfigure}

\hfill

\begin{subfigure}[b]{5in}
\centering
\begin{tikzpicture}[scale=1.3, every node/.style={transform shape}]
\filldraw[fill=gray, draw=black]  (0,0) rectangle (0.70,0.70);
\draw[text = white, font=\tiny] (0.35,0.35) node {2N-2};
\draw (0.35,0.7)-- ++(0.45,0.45);
\draw (0.65,1.0) -- (0.95,1.0);
\draw (0.80,1.15)-- ++(0.45,-0.45);
\filldraw[fill=gray, draw=black]  (0.90,0) rectangle ++(0.70,0.70);
\draw[text = white, font=\tiny] (1.25,0.35) node {2N-2};
\draw (0.80,1.15) -- (0.80,1.45);
\filldraw[fill=white, draw=black]  (0.45,1.45) rectangle ++(0.70,0.70);
\draw[text = black, font=\tiny] (0.80,1.80) node {2N};
\draw[text = black] (1.35,1.80) node {$\supset$};
\draw(1.90,1.80) circle (0.35 cm);
\draw[text = black, font=\tiny] (1.90,1.80) node {2N-1};
\draw (2.25,1.80)-- ++(0.5,0);
\filldraw[fill=gray, draw=black] (3.1,1.80) circle (0.35 cm);
\draw[text = white, font=\tiny] (3.1,1.80) node {2N-4};
\draw (3.45,1.80)-- ++(0.5,0);
\draw(4.3,1.8) circle (0.35 cm);
\draw[text = black, font=\tiny] (4.3,1.80) node {2N-3};
\draw (4.65,1.80) -- ++(0.5,0) ++(0.1,0) -- ++(0.2,0) ++(0.1,0) -- ++(0.2,0) ++(0.1,0) --++(0.2,0) ++(0.1,0) -- ++(0.5,0);
\draw [fill=gray, draw=black] (6.9, 1.8) circle (0.35 cm);
\draw[text = white, font=\tiny] (6.9,1.8) node {2};
\draw (7.25,1.8) -- ++(0.5,0);
\draw(8.1,1.8) circle (0.35);
\draw[text = black, font=\tiny] (8.1,1.8) node {3};
\end{tikzpicture}
\caption{Dual frame with $\tilde{T}_{SO(2N)}$ block}
\end{subfigure}
\caption{The linear quiver dual to $\tilde{T}_{SO(2N)}$ coupled to a superconformal tail}
\label{fig:dTSp}
\end{figure}
We now consider the linear quiver given by gauge groups $SO(2N) \times USp(2N-2) \times \cdots \times USp(2N-2) \times SO(2N) $. There are a total of $2N-3$ gauge groups and each end has $USp(2N-2)$ flavor symmetry.  This is dual to a $\tilde{T}_{SO(2N)}$ block coupled to a superconformal tail $SO(2N-1) \times USp(2N-4) \times \cdots \times USp(2) \times SO(3)$ where the $SO(2N-1)$ node of the tail is a sub-group of the $SO(2N)$ flavor symmetry of $\tilde{T}_{SO(2N)}$. See figure \ref{fig:dTSp}. The two $USp(2N-2)$ flavor symmetries of the $\tilde{T}_{SO(2N)}$ block can then be identified with the flavor symmetry at either end of the linear quiver. It is then straight forward to use the $F$-term relations of the linear quiver to establish the chiral ring relation 
\begin{equation}
\tr \Omega \mu_1 \Omega \mu_1 = \tr \Omega \mu_2 \Omega \mu_2 \ , 
\end{equation}
where $\mu_1$ and $\mu_2$ are the dimension 2 operators transforming as the adjoint of $USp(2N-2)$ flavor symmetries of $\tilde{T}_{SO(2N)}$.  

We can also consider the superconformal linear quiver of $2N-2$ nodes given by $SO(2N) \times USp(2N-2) \times \cdots \times USp(2N-2)$. The quiver then ends in a $USp(2N-2)$ flavor symmetry on the left and a $SO(2N)$ symmetry on the right. This theory can be shown to be S-dual to a $\tilde{T}_{SO(2N)}$ block coupled to a superconformal tail whose nodes are $USp(2N-2) \times SO(2N-1) \times USp(2N-4) \times \cdots \times SO(3) $. The $USp(2N-2)$ node of the tail is obtained by gauging one of the two $USp(2N-2)$ flavor symmetries of the $\tilde{T}_{SO(2N)}$ block. We will also need to couple a half-hyper to this node in order to ensure that its $\beta$-function vanishes. These theories can be visualized as in figure \ref{fig:dTSOSP}. 
\begin{figure}
\centering
\begin{subfigure}[b]{5in}
\centering
\begin{tikzpicture}[scale=1.3, every node/.style={transform shape}]
\filldraw[fill=gray, draw=black]  (0,0) rectangle (0.70,0.70);
\draw[text = white, font=\tiny] (0.35,0.35) node {2N-2};
\draw (0.70,0.35)--(1.2,0.35);
\filldraw[fill=white, draw=black] (1.55,0.35) circle (0.35 cm);
\draw[text = black, font=\tiny] (1.55,0.35) node {2N};
\draw (1.9,0.35)--(2.4,0.35);
\filldraw[fill=gray, draw=black] (2.75,0.35) circle (0.35 cm);
\draw[text = white, font=\tiny] (2.75,0.35) node {2N-2};
\draw (3.10,0.35) -- ++(0.5,0) ++(0.1,0) -- ++(0.2,0) ++(0.1,0) -- ++(0.2,0) ++(0.1,0) --++(0.2,0) ++(0.1,0) -- ++(0.5,0);
\filldraw[fill=gray, draw=black]  (5.45,0.35) circle (0.35 cm);
\draw[text = white, font=\tiny] (5.45,0.35) node {2N-2};
\draw (5.8,0.35) -- ++(0.5,0);
\filldraw[fill=white, draw=black] (6.15,0) rectangle ++(0.7, 0.7);
\draw[text = black, font=\tiny] (6.45,0.35) node {2N};
\end{tikzpicture}
\caption{Linear quiver with $USp(2N-2)$ and $SO(2N)$ ends} 
\end{subfigure}

\hfill
\hfill

\begin{subfigure}[b]{5in}
\centering
\begin{tikzpicture}[scale=1.3, every node/.style={transform shape}]
\filldraw[fill=gray, draw=black]  (0,0) rectangle (0.70,0.70);
\draw[text = white, font=\tiny] (0.35,0.35) node {2N-2};
\draw (0.35,0.7)-- ++(0.45,0.45);
\draw (0.65,1.0) -- (0.95,1.0);
\draw (0.80,1.15)-- ++(0.45,-0.45);
\filldraw[fill=white, draw=black]  (0.90,0) rectangle ++(0.70,0.70);
\draw[text = black, font=\tiny] (1.25,0.35) node {2N};
\draw (0.80,1.15) -- (0.80,1.45);
\filldraw[fill=gray, draw=black]  (0.80,1.80) circle (0.35 cm);
\draw[text = white, font=\tiny] (0.80,1.80) node {2N-2};
\draw (0.80,2.15) -- +(0,0.5);
\draw (0.55,2.65) rectangle ++(0.5,0.5);
\draw[text = black, font=\tiny] (0.80,2.9) node {1};
\draw(1.15,1.80)--(1.55,1.80);
\draw(1.90,1.80) circle (0.35 cm);
\draw[text = black, font=\tiny] (1.90,1.80) node {2N-1};
\draw (2.25,1.80)-- ++(0.5,0);
\filldraw[fill=gray, draw=black] (3.1,1.80) circle (0.35 cm);
\draw[text = white, font=\tiny] (3.1,1.80) node {2N-4};
\draw (3.45,1.80)-- ++(0.5,0);
\draw(4.3,1.8) circle (0.35 cm);
\draw[text = black, font=\tiny] (4.3,1.80) node {2N-3};
\draw (4.65,1.80) -- ++(0.5,0) ++(0.1,0) -- ++(0.2,0) ++(0.1,0) -- ++(0.2,0) ++(0.1,0) --++(0.2,0) ++(0.1,0) -- ++(0.5,0);
\draw [fill=gray, draw=black] (6.9, 1.8) circle (0.35 cm);
\draw[text = white, font=\tiny] (6.9,1.8) node {2};
\draw (7.25,1.8) -- ++(0.5,0);
\draw(8.1,1.8) circle (0.35);
\draw[text = black, font=\tiny] (8.1,1.8) node {3};
\end{tikzpicture}
\caption{Dual frame with $\tilde{T}_{SO(2N)}$ block}
\end{subfigure}
\caption{The linear quiver dual to $\tilde{T}_{SO(2N)}$ coupled to a superconformal tail}
\label{fig:dTSOSP}
\end{figure}

Now if $\mu_1^{i j}$ is the dimension 2 operator of  $\tilde{T}_{SO(2N)}$ theory transforming in adjoint representation of $USp(2N-2)$ flavor symmetry while $\mu_{3\alpha\beta}$ is the dim. 2 operator transforming as the adjoint of the $SO(2N)$ flavor symmetry then we identify their duals in the linear quiver to be such that 
\begin{align}
\mu_1^{i j} =&~ q_{1\alpha}^{\phantom{1}i}q_{1\alpha}^{\phantom{1}j} \ , \\
\mu_{3\alpha\beta} =&~ \Omega_{i j}q_{2N-1,\alpha}^{\phantom{1}i}q_{2N-1,\beta}^{\phantom{1}j} \ . 
\end{align}
The $F$-term relations of the linear quiver are 
\begin{align}
\Omega_{i j}(q_{1\alpha}^{\phantom{1} i} q_{1\beta}^{\phantom{1} j}+q_{2\alpha}^{\phantom{2} i} q_{2\beta}^{\phantom{2} j}) &=0 \ , \nn \\
q_{2\alpha}^{\phantom{1} i} q_{2\alpha}^{\phantom{2} j}+q_{3\beta}^{\phantom{3} i} q_{3\beta}^{\phantom{3} j} &=0 \ , \\
\Omega_{i j}(q_{3\alpha}^{\phantom{1} i} q_{3\beta}^{\phantom{1} j}+q_{4\alpha}^{\phantom{2} i} q_{4\beta}^{\phantom{2} j}) &=0 \ ,  \nn  \\
\vdots \notag
\end{align} 
Using these we can then write 
\begin{align}
\tr \Omega \mu_1 \Omega \mu_1& =\Omega_{i j} q_{1\alpha}^{\phantom{1}j}q_{1\alpha}^{\phantom{1}k}\Omega_{k l} q_{1\beta}^{\phantom{1}l}q_{1\beta}^{\phantom{1}i} \nn \\
&= \Omega_{k l}\Omega_{i j} q_{2\beta}^{\phantom{1}i}  q_{2\alpha}^{\phantom{1}j}q_{2\alpha}^{\phantom{1}k}q_{2\beta}^{\phantom{1}l} \nn \\
&= \Omega_{k l}\Omega_{i j} q_{3\beta}^{\phantom{1}i}  q_{3\alpha}^{\phantom{1}j}q_{3\alpha}^{\phantom{1}k}q_{3\beta}^{\phantom{1}l} \nn \\
&= (\Omega_{i j} q_{3\beta}^{\phantom{1}i}  q_{3\alpha}^{\phantom{1}j})(\Omega_{k l}q_{3\alpha}^{\phantom{1}k}q_{3\beta}^{\phantom{1}l})  \\
\vdots \notag \nn \\
&= (\Omega_{i j} q_{2N-1\beta}^{\phantom{1}i}  q_{2N-1\alpha}^{\phantom{1}j})(\Omega_{k l}q_{2N-1\alpha}^{\phantom{1}k}q_{2N-1\beta}^{\phantom{1}l})\nn \\
&= \tr \mu_3^2 \nn \ . 
\end{align}
Thus we establish that for $\tilde{T}_{SO(2N)}$ theories, the following chiral ring relation holds 
\begin{equation}
\tr \Omega \mu_1 \Omega \mu_1 = \tr \Omega \mu_2 \Omega \mu_2 = \tr \mu_3^2 \ . 
\end{equation}

\bibliographystyle{ytphys}
\bibliography{refs}

\end{document}